\overfullrule=0pt

\global\newcount\secno \global\secno=1
\global\newcount\propno \global\propno=1
\global\newcount\eqnum \global\eqnum=1

\def\prop#1{\xdef #1{\the\secno.\the\propno}
            \global\advance\propno by1  }
\def\Eqno#1{\xdef #1{\the\secno.\the\eqnum} \eqno(\the\secno.\the\eqnum)
            \global\advance\eqnum by1
            \eqlabeL #1}

\def\propApp#1{\xdef #1{\App.\the\propno}
            \global\advance\propno by1  }
\def\EqnoApp#1{\xdef #1{\App.\the\eqnum} \eqno(\App.\the\eqnum)
            \global\advance\eqnum by1
            \eqlabeL #1}

\def\proplabeL#1{\leavevmode\vadjust{\llap{\smash
{\line{{\escapechar=` \hfill\llap{\sevenrm \string#1$\;\;$}}}}}} \hskip-6pt}
\def\eqlabeL#1{{\escapechar-1\rlap{\sevenrm\hskip.05in\string#1}}}

\def\proplabeL#1{} 
\def\eqlabeL#1{} 

\def\mapright#1{\smash{\mathop{\longrightarrow}\limits^{#1}}}

\def\mapdown#1{\Big\downarrow\rlap{$\vcenter{\hbox{$\scriptstyle#1$}}$}}

\def\mapright#1{\smash{\mathop{\longrightarrow}\limits^{#1}}}

\def\mapdown#1{\big\downarrow\rlap{$\vcenter{\hbox{$\scriptstyle#1$}}$}}

\def\simrightarrow{\smash{\mathop{\rightarrow}\limits^{\sim}}}

\def\simmaprightarrow#1{\smash{\mathop{\longrightarrow}\limits^{\sim}_{#1}}}
\def\simmapleftarrow#1{\smash{\mathop{\longleftarrow}\limits^{\sim}_{#1}}}

\def\simunder#1{ \; {\;\atop \widetilde{ \;^{^{#1}} }  } \; }

\def\G{\Cal G}

\def\GGamma{\boldsymbol{\Gamma}}

\def\rqed{$\hfill\square$}

\font\bigcmr=cmr10 scaled \magstep 2

\def\mainI{4.10}
\def\mainII{4.11}
\def\mainInp{5.7}
\def\mainIInp{5.8}
\def\summary{5.12}
\def\KLI{4.6}
\def\KLII{5.1}
\input amstex.tex

\def\Proclaim#1#2{ \prop{#2} \proclaim{#1 #2} \proplabeL{#2} }
\def\ProclaimApp#1#2{ \propApp{#2} \proclaim{#1 #2} \proplabeL{#2} }

\documentstyle{amsppt}


\vskip0.3cm

\topmatter
\title \nofrills  \bigcmr
c=2 Rational Toroidal Conformal Field Theories \\
via the Gauss Product
\endtitle
\author \rm
Shinobu Hosono, Bong H. Lian, Keiji Oguiso, Shing-Tung Yau
\endauthor
\address
where is
\endaddress

\abstract  \nofrills {\bf Abstract.}
We find a concise relation between the moduli $\tau, \rho$ of
a rational Narain lattice $\Gamma(\tau,\rho)$ and the corresponding
momentum lattices of left and right chiral algebras via the Gauss product.
As a byproduct, we find an
identity which counts the cardinality of a certain double coset
space defined for isometries between the discriminant forms of rank two
lattices.
\endabstract

\leftheadtext{ S. Hosono, B.H. Lian, K. Oguiso, S.-T. Yau  }
\rightheadtext{$c=2$ rational toroidal CFT}
\endtopmatter

\def\efsize#1{{\eightpoint#1 }}

\centerline{\bf Table of Contents }

{  
\leftskip1cm\rightskip1cm
\item{\S 0.} Introduction and main results
\item{\S 1.} Classical results on quadratic forms
\item{\S 2.} Narain lattices of toroidal compactifications
\item{\S 3.} Rational conformal field theory
\item{\S 4.} $c=2$ rational toroidal CFT -- primitive case
\item{\S 5.} $c=2$ rational toroidal CFT -- non-primitive case
\par\noindent
{Appendix A.} Gauss product on $CL(D)$
\par\noindent
{Appendix B.} $O(A_\Gamma)$ for a primitive lattice $\Gamma$
\par\noindent
{Appendix C.} The coset space $O(d,\bold R)\times O(d,\bold R) \setminus
                   O(d,d;\bold R)$ and $O(d,d;\bold Z)$
\par}

\head
{\bf \S 0. Introduction and main results }
\endhead

\global\secno=0
\global\propno=1
\global\eqnum=1

\noindent
{\bf (0-1) Introduction -- some background --}
Since the 80's,  string
compactifications on real $d$-dimensional tori $T^d$ have been a source of
several important ideas, such as orbifold [DHVW],
T-duality [KY], etc., which later have been successfully generalized to string
compactifications on more non-trivial geometries of Calabi-Yau manifolds
(see, e.g. [GSW][GY][Po] for references).
One thing we learned from these important
developments is that certain properties of string theory (conformal
field theory) often translate into deep and interesting geometrical
insights when interpreted
in the language of $\sigma$-models (geometry). Gepner's correspondence
between N=2 superconformal field theory (SCFT) and $\sigma$-models
on Calabi-Yau manifolds
is one such well-known example. Mirror Symmetry is yet another.
In this paper, we will study an important property, known as
{\it rationality}, of conformal string compactifications on $T^d$.
That rationality translates into
interesting questions on the geometry side was brought
to light in recent works of [Mo][GV].
For example it was found in [Mo] that rationality
of a CFT on an elliptic curve $E_\tau$
$(\cong T^2)$ implies that $E_\tau$ has
non-trivial endomorphisms, i.e. $E_\tau$ is of CM (complex multiplication)
type. Several other deep questions on rationality of the
string compactification on general Calabi-Yau manifolds have also been raised
in [Mo][GV], although their answers are still conjectural.

In this paper we will restrict our attention to CFTs on $T^2$.
In this case, as shown by G. Moore [Mo],
one has the following characterization of rationality:
a CFT on $T^2(=E_\tau)$ is rational if and only if the parameters
$\tau, \rho$ are elements of the imaginary quadratic field
$\bold Q(\sqrt{D})$ for some $D<0$, where $\tau$ is the
complex structure and $\rho=B+\sqrt{-1} Vol(E_\tau)$ is the complexified
K\"ahler modulus. This allows one, in principle, to parameterize rational
toroidal CFTs by such pairs $\tau,\rho$. In this paper,
we further the investigation along this line
by studying precisely {\it how} this parameterization can
be realized. As a result,
we find a precise correspondence between
a Narain lattice $\Gamma(\tau,\rho)$ and
a RCFT triple $(\Gamma_l,\Gamma_r,\varphi)$, which consists of a pair of
momentum lattices for chiral algebras plus a gluing map $\varphi$.
In our view, this is one step toward understanding the geometry of
rational CFTs on $T^2$, and more generally, on Calabi-Yau manifolds.
The correspondence above is realized precisely
by means of the composition law of Gauss on
primitive binary quadratic forms {\it plus} an important extension
to the non-primitive forms. The latter is crucial if one wishes
to allow arbitrary discriminants $D$.

Our result is clearly relevant to the
classification problem of rational CFTs with $c=2$ (see for
example [DW] and references therein for developments on this problem),
which generalizes the old and well-known classification of $c=1$ rational CFTs
given in [Gi][Ki1]. However we restrict ourselves
in this paper to rational CFTs on $T^2$, which are
among the best known and well-studied examples in string
theory (for which we refer the readers to
[HMV][GSW][Po][Ki2] and references therein).

\vskip0.5cm
\noindent
{\bf (0-2) Main results.} Our main results are:

\vskip0.1cm

{\bf Theorem {\mainI}}, {\bf Theorem {\mainII}},
{\bf Theorem {\mainInp}} and {\bf Theorem {\mainIInp}},

\vskip0.1cm
\noindent
where we obtain a complete description of rational toroidal conformal field
theories in terms of the Gauss product on the classes of binary
quadratic forms. The general results are summarized and
restated a little differently in
{\bf Summary {\summary}.}

\vskip0.1cm

In section 1, we briefly summarize the classical theory of
binary quadratic forms
and the Gauss product. The basics of the Gauss product are
summarized in Appendix A, where we extend the product to non-primitive
quadratic forms. In section 2, we introduce Narain lattices
and their moduli space. We also describe the so-called
$T$-duality group. This section is meant for setting up notations
and reviewing some well-known results
(see e.g. [GSW][Po][Ki2] and references therein for the original works).
In section 3, we define RCFTs on $T^2$ (rational toroidal CFT) and summarize
their characterizations given in [HMV]
(see also [Mo][Wa][GV] for a more recent perspective).
We then state our classification problem (after Proposition 3.4),
and study the first part of this problem using some results
of [Ni]. These results were first used in [Mo] to study
of rational toroidal CFTs and rational CFTs
on singular K3 surfaces, i.e. K3 surfaces of maximal Picard 
number 20 [SI]. 
In section 4, we discuss our classification in the case when
the relevant lattices are primitive (which correspond to primitive
quadratic forms). In section 5, we extend it to
the general non-primitive case.
In subsection (5-4), we will see that
the diagonal RCFTs obtained in [GV]
fit into our list of RCFTs in a natural way.
A summary of the classification is given in
{\bf Summary {\summary}.}

In our classification, the classical theory of the binary quadratic
forms of Dirichlet and Gauss will come into
play in an interesting and essential way
({\bf Lemma {\KLI}} and {\bf Lemma {\KLII}}). The reader can get
a quick glance of this
in examples given in subsection (0-3) below (primitive case),
and also in (5-5) (non-primitive case).

\vskip0.7cm
\noindent
{\bf (0-3) Example ($D=-39$).} It will be helpful to
see how our classification works in this example now with
details given later. Here we present two tables:
One is the product table of the class group $Cl(D)$
(cf. Theorem 1.1) of discriminant $D=-39$, and the other is
the table listing the RCFT data $(\Gamma_l,\Gamma_r,\varphi)$ against
the Narain lattices $\Gamma(\tau_{\Cal C_i},\rho_{\Cal C_j})$.

To make the first table, let us note that
the class group $Cl(D)$ here consists of the following four $SL_2 \bold Z$
equivalence classes of binary quadratic forms:
$$
\Cal C_1 =[ Q(1,1,10) ] \;\;,\;\;
\Cal C_2 =[ Q(2,1,5) ] \;\;,\;\;
\Cal C_3 =[ Q(2,-1,5) ] \;\;,\;\;
\Cal C_4 =[ Q(3,3,4) ] \;\;.\;\;
$$
The notation $Q(a,b,c)$ abbreviates the quadratic form
$f(x,y)=ax^2+bxy+cy^2$. A quadratic form can be identified with 
a lattice with a chosen ordered basis in which
the bilinear form is given by the matrix
$\left( \smallmatrix 2a & b \\ b & 2c \\ \endsmallmatrix \right)$.
Under this identification, an $SL_2 \bold Z$ equivalence class 
$\Cal C$ of quadratic forms is nothing but
an isomorphism class of lattices equipped with orientations, while
a $GL_2 \bold Z$ equivalence class $\bar\Cal C$ of quadratic forms
is nothing but an isomorphism class of lattices without orientations.
Now we write the product table of $Cl(D)$:

\def\oneColumn#1{\matrix #1 \hfill \\ \endmatrix}
\def\twoColumn#1#2{\matrix #1 \hfill \\ #2 \hfill \\ \endmatrix}
\def\twoLow#1#2{\matrix #1 & #2 \hfill \\ \endmatrix}

\vskip0.3cm
\centerline{
\vbox{\offinterlineskip
\def\vspa{\omit & \omit &height1pt& \omit && \omit  && \omit &\cr}
\halign{ \strut
       #&  $\;$  $#$  \hfil
&{\vrule# width1pt} &  $\;$ \hfil $#$ \hfil
&      #&  $\;$ \hfil $#$ \hfil
&      #&  $\;$ \hfil $#$ \hfil
&      #&  $\;$ \hfil $#$ \hfil
&      #
\cr
&   &&  \Cal C_1 && \Cal C_2 && \Cal C_3 && \Cal C_4 &\cr
\noalign{\hrule height1pt}
\vspa
&  \Cal C_1
  &&  \Cal C_1
  &&  \Cal C_2
  &&  \Cal C_3
  &&  \Cal C_4 &\cr
\vspa
& \Cal C_2
  &&  \Cal C_2
  &&  \Cal C_4
  &&  \Cal C_1
  &&  \Cal C_3  &\cr
\vspa
& \Cal C_3
  &&  \Cal C_3
  &&  \Cal C_1
  &&  \Cal C_4
  &&  \Cal C_2  &\cr
& \Cal C_4
  &&  \Cal C_4
  &&  \Cal C_3
  &&  \Cal C_2
  &&  \Cal C_1  &\cr
} }   }
\vskip0.5cm
{\leftskip1cm \rightskip1cm
\noindent
{\bf Table 1.}  Table of Gauss product ($D=-39$). \par }

\vskip0.3cm
Table 2 lists the data for the RCFTs on $T^2$. The data that determines
an RCFT consists of its momentum lattices $\Gamma_l,\Gamma_r$
with determinant $-D$, together with an isometry of their discriminant
groups $\varphi$. This isometry ``glues'' together the left and right
sector of the RCFT.
Equivalently, the data $\Gamma_l,\Gamma_r,\varphi$ can also be described
in terms of a Narain lattice $\Gamma(\tau,\rho)$ which contains
$\Gamma_l,\Gamma_r$.
The correspondence between triples $(\Gamma_l, \Gamma_r, \varphi)$ and
Narain lattices $\Gamma(\tau,\rho)$,
$\tau,\rho \in \bold Q(\sqrt{D})$, are shown in Table 2.
The key observation here is that we have $\Cal C=C_i*C_j^{-1},
\Cal C'=\Cal C_i*\Cal C_j$ for the triple $(\Gamma_{\bar \Cal C},
\Gamma_{\bar\Cal C'},\varphi)$ that corresponds to a Narain Lattice
$\Gamma(\tau_{\Cal C_i},\rho_{\Cal C_j})$.

To describe the correspondence more precisely, let us
associate to each quadratic form $Q(a,b,c)$ the complex number
$\tau_{Q(a,b,c)}=\rho_{Q(a,b,c)}=\frac{b+\sqrt{D}}{2a} \;
(D=b^2-4ac)$. Note that since $D<0$, these complex numbers lie in the
upper half plane $\bold H_+$. Given an $SL_2\bold Z$ equivalence class
of quadratic forms $\Cal C=[Q(a,b,c)]$, let $\tau_{\Cal C}=\rho_{\Cal C}$
be the
$SL_2 \bold Z$ orbit of $\tau_{Q(a,b,c)}$.
Given a $GL_2\bold Z$ equivalence class $\bar \Cal C$, we denote by
$\Gamma_{\bar \Cal C}$ a lattice in the corresponding isomorphism
class $\bar \Cal C$ of lattices.
Then Table 2 also describes the data of the RCFTs in terms of
Narain lattices $\Gamma(\tau_{\Cal C},\rho_{\Cal C'})$.

\vskip0.3cm

\centerline{
\vbox{\offinterlineskip
\def\vspa{\omit & \omit &height1pt& \omit && \omit  && \omit &\cr}
\halign{ \strut
       #&  $\;$  $#$  \hfil
&{\vrule# width1pt} &  $\;$ \hfil $#$ \hfil
&      #&  $\;$ \hfil $#$ \hfil
&      #&  $\;$ \hfil $#$ \hfil
&      #
\cr
&   &&  \rho_{\Cal C_1}
    &&  \twoLow{\rho_{\Cal C_2} \hskip1.5cm }{\rho_{\Cal C_3}}
    &&  \rho_{\Cal C_4}
    &\cr
\vspa
\vspa
\noalign{\hrule height1pt}
\vspa
& \oneColumn{\tau_{\Cal C_1}}
  &&  \boxed{\, (\Gamma_{\bar\Cal C_1},\Gamma_{\bar\Cal C_1},id) \,}
  &&  \boxed{\;\;\;\twoLow{(\Gamma_{\bar\Cal C_3},\Gamma_{\bar\Cal C_2},id)}
                 {(\Gamma_{\bar\Cal C_2},\Gamma_{\bar\Cal C_3},id)} \;\;\; }
  &&  \boxed{ \; (\Gamma_{\bar\Cal C_4},\Gamma_{\bar\Cal C_4},id) \;} &\cr
\vspa
& \displaystyle{{\tau_{\Cal C_2}} \atop {\tau_{\Cal C_3}}}
  &&  \boxed{\twoColumn{(\Gamma_{\bar\Cal C_2},\Gamma_{\bar\Cal C_2},id)}
                   {(\Gamma_{\bar\Cal C_3},\Gamma_{\bar\Cal C_3},id)}}
&& \boxed{\matrix {(\Gamma_{\bar\Cal C_1},\Gamma_{\bar\Cal C_4},\varphi_1)} &
      {(\Gamma_{\bar\Cal C_4},\Gamma_{\bar\Cal C_1},\varphi_1^{-1})} \\
      {(\Gamma_{\bar\Cal C_4},\Gamma_{\bar\Cal C_1},\varphi_1^{-1})} &
      {(\Gamma_{\bar\Cal C_1},\Gamma_{\bar\Cal C_4},\varphi_1)} \\ \endmatrix}
  &&  \boxed{
      \twoColumn{(\Gamma_{\bar\Cal C_3},\Gamma_{\bar\Cal C_3},\varphi_2)}
                {(\Gamma_{\bar\Cal C_2},\Gamma_{\bar\Cal C_2},\varphi_2)}}
     &\cr
\vspa
& \oneColumn{\tau_{\Cal C_4}}
  &&  \boxed{\, (\Gamma_{\bar\Cal C_4},\Gamma_{\bar\Cal C_4},id) \,}
  &&  \boxed{\;\; \twoLow{
               (\Gamma_{\bar\Cal C_2},\Gamma_{\bar\Cal C_3},\varphi_2)}
             {(\Gamma_{\bar\Cal C_3},\Gamma_{\bar\Cal C_2},\varphi_2)} \;\;}
  &&  \boxed{\,(\Gamma_{\bar\Cal C_1},\Gamma_{\bar\Cal C_1},\varphi_3) \,}
      &\cr
} }  }

\vskip0.5cm
{\leftskip1cm \rightskip1cm
\noindent
{\bf Table 2.}  Table of RCFT data. RCFT data $(\Gamma_l,\Gamma_r,\varphi)$
are listed against the Narain lattices
$\Gamma(\tau_{\Cal C_i}, \rho_{\Cal C_j})$. Boxed entries define the same
RCFT up to worldsheet parity involution. \par }

\vskip0.3cm

The lattices $\Gamma_{\bar\Cal C_i}$ in Table 2 are not all inequivalent.
In fact, it is easy to verify that there are only three equivalence
classes $\bar\Cal C_1,
\bar{\Cal C_2}=\bar{\Cal C_3}, \bar\Cal C_4$. For brevity, we
do not describe here the gluing data
$\varphi_1, \varphi_2,\varphi_3$ explicitly, but we will
discuss their general construction later.

Note that the first column (or row) corresponds
to the so-called diagonal modular invariants of RCFT, whose characterization
has been obtained recently in [GV](see section (5.4)).
We observe that this fits naturally into our general
classification (see Proposition 5.7).

\vskip0.7cm
\noindent
{\bf (0-4) Average one formula.}
As a corollary
to our Main Theorems, we obtain the following ``average one'' formula
for definite lattices. This formula connects, in an interesting way,
lattice problems arising from RCFTs to the class group of binary
quadratic forms (see Corollary 4.13).

\Proclaim{Theorem}{\averageOne} Let $\Cal L^p(D)$ be the set of isomorphism
classes of primitive, definite, even, integral lattices of
determinant $-D$ and rank 2. Also let
$A_\Gamma =(\Gamma^*/\Gamma, q_\Gamma)$ be the
discriminant group $\Gamma^*/\Gamma$ equipped with the quadratic form
$q_\Gamma: \Gamma^*/\Gamma \rightarrow \bold Q/2\bold Z$. Then the following
formula holds:
$$
\frac{1}{|\text{Sym}^2 \Cal L^p(D)|}
\sum_{(\Gamma,\Gamma')\in \text{Sym}^2 \Cal L^p(D)}
\left| O(\Gamma)\setminus \text{Isom}(A_\Gamma,A_{\Gamma'}) /O(\Gamma')
\right|
=1 \;\;,
\Eqno{\averageOneFormula}
$$
where $ \text{Isom}(A_\Gamma,A_{\Gamma'})$ is the set of isometries
$\varphi: A_\Gamma \;\simrightarrow\; A_{\Gamma'}$ and the double quotient is
defined by the natural actions $\varphi \mapsto \bar h\cdot
\varphi \cdot \bar g^{-1}$ of lattice isometries $g \in O(\Gamma),\;
h \in O(\Gamma')$.
\endproclaim

\noindent
{\bf Remark.} Here $\text{Sym}^2 \Cal L^p(D)$ denotes the set of symmetric
pairs of elements of $\Cal L^p(D)$.
The set $\Cal L^p(D)$ is identified with the set $\widetilde{Cl}(D)$ of
$GL_2\bold Z$ equivalence classes of the quadratic forms.
When the determinant $-D(>2)$ is a prime,
the elements of $\widetilde{Cl}(D)\cong \Cal L^p(D)$ are all isogeneous (see
e.g. [Za, \S 12]), and we have $\text{Isom}(A_\Gamma,A_{\Gamma'})=\{\pm 1\}$.
Therefore $|O(\Gamma)\setminus \text{Isom}(A_\Gamma,A_{\Gamma'})/
O(\Gamma')|=1$ for all $(\Gamma,\Gamma') \in Sym^2\Cal L^p(D)$.
That verifies (\averageOneFormula) immediately in this special case.
In general  $\widetilde{Cl}(D)\cong \Cal L^p(D)$ contains more than one
isogeny classes, in which case
the ``average one formula'' is
a very interesting generalization.
Note that $\text{Isom}(A_\Gamma,A_{\Gamma'})$ is empty
if $\Gamma$ and $\Gamma'$ are not isogeneous (see [Ni]).
The average one formula (\averageOneFormula)
for indefinite lattices $\Cal L^p(D)\,(D>0)$ also follows from the
counting problem of general {\it $V$-rational} Narain lattices (see
[HLOY2]).

\vskip0.3cm
\noindent
{\bf Acknowledgements.} 
We would like to thank B. Gross for interesting discussions
on quadratic forms.  We thank S. Gukov, A. Strominger,
and C. Vafa for helpful communications. The first named author would
like to thank K. Wendland for discussions and sending him her 
Ph.D thesis. The first and third named authors also
thank the Department of Mathematics of Harvard University,
where this work started, for
its hospitality and financial support during their visit.
We also would like to thank the referee for his critical reading of
the manuscript.

\vskip1cm
\head
{\bf \S 1. Classical results on quadratic forms }
\endhead

\global\secno=1
\global\propno=1
\global\eqnum=1

In this section we summarize some classical results on (positive definite)
binary quadratic forms following [Za][Ca].
These results turn out to be crucial for our classification problem.

\vskip0.7cm
\noindent{\bf (1-1) Quadratic forms.} Let us denote by $Q(a,b,c)$ an
integral quadratic form in two variables;
$$
Q(a,b,c): f(x,y)=a x^2 + b x y + c y^2 \quad (a,b,c \in \bold Z).
$$
Unless stated otherwise, a quadratic form will mean
a positive definite, integral, quadratic form
in two variables. A quadratic form $Q(a,b,c)$ is
called {\it primitive} if $gcd(a,b,c)=1$. The group $SL_2\bold Z$ acts on
quadratic forms $A:Q(a,b,c)\mapsto Q(a',b',c')$, $(A \in SL_2\bold Z)$, by
$$
\,^tA \left( \matrix
2 a & b \cr
b & 2c  \cr \endmatrix\right) A
=
\left(\matrix
2 a' & b' \cr
b' & 2c'  \cr \endmatrix\right)  \;\;.
\Eqno{\GLtwoTransf}
$$
Since $A=\pm id$ acts trivially, only the group $PSL_2\bold Z$ acts
effectively.
The action leaves the discriminant $D:=b^2-4ac (<0)$ unchanged.
Two quadratic forms are said to be {\it properly equivalent} if
they are in the same $SL_2\bold Z$ orbit.
We consider the set of properly equivalent classes and denote it by
$$
Cl(D):=\{\; Q(a,b,c): \text{primitive quadratic form},
D=b^2-4ac<0, \; a>0 \; \}/
\sim_{ SL_2 \bold Z }.
$$
It has been known, since Gauss, that $Cl(D)$ is a finite set.
Its cardinality is called a {\it class number}, denoted by $h(D)=|Cl(D)|$.
We often write the set of classes by
$$
Cl(D)=\{ \Cal C_1, \Cal C_2, \cdots, \Cal C_{h(D)} \} \;\;.
$$

Similarly the group $GL_2\bold Z$ acts on quadratic forms by the same 
formula as (\GLtwoTransf), and two quadratic
forms are said to be {\it improperly equivalent} if they are in the same
$GL_2\bold Z$ orbit. We consider the set of improperly equivalent classes
and denote it by
$$
\widetilde{Cl}(D):=\{ \; Q(a,b,c);
\text{ primitive  quadratic form}, D=b^2-4ac<0, \; a>0
\;\}/\sim_{GL_2\bold Z}.
$$
There is obviously a natural surjection
$q: Cl(D) \rightarrow \widetilde{Cl}(D)$,
$\Cal C\mapsto\bar\Cal C$, and $q^{-1}(\bar\Cal C)$ has either one or
two classes.

\vskip0.7cm
\noindent
{\bf (1-2) Even lattices of rank two.}
An abstract (integral) lattice $\Gamma=(\Gamma, (*,*))$ is
a $\bold Z$-module equipped with
non-degenerate bilinear form $(*,*): \Gamma \times \Gamma \rightarrow
\bold Z$. Let $\Gamma$ be a lattice.
It is called {\it even} (respectively positive definite) if
$( x,x ) \in 2\bold Z$  ($x\neq 0 \Rightarrow (x,x) >0 $)
for $x \in \Gamma$.
We denote by $\Gamma(n)$ the lattice whose
bilinear form is given by $n$-times the bilinear form of $\Gamma$, i.e.
$\Gamma(n)=(\Gamma, n(*,*))$. We say that $\Gamma$ is {\it primitive}
if $\Gamma=\Gamma'(n)$ for some even lattice
$\Gamma'$ implies that $n= \pm1$.
We denote the dual lattice $\Gamma^*(:=\text{Hom}(\Gamma,\bold Z))$ 
and have $|\text{det}\, \Gamma|=|\Gamma^*/\Gamma|$, which is called the
determinant of the lattice $\Gamma$. We consider the following set of
isomorphism classes of lattices;
$$
\align
\Cal L^p(D):= \{\; \Gamma: &\text{ even, primitive, positive definite
lattices of rank two} \\
& \text{ and } D=- \text{det} \Gamma  \;\}/GL_2 \bold Z
\endalign
$$
In this paper, both $\GGamma$, $[\Gamma]$ denote the isomorphism class
of a lattice $\Gamma$.

For an even lattice $\Gamma$ of determinant $-D$ and rank two, choosing a
basis $u_1,u_2$, we may associate a symmetric matrix,
$$
\left(\matrix
2 a & b \cr
b & 2c  \cr \endmatrix \right)
:=
\left(\matrix (u_1,u_1) & (u_1,u_2) \cr
              (u_2,u_1) & (u_2, u_2) \cr \endmatrix\right) \;\;,
\Eqno{\Lintersec}
$$
with $a,b,c \in \bold Z$. If $\Gamma$ is positive definite, then
$4ac-b^2=-D>0$ and $a>0$. Moreover if $\Gamma$ is primitive, then
$gcd(a,b,c)=1$. Therefore for an even, positive definite, primitive
lattice $\Gamma$, we may associate a primitive quadratic form $Q(a,b,c)$
by choosing a basis of $\Gamma$. It is clear that changing the basis
$u_1, u_2$ results in a quadratic form which is $GL_2\bold Z$ equivalent
to $Q(a,b,c)$.  Similarly, since isomorphic lattices have the same
$GL_2\bold Z$ orbit of the matrix (\Lintersec), they also correspond to
the same $GL_2\bold Z$ equivalence class of quadratic forms. Conversely,
it is also clear that a $GL_2\bold Z$ equivalence class $\bar\Cal C=
\overline{[Q(a,b,c)]} \in \widetilde{Cl}(D)$ defines a unique
isomorphism class of
lattices $\Gamma=\bold Z u_1\oplus \bold Z u_2$ by (\Lintersec).
Therefore we can identify, and will do so hereafter, the set
$\Cal L^p(D)$ with the set $\widetilde{Cl}(D)$ under the natural
one to one correspondence
$$
\bar\Cal C=\overline{[Q(a,b,c)]} \leftrightarrow \GGamma=[\Gamma]
\text{ such that } (\Lintersec) \text{ holds.}
\Eqno{\QtoLattice}
$$
By this correspondence we often write explicitly,
$$
\widetilde{Cl}(D)=\{ \GGamma_1,\GGamma_2,\cdots,\GGamma_{\widetilde h(D)} \}
\qquad (\widetilde h(D)=|\widetilde{Cl}(D)|) .
$$
Also it will be useful to write $\widetilde{Cl}(D)=\{\, q(\Cal C) |
\Cal C \in Cl(D) \,\}$, in terms of the natural surjective map
$q: Cl(D) \rightarrow \widetilde{Cl}(D)$.

\vskip0.7cm
\noindent{\bf (1-3) Class group.} We summarize the following nice property
of the set $Cl(D)$:

\Proclaim{Theorem}{\gaussCl}{\bf (Gauss)} The set $Cl(D)$ has
a commutative, associative, composition law (Gauss product) which makes $Cl(D)$
an abelian group with the unit being the class represented
by $Q(1,0,-\frac{D}{4}): x^2-\frac{D}{4}y^2$ for $D \equiv 0 \text{mod } 4$
and $Q(1,1,\frac{1-D}{4}): x^2+xy+\frac{1-D}{4}y^2$ for $D\equiv 1
\text{mod } 4$.
\endproclaim

Note that since $D$ has the shape $b^2-4ac$, we have $D \equiv 0$ or $1$
mod $4$.
The precise definition of the Gauss product is summarized in Appendix A.
As it is explained there (Proposition A.7), for any given
class $\Cal C\in Cl(D)$, we have
$$
q^{-1}(q(\Cal C))=\{ \, \Cal C, \Cal C^{-1} \,\}
\;\;.
\Eqno{\fiberC}
$$
Hence if $\Cal C^{-1}\not= \Cal C$, then $\Cal C^{-1}$ and
$\Cal C$ are improperly equivalent, and thus we have the following relation:
$$
\widetilde h(D)=\frac{1}{2}\left( h(D)+\# \{ \Cal C \in Cl(D) \;|\;
\Cal C^{-1}= \Cal C \;\} \right) \;\;.
\Eqno{\cardTildhD}
$$

\vskip0.7cm
\noindent
{\bf (1-4) Non-primitive quadratic forms.}
In our classification problem, lattices which are not necessarily
primitive will appear. We define the following sets of equivalence classes:
$$
CL(D):=\{\; Q(a,b,c): \text{quadratic form},
D=b^2-4ac<0, \; a>0  \; \}/
\sim_{ SL_2 \bold Z } \;\;,
$$
and
$$
\widetilde{CL}(D):=\{\; Q(a,b,c): \text{quadratic form},
D=b^2-4ac<0, \; a>0 \; \}/
\sim_{GL_2 \bold Z } \;\;.
$$
Obviously these two sets respectively include $Cl(D)$ and
$\widetilde{Cl}(D)$. If $f(x,y)$ is a quadratic form of
discriminant $D$, then $f(x,y)=\lambda g(x,y)$ for some positive integer
$\lambda$,
and some primitive quadratic
form $g(x,y)$ of discriminant $D/\lambda^2$. From finiteness of
$Cl(D/\lambda^2)$, it follows that $CL(D)$ is finite.
As before, we have a natural
correspondence between improper equivalence classes of
quadratic forms and isomorphism classes of lattices.
Under this correspondence, we
identify the set $\widetilde{CL}(D)$ with
$$
\Cal L(D):=  \{\; \Gamma: \text{even, positive definite
lattice of rank two and } -D=|\Gamma^*/\Gamma|  \;\}/\text{isom.}\,.
$$
There is also a natural map $q:CL(D) \rightarrow \widetilde{CL}(D)$, and
we have
$$
q^{-1}(q(C))=\{ \, C, \sigma C \,\} \;\;,
$$
where $\sigma \in GL_2\bold Z$ is an involution which is not
in $SL_2\bold Z$.  In this paper, we fix the involution $\sigma$ so
that it acts on quadratic forms by
$$
\sigma: Q(a,b,c) \mapsto Q(a,-b,c) \;\;.
\Eqno{\invsigma}
$$
Note that if $C\in Cl(D)$, then
$\sigma C=C^{-1}$ (see Proposition A.7 in Appendix A).
As in (\cardTildhD), we have:
$$
|\widetilde{CL}(D)|=\frac{1}{2} \left(
|CL(D)|+\# \{ C \in CL(D) \,|\, \sigma C=C \,\} \right) \;\;.
\Eqno{\cardCLD}
$$

\vskip0.7cm
\noindent
{\bf (1-5) Reduced forms.}
Each class $\Cal C\in CL(D)$ has a special representative, which is
called {\it the reduced form}. To describe its construction, let us
consider an element
$$
S_n=\left( \matrix n & 1 \cr -1 & 0 \cr \endmatrix \right) \in SL_2\bold Z .
$$
By (\GLtwoTransf) we have
$$
S_0: Q(a,b,c) \mapsto Q(c,-b,a) \;\;,\;\;
S_nS_0: Q(a,b,c) \mapsto Q(a,b-2na,c-nb+n^2a) \;\;.
$$
Using these two operations, we see that any given class $C\in CL(D)$
contains a quadratic form $Q(a,b,c)$ with $a \leq c$ and $-a < b \leq a$, and
for which we have $-D=4ac-b^2 \geq 4a^2-a^2=3a^2$, i.e.
$$
0< a \leq \sqrt{-D/3}  \;\;.
$$
Moreover, if $a=c$, we can make $0 \leq b \leq a=c$.
To summarize we have

\Proclaim{Proposition}{\reducedF} The set $CL(D)$ is finite.
Each class $C\in CL(D)$ can be represented by a unique form
$Q(a,b,c)$, called the reduced form, satisfying
\item{1)}
$ -a<b\leq a \;,\;     0<a\leq \sqrt{-D/3} \;,\;
 c=\frac{b^2-D}{4a} \in \bold Z \;, \;\; a<c \;$, or
\item{2)}
$0\leq b \leq a \;, \; 0<a\leq \sqrt{-D/3} \;,\;
a=c=\frac{b^2-D}{4a}  \;\;$.
\endproclaim

\vskip0.3cm

Note that the uniqueness of the reduced form $Q(a,b,c) \in \Cal C$ for
a given class $\Cal C \in CL(D)$ is clear from the fact that $SL_2\bold Z$
is generated by $S_0$ and $S_1$.

\vskip1cm
\head {\bf \S 2. Narain lattices and toroidal compactifications}
\endhead

\global\secno=2 \global\propno=2 \global\eqnum=2

\noindent {\bf (2-1) Narain lattice.} Let $\bold R^{d,d}$ be the
vector space $\bold R^{2d}$ equipped with the bilinear form
$\langle x,x \rangle_{\bold R^{d,d}}=
x_1^2+\cdots+x_d^2-x_{d+1}^2-\cdots-x_{2d}^2$, of signature
$(d,d)$. Let $\bold R^{d,0}\subset\bold R^{d,d}$ be the subspace
of vectors $x$ with $x_{d+1}=\cdots=x_{2d}=0$. Likewise for $\bold
R^{0,d}$. Then we have an orthogonal decomposition with respect to 
$\langle \;,\,\rangle_{\bold R^{d,d}}$:
$$
\bold R^{d,d}=\bold R^{d,0}\oplus \bold R^{0,d} \; .
\Eqno{\fixRdd}
$$
Let $U$ be the rank two lattice $\bold Z e\oplus\bold Z f$ with
bilinear form given by $\langle e,e\rangle=\langle f,f\rangle=0$,
$\langle e,f\rangle=1$.

\Proclaim{Definition}{\NarainL} {\bf (Narain lattice)}
\item{1)} A Narain lattice is a subgroup $\Gamma\subset\bold R^{d,d}$
of rank $2d$
such that $\langle,\rangle|_\Gamma$ is even unimodular. A Narain
embedding is an isometric embedding
$$
\Phi : U^{\oplus d} \hookrightarrow \bold R^{d,d} \;\;.
$$
\item{2)} Two Narain lattices $\Gamma,\Gamma'$ are said to be equivalent
if $\Gamma'=g\Gamma$ for some $g\in O(d,d;\bold R)$ preserving the
decomposition (\fixRdd).
\endproclaim

Since every abstract even unimodular lattice of signature $(d,d)$
is isomorphic to $U^{\oplus d}$, every Narain lattice is the image
of a Narain embedding.

\Proclaim{Definition}{\CFTonTd} {\bf (Partition function of CFT on
$T^d$)} Given a Narain lattice $\Gamma$, we define its partition
function
$$
Z^{\Gamma}(q,\bar q)={1\over \eta(q)^d \bar \eta(q)^d }
\sum_{p=p_l+p_r \in \Gamma}
q^{{1\over2}|p_l^2|} \bar q^{{1\over2}|p_r^2|} \;,
$$
and its parity invariant form
$$
\tilde Z^{\Gamma}(q,\bar q)={1\over \eta(q)^d \bar \eta(q)^d }
\sum_{p=p_l+p_r \in \Gamma} \frac{1}{2} \left(
q^{{1\over2}|p_l^2|} \bar q^{{1\over2}|p_r^2|}  +
\bar q^{{1\over2}|p_l^2|} q^{{1\over2}|p_r^2|}  \right)\;,
$$
where $p=p_l+p_r$ is the decomposition according to ({\fixRdd})
and $|p_l^2|=
|\langle p_l,p_l \rangle_{\bold R^{2,2}}|$
$=|\langle p_l,p_l \rangle_{\bold R^{2,0}}|$,
$|p_r^2|=|\langle p_r,p_r \rangle_{\bold R^{2,2}}|$
$=|\langle p_r,p_r \rangle_{\bold R^{0,2}}|$ .
$\eta(q)=q^{1\over24}\prod_{n>0} (1-q^n) \; (q=e^{2\pi
\sqrt{-1}t}$ with $t$ in the upper half plane) is the Dedekind
eta function.
\endproclaim

\noindent {\bf Remark.} 1) If $g\in O(d,d;\bold R)$ preserves the
decomposition (\fixRdd), then it is obvious that $\Gamma,g\Gamma$
have the same partition function, for any Narain lattice $\Gamma$.
In other words, equivalent Narain lattices have the same partition
function.

\noindent
2) In [Na], the partition function is computed by means of a path
integral in a sigma model with target space a flat torus $T^d$. A
Narain lattice $\Gamma$ plays the role of the momentum-winding lattice.
The partition function is modular invariant, i.e. invariant under
$t\rightarrow t+1$ and $t\rightarrow -1/t$, as a result of that
$\Gamma$ is even and unimodular.  See, e.g., [Po, \S 8.4, p.252]
for a proof of modular invariance. We will return to modular
invariance later.

\noindent
3) We introduce the parity invariant form $\tilde Z^\Gamma(q,\bar q)$
since this fits well in our diagram given in Theorem 4.10, and also
the bijection that we will prove in Proposition 4.12. The parity invariance
refers to the invariance under the worldsheet parity involution (which
is equivalent to the involution
$\pi_2: (\tau,\rho) \mapsto (\tau,-\bar\rho)$, cf. Lemma 4.9).

\vskip0.7cm
\noindent
{\bf (2-2) Moduli space of Narain lattices.}
Here we summarize some known facts about the space which
parametrizes the equivalence classes of Narain lattices. This
space is known as {\it a Narain moduli space}.

Let us define the group $O(d,d;\bold R)$ explicitly by
$$
O(d,d;\bold R)=\bigg\{\, X \in Mat(2d,\bold R) \,|\, \,^tX \left(
\matrix \bold 1_d & 0 \cr
                            0  & -\bold 1_d \cr \endmatrix \right) X
=\left( \matrix \bold 1_d & 0 \cr
                            0  & -\bold 1_d \cr \endmatrix \right) \;\bigg\},
\Eqno{\OrthoR}
$$
where $\bold 1_d$ represents the unit matrix of size $d$. Let
$O(d;\bold R)\times O(d;\bold R)\subset O(d,d;\bold R)$ denote the
subgroup preserving the decomposition (\fixRdd). Put
$$
\bold E:=\frac{1}{\sqrt2}\left(\matrix  \bold 1_d & \bold 1_d \cr
                    \bold 1_d & -\bold 1_d \cr \endmatrix \right).
$$
Let $\Gamma_0$ be the $\bold Z$-span of the column vectors of
$\bold E$. Then we may verify that
$\Gamma_0$ is even unimodular of rank $2d$ in
$\bold R^{d,d}$, hence a Narain lattice. Let $O'(d,d;\bold
Z)\subset O(d,d;\bold R)$ denote the subgroup preserving
$\Gamma_0$. Note that $\bold E^{-1}=\bold E$. Then, 
it is straightforward to determine the subgroup to be
$$
O'(d,d;\bold Z)=\bold E O(d,d;\bold Z) \bold E \;,
$$ 
with
$$
O(d,d;\bold Z):=\bigg\{\, Y \in Mat(2d,\bold Z) \,|\,
\,^tY \left( \matrix      0 & \bold 1_d  \cr
                    \bold 1_d & 0 \cr \endmatrix \right) Y
=\left( \matrix 0 & \bold 1_d  \cr
               \bold 1_d & 0  \cr \endmatrix \right) \;\bigg\}.
$$

\Proclaim{Proposition}{\ModuliNarain} {\bf (Narain moduli space)}
The equivalence classes of Narain lattices
are parametrized by the coset space
$$\
\Cal M_{d,d}:=O(d;\bold R)\times O(d;\bold R) \setminus
O(d,d;\bold R) / O'(d,d;\bold Z)  \;\;. \Eqno{\ModuliM}
$$
\endproclaim

\demo{Proof} Given a Narain lattice $\Gamma\subset\bold R^{d,d}$,
there exists an isomorphism $g:\Gamma_0\rightarrow\Gamma$, since
even, unimodular indefinite lattices are unique up to isomorphisms.
This extends uniquely to an isometry of $\bold R^{d,d}$
which we denote by the same
$g\in O(d,d;\bold R)$. We associate to $\Gamma$, the left coset
$g O'(d,d;\bold Z)\in O(d,d;\bold R)/O'(d,d;\bold Z)$. If
$g':\Gamma_0\rightarrow\Gamma$ is another isomorphism, then
$g^{-1}\cdot g'$ is an isometry of $\Gamma_0$, i.e. an element of
$O'(d,d;\bold Z)$. It follows that $g'O'(d,d;\bold Z)=gO'(d,d;\bold
Z)$. This shows that the correspondence $\Gamma\mapsto
gO'(d,d;\bold Z)$ is well-defined.

This correspondence is 1-1: if $\Gamma\mapsto gO'(d,d;\bold Z)$ and
$\Gamma'\mapsto g'O'(d,d;\bold Z)$ for some $g,g'\in O(d,d;\bold
R)$ with $g'O'(d,d;\bold Z)=gO'(d,d;\bold Z)$, then $g^{-1}\cdot
g'\in O'(d,d;\bold Z)$, i.e. $g^{-1}\cdot g'\Gamma_0=\Gamma_0$. It
follows that $\Gamma'=g'\Gamma_0=g\Gamma_0=\Gamma$. The
correspondence is also onto: if $g\in O(d,d;\bold R)$, then
$\Gamma=g\Gamma_0$ is a Narain lattice with $\Gamma\mapsto
gO'(d,d;\bold Z)$. This shows that $O(d,d;\bold R)/O'(d,d;\bold Z)$
parameterizes all Narain lattices.

Let $\Gamma$ be a Narain lattice, and let $\Gamma\mapsto
gO'(d,d;\bold Z)$. If $h\in O(d;\bold R)\times O(d;\bold R)$, then
$hg:\Gamma_0\rightarrow h\Gamma$ is an isomorphism, and so
$h\Gamma\mapsto hgO'(d,d;\bold Z)$. This shows that the
correspondence $\Gamma\mapsto gO'(d,d;\bold Z)$ is compatible with
the left action of $O(d;\bold R)\times O(d;\bold R)$. This induces
a 1-1 correspondence between equivalence classes of Narain
lattices and $\Cal M_{d,d}$. \rqed
\enddemo

\noindent {\bf Remark.} We introduce some notations here.

\noindent
1) Introduce the following  conjugate of $O(d,d;\bold R)$,
$$
O'(d,d;\bold R):=
\bold E \, O(d,d; \bold R) \, \bold E \;\;.
\Eqno{\twoOrtho}
$$

\noindent
2) Put $e_i=(0,..,0,e,0,..,0),~f_i=(0,..,0,f,0,..,0)\in U^{\oplus
d}$, where $e,f$ lies in the $i$-th slot. Given a Narain embedding
$\Phi: U^{\oplus d}\hookrightarrow \bold R^{d,d}$, we define the following 
matrix:
$$
\aligned W(\Phi)&:=\frac{1}{\sqrt{2}} \left( \Phi(e_1+f_1) \;
\cdots \; \Phi(e_d+f_d) \; \Phi(e_1-f_1) \; \cdots \;
\Phi(e_d-f_d) \right)  \cr &= \left( \Phi(e_1) \cdots \Phi(e_d) \;
\Phi(f_1) \cdots \Phi(f_d) \right)~\bold E.
\endaligned
\Eqno{\OrthoPhi}
$$
It is easy to check that $W(\Phi)\in O(d,d; \bold R)$. Conversely,
for any $X\in O(d,d;\bold R)$, there is a unique Narain embedding
$\Phi$ such that $W(\Phi)=X$. In other words, $O(d,d;\bold R)$
parameterizes all Narain embeddings by the identification $\Phi\equiv W(\Phi)$.

\noindent
3) Likewise, $O(d,d;\bold R)/O'(d,d;\bold Z)$
parameterizes all Narain lattices $\Gamma=\text{Im}\,\Phi$
by the identification $\Gamma\equiv W(\Phi)O'(d,d;\bold Z)$.
\rqed

\vskip0.5cm \noindent {\bf Example ($d=1$).}
We have
$$
O'(1,1;\bold R)=\bigg\{ \; \left( \matrix R & 0 \cr 0 & 1/R \cr
\endmatrix \right) \;| \; R\not=0 \;\bigg\} \sqcup \bigg\{ \;
\left( \matrix 0 & R  \cr 1/R  & 0 \cr \endmatrix \right) \;| \;
R\not=0 \;\bigg\} , \Eqno{\OpOne}
$$
where the each factor consists of two connected components. After
the conjugation by (\twoOrtho), the subgroup
$O(1;\bold R)\times O(1;\bold R) \subset O(1,1;\bold R)$ becomes
a subgroup $G\times G$ in $O'(1,1;\bold R)$ consisting of
the four elements:
$\pm \left( \smallmatrix 1 & 0 \cr 0 & 1 \cr \endsmallmatrix
\right)$, $\pm \left( \smallmatrix 0 & 1 \cr 1 & 0 \cr
\endsmallmatrix \right)$. Then we get
$$
G\times G \setminus O'(1,1;\bold R) = \bigg\{ G \times G \cdot
 \left( \matrix R & 0 \cr 0 & 1/R \cr \endmatrix \right) \;\big|\;
R>0 \; \bigg\}. \Eqno{\orbitOne}
$$
Thus the four components of $O'(1,1;\bold R)$ are collapsed to
one. The subgroup $O(1,1;\bold Z) \subset
O'(1,1;\bold R)$ consists of four elements obtained from (\OpOne) by
setting $R=\pm 1$. It is easy to check that each of them acts on
(\orbitOne) either trivially or by $R \rightarrow 1/R$, which is a
well-known duality transformation. This shows that the Narain
moduli space $\Cal M_{1,1}$ has the cross section $\{\bold E\left(
\smallmatrix R & 0 \cr 0 & 1/R \cr \endsmallmatrix \right)\bold E
\;|\;1\geq R>0 \} \subset O(1,1;\bold R)$. By Proposition
\ModuliNarain, we get all the (inequivalent) Narain lattices
$\Gamma$ by letting this set act on the fixed Narain lattice
$\Gamma_0$.
Since $\Gamma_0$ is spanned by the columns of the matrix $\bold
E$, it follows that such a $\Gamma$ is spanned by the columns of the
matrix
$$
\bold E\left( \smallmatrix R & 0 \cr 0 & 1/R \cr \endsmallmatrix
\right)\bold E \cdot \bold E = \frac{1}{\sqrt2} \left( \matrix R & 1/R
\cr R & -1/R \cr \endmatrix \right) .
$$
By (\fixRdd), each vector $p\in\Gamma$ decomposes over $\bold R$ as
$$
p= \frac{1}{\sqrt2}\left( \matrix m R + n/R \cr mR - n/R \cr
\endmatrix \right) = \frac{1}{\sqrt2}\left( \matrix m R + n/R \cr
0 \cr \endmatrix \right) + \frac{1}{\sqrt2}\left( \matrix 0 \cr mR
- n/R \cr \endmatrix \right),
$$
where $m,n\in\bold Z$. Finally, the partition function of $\Gamma$
is
$$
Z^{\Gamma}(q,\bar q)={1\over \eta(q) \overline{{\eta(q)}} }
\sum_{m,n \in \bold Z} q^{{1\over4}({m\over R}+n R)^2} \bar
q^{{1\over4}({m\over R}-n R)^2} \;\;.
$$
This is the well-known partition function of toroidal
compactification on $T^1=S^1$ with radius $R$. \rqed

\vskip0.7cm
\noindent
{\bf (2-3) Another parameterization of Narain lattices.}
There is a convenient parameterization of the
homogeneous space $O(d;\bold R)\times O(d; \bold R) \setminus
O(d,d;\bold R)$ (precisely its conjugate by $\bold E$,
see, e.g. [Na][Ki2][NW]).  To summarize this,
let us define a matrix in $O'(d,d;\bold R)$ of the form
$$
W'(\Lambda,B):= \left( \matrix \,^t\Lambda^{-1} & 0 \cr 0 &
\Lambda \cr \endmatrix\right) \left(\matrix \bold 1_d & - B \cr 0
& \bold 1_d \cr \endmatrix\right) \;\;,
$$
where $\Lambda \in GL(d,\bold R)$, and $B \in Mat(d,\bold R)$ is an
antisymmetric matrix, i.e. $\,^t B=-B$. 
In the following, we denote by $A(d,\bold
R)$ the set of all antisymmetric real matrices.

\Proclaim{Proposition}{\cosetpara} The coset space $G\times G
\setminus O'(d,d;\bold R)$, where $G\times G:=
\bold E\,( O(d,\bold R) \times O(d,\bold R)\,) \bold E$, 
can be represented by;
$$
G\times G\setminus O'(d,d;\bold R)=\{ G_{diag}\cdot W'(\Lambda,B)
\;|\; \Lambda \in GL(d,\bold R), \; B \in A(d,\bold R) \}\;,
$$
where
$$
G_{diag}\cdot W'(\Lambda,B):= \left\{ \left( \matrix g & 0 \cr 0 &
g \cr \endmatrix \right) W'(\Lambda, B) \;|\; g \in O(d,\bold R)
\;\right\}\;. \Eqno{\orbitDiag}
$$
Moreover, when $G_{diag}\cdot W'(\Lambda,B)=G_{diag}\cdot
W'(\Lambda',B')$, we have $O(d,\bold R)\cdot \Lambda=
O(d,\bold R)\cdot \Lambda',$ $B=B'$.
So, if we fix a ``gauge'' for the $O(d,\bold R)$ action
on $GL(d,\bold R)$ (i.e.
a cross section for the orbit space $GL(d,\bold R)/O(d,\bold R)$),
then $\Lambda$ and $B$ are uniquely determined for each
orbit in the coset space $G\times G \setminus O'(d,d;\bold R)$.
\endproclaim

An elementary proof of this is given in Appendix C. Here we point out
that the $\Lambda \in GL(d,\bold R)$ may be identified with
the lattice of the target torus, i.e. $T^d=\bold
R^d/L(\Lambda)$ with $L(\Lambda)$ being the lattice generated by the
column vectors of $\Lambda$. Under this identification, the matrix
$\,^t\Lambda^{-1}$ defines the dual torus $(T^d)^\vee$. Also the
antisymmetric matrix $B$ represents the so-called $B$-field, which
generalizes the electro-magnetic field in particle theory to
string theory. By Proposition {\cosetpara}, the coset $O(d,\bold
R)\times O(d,\bold R) \setminus O'(d,d;\bold R)$ is in one-to-one
correspondence with the set $(\,GL(d,\bold R)/O(d,\bold R)\,) \times
A(d,\bold R)$.

For $d=2$, the parametrization
(\orbitDiag) can be made even more explicit.
We fix a ``gauge'' for $GL(2,\bold R)/O(2,\bold R)$ by choosing
a special cross section;
$$
\Lambda_0= \sqrt{\frac{\rho_2}{\tau_2}} \left( \matrix 1 & \tau_1
\cr
                    0 & \tau_2 \cr  \endmatrix \right)\in GL(2,\bold R)
\;\; (\rho_2, \tau_2 >0) \;\;.\;\;
$$
Note that $\text{det}\, \Lambda_0 = \rho_2$ is the volume of
the torus defined by the lattice $L(\Lambda_0)$.
Letting the $B$-field be $B_{12}=:\rho_1$, we arrive at the
following parametrization
$$
W'(\tau,\rho):= \frac{1}{\sqrt{\rho_2\tau_2}} \left( \matrix
\tau_2 & 0 & 0 & 0 \cr -\tau_1 & 1 & 0 & 0 \cr 0 & 0 & \rho_2 &
\tau_1 \rho_2 \cr 0 & 0 & 0 & \tau_2\rho_2 \cr \endmatrix \right)
\left( \matrix 1 & 0 & 0 & - \rho_1 \cr
               0 & 1 & \rho_1 & 0 \cr
               0 & 0 & 1 & 0 \cr
               0 & 0 & 0 & 1 \cr \endmatrix \right) \;\;,
\Eqno{\paraW}
$$
where $\tau=\tau_1+\sqrt{-1}\tau_2, \rho=\rho_1+\sqrt{-1}\rho_2
\;(\tau_2,\rho_2 >0)$. It is clear that $\tau$ describes the
complex structure of the torus $T^2=\bold
R^2/L(\Lambda_0)(=:E_\tau)$. The parameter $
\rho=\rho_1+\sqrt{-1}\rho_2=B_{12}+\sqrt{-1}Vol(E_\tau)$ is the
so-called complexified K\"ahler modulus of $E_\tau$. It is also
clear that both parameters $\tau$ and $\rho$
can take arbitrary values in the upper half plane $\bold H_+$, i.e.
$(\tau,\rho)$ ranges over all of $\bold H_+ \times \bold H_+$.

The right action $O(2,2;\bold Z)$ on the coset space $G\times G
\setminus O'(2,2;\bold R)$ is known
in the physics literatures as the $T$-duality group.

\Proclaim{Proposition}{\TdualG}{\bf (Duality transformations)}
\item{1)} The following elements generate the group $O(2,2;\bold Z)$:
$$
\aligned
& S_1=\left( \matrix 0 & -1 & 0 & 0 \\
                   1 &  0 & 0 & 0 \\
                   0 &  0 & 0 & -1 \\
                   0 &  0 & 1 & 0 \\ \endmatrix \right),\,
T_1=\left( \matrix 1 &  0 & 0 & 0 \\
                   1 &  1 & 0 & 0 \\
                   0 &  0 & 1 & \hskip-5pt -1 \\
                   0 &  0 & 0 &  1 \\ \endmatrix \right),\,
R_1=\left( \matrix 0 &  0 & \hskip-5pt -1 & 0 \\
                   0 &  1 & 0 & 0 \\
                  -1 &  0 & 0 & 0 \\
                   0 &  0 & 0 &  1 \\ \endmatrix \right) ,
\cr
& S_2=\left( \matrix 0 & 0 & 0 & 1 \\
                   0 &  0 & \hskip-5pt -1 & 0 \\
                   0 &  1 & 0 &  0 \\
                   -1 &  0 & 0 & 0 \\ \endmatrix \right) ,\,
T_2=\left( \matrix 1 &  0 & 0 & 1 \\
                   0 &  1 & \hskip-5pt -1 & 0 \\
                   0 &  0 & 1 &  0 \\
                   0 &  0 & 0 &  1 \\ \endmatrix \right) ,\,
R_2=\left( \matrix  -1&  0 & 0 & 0 \\
                   0 &  1 & 0 & 0 \\
                   0 &  0 & \hskip-5pt -1& 0 \\
                   0 &  0 & 0 & 1 \\ \endmatrix \right) .
\cr
\endaligned
$$
\item{2)} The above generators act on the orbits
$(G\times G) \cdot W'(\tau,\rho)$ from the right by
$$
\aligned & S_1: (\tau,\rho) \rightarrow (-1/\tau,\rho) \;\;,\;\;
T_1: (\tau,\rho) \rightarrow (\tau+1,\rho) \;\;,\;\;
R_1: (\tau,\rho) \rightarrow (\rho,\tau) \\
& S_2: (\tau,\rho) \rightarrow (\tau,-1/\rho) \;\;,\;\; T_2:
(\tau,\rho) \rightarrow (\tau,\rho+1) \;\;,\;\;
R_2: (\tau,\rho) \rightarrow (-\bar\tau,-\bar\rho). \\
\endaligned
$$
\item{3)} There is a group isomorphism,
$$
O(2,2;\bold Z) \cong P(SL_2 \bold Z \times SL_2\bold Z) \rtimes
(\bold Z_2 \times \bold Z_2) \;,
$$
where $P(SL_2 \bold Z \times SL_2\bold Z)$ represents the quotient 
group of $SL_2 \bold Z \times SL_2\bold Z$ 
by the involution: $(g,h) \mapsto (-g,-h)$, and $\rtimes$ denotes the 
semi-direct product. 
\endproclaim

\demo{Sketch of Proof} We will give a proof of 1) in Appendix C.
By Proposition {\cosetpara} and the parametrization of the orbit
(\paraW), assertion 2) is derived by straightforward
calculations. For a proof of 3), we use an explicit surjective
group homomorphism, which is constructed in Appendix C,
$\phi_{\bold Z}: SL_2\bold Z \times SL_2 \bold Z
\rightarrow O(2,2;\bold Z)\cap O_0'(2,2;\bold R)$ with $Ker(\phi_{\bold Z})
=\{(\bold 1_2,\bold 1_2),(-\bold 1_2,-\bold 1_2)\}$, where $O_0'(2,2;\bold R)$
is the connected component of the identify (see Proposition C.2).
Thus we find a subgroup
$P(SL_2\bold Z \times SL_2 \bold Z)$ in $O(2,2;\bold Z)$. As is clear
from the argument for the decomposition (C.5), this
subgroup is a normal subgroup with $P(SL_2\bold Z \times SL_2 \bold Z)
\setminus O(2,2;\bold Z)=\{ \bold 1_4, R_1,R_2,R_1R_2\}
\cong \bold Z_2\times \bold Z_2$.
\rqed
\enddemo

\vskip0.7cm
\noindent
{\bf (2-4) Narain lattice ($d=2$).}
We now use
the explicit parameterization $W'(\tau,\rho)$ in (\paraW) of the coset space
$G \times G \setminus O'(2,2;\bold R)$, to produce
a convenient parameterization $\Gamma(\tau,\rho)$ of Narain lattices.
By (\OrthoPhi), the element
$W'(\tau,\rho) \in O'(2,2;\bold R)$ determines a unique
Narain embedding $\Phi_{\tau,\rho}: U^{\oplus 2}
\hookrightarrow \bold R^{2,2}$
via $W(\Phi_{\tau,\rho})=\bold E W'(\tau,\rho)\bold E$.
We will write down $\Phi_{\tau,\rho}$ explicitly.
Consider
$$
W(\Phi_{\tau,\rho})\bold E
=
\bold E W'(\tau,\rho)
=: ( \bold e_1(\tau,\rho) \; \bold e_2(\tau,\rho) \; \bold
e_3(\tau,\rho) \; \bold e_4(\tau,\rho) ).
\Eqno{\BoldEs}
$$
Then we have $\langle \bold e_i(\tau,\rho), \bold
e_{j+2}(\tau,\rho) \rangle_{\bold R^{2,2}}=\delta_{ij}, \langle
\bold e_i(\tau,\rho), \bold e_{j}(\tau,\rho) \rangle_{\bold
R^{2,2}}= \langle \bold e_{i+2}(\tau,\rho),$ $\bold
e_{j+2}(\tau,\rho) \rangle_{\bold R^{2,2}}=0 \; (1\leq i,j \leq
2)$. By (\OrthoPhi), the Narain embedding $\Phi_{\tau,\rho}$
is given by by $\Phi_{\tau,\rho}(e_i) =
\bold e_i(\tau,\rho)\;, \Phi_{\tau,\rho}(f_j) = \bold
e_{j+2}(\tau,\rho)$,  and its image is the Narain lattice
$$
\Gamma(\tau,\rho) := \bold Z \, \bold e_1(\tau,\rho) \oplus \bold Z
\, \bold e_2(\tau,\rho) \oplus \bold Z \, \bold e_3(\tau,\rho)
\oplus \bold Z \, \bold e_4(\tau,\rho)\subset\bold R^{2,2}. \Eqno{\NarainEs}
$$
Explicitly the $e_i(\tau,\rho)$ are
$$
\frac{1}{\sqrt{2\tau_2\rho_2}} \left( \matrix \tau_2 \\ -\tau_1 \\
\tau_2 \\ -\tau_1 \\ \endmatrix \right) ,
\frac{1}{\sqrt{2\tau_2\rho_2}} \left( \matrix 0 \\ 1 \\ 0 \\ 1 \\
\endmatrix \right) , \frac{1}{\sqrt{2\tau_2\rho_2}} \left( \matrix
\rho_2 \\ \rho_1 \\ -\rho_2 \\ \rho_1 \\ \endmatrix \right) ,
\frac{1}{\sqrt{2\tau_2\rho_2}} \left( \matrix
-\rho_1\tau_2+\tau_1\rho_2  \\
\rho_1\tau_1+\tau_2\rho_2  \\
-\rho_1\tau_2-\tau_1\rho_2  \\
\rho_1\tau_1-\tau_2\rho_2  \\
\endmatrix \right).
\Eqno{\NarainPara}
$$

The following properties of the Narain lattice $\Gamma(\tau,\rho)$
are immediate from Propositions {\cosetpara}, {\TdualG}:

\Proclaim{Proposition}{\DualityInvariant}
\item{1)} $\Gamma(\tau,\rho),\Gamma(\tau',\rho')$ are equivalent if and only
if $(\tau,\rho)$ and $(\tau',\rho')$ are related by a
duality transformation on $\bold H_+^2$.
\item{2)} Every Narain lattice $\Gamma\subset\bold R^{2,2}$ is
equivalent to $\Gamma(\tau,\rho)$ for some $\tau,\rho\in\bold H_+$.
\endproclaim

\vskip1cm
\head
{\bf \S 3. Rational conformal field theory}
\endhead

\global\secno=3
\global\propno=3
\global\eqnum=3

\noindent
{\bf (3-1) Algebraic CFT.}
Algebraically, a conformal field theory (CFT)
is  described by its so-called {\it chiral algebras} and their
representations. For a mathematical exposition see [FLM][LZ][Kac][MN].
For a physical exposition see [GSW][Po]. We
give a rough schematic description of this theory here,
but will be more precise when we come to CFTs on $T^2$.
Examples of chiral algebras $\Cal A$
often come from infinite dimensional Lie algebras and
certain generalizations such as the $W$-algebras. The basic
setup of a CFT includes (1) two chiral
algebras $\Cal A_L, \Cal A_R$, which are called
respectively the {\it left} and
the {\it right} chiral algebras; (2) a class of representations
$\Cal H_{L,j} \; (j \in \Lambda_L)$ and $\Cal H_{R,k} \;
(k \in \Lambda_R)$, where $\Lambda_{L,R}$ are some index sets;
(3) the characters of the
representations $ch_{L,j}(q)=Tr_{\Cal H_{L,j}}q^{d_L}$,
$\overline{ch_{R,j}(q)}=Tr_{\Cal H_{R,j}}\bar q^{d_L}$, where $d_L, d_R$ are
the scaling operators in the chiral algebras and $q=e^{2\pi \sqrt{-1} t},
\bar q=e^{-2\pi \sqrt{-1} \bar t}$; (4)
({\it partition function}), a real analytic modular
invariant (i.e. invariant under the transformations
$t \rightarrow t+1, t \rightarrow -1/t$) function of the shape
$$
Z(q,\bar q)=
\sum_{i \in \Lambda_L, j \in \Lambda_R}
N_{ij} ch_{L,i}(q) \overline{ch_{R,j}(q)} \; ,
$$
where $N_{ij}$ are some positive integers.
A CFT is called {\it rational} iff its chiral algebras have
only finitely many irreducible representations, in which case
the index sets $\Lambda_L, \Lambda_R$ are the
finite lists of representations.

We now consider CFTs on $T^2$. This classes of CFTs
are parameterized by equivalence classes of Narain lattices.
For a generic Narain lattice $\Gamma$, the chiral algebras are
generated by vertex operators of a Heisenberg algebra 
(also known as ``$U(1)$ currents'' in physics).
The chiral algebras also contain the $c=2$ Virasoro algebra as a subalgebra.
The lattice $\Gamma$ plays the role of the momentum-winding lattice.
Each momentum-winding vector $p=p_l+p_r\in\Gamma$ corresponds to
a pair of irreducible representations labeled by
$p_l,p_r$ (``$U(1)$ charges'' of the $U(1)$ currents).
The partition function is then given by $Z^\Gamma(q,\bar q)$,
as in Definition (\CFTonTd).
When the CFT becomes rational, something very interesting happens.
First, the chiral algebras become
significantly enlarged. Second, the infinite list of representations
(of the small chiral algebras),
indexed by $p\in\Gamma$, reconstitute themselves,
and then break up into finitely many irreducible representation
of the enlarged chiral algebras.

The main task of this paper is to describe Narain lattices $\Gamma$
which yield rational CFT in terms of the momentum lattices of the left
and right handed chiral algebras, and count them.

\vskip0.7cm
\noindent{\bf (3-2) Discriminant $A_\Gamma$.}
Let $\Gamma$ be an even, positive definite lattice. The quotient
$\Gamma^*/\Gamma$ is a group called the {\it discriminant group} of
$\Gamma$ (, see [Ni] for details). Let us fix a
basis $u_1,u_2$ of $\Gamma$, and denote its dual basis by
$u_1^*, u_2^*$. Then the intersection form (\Lintersec) relates these two
bases by
$$
(u_1^* \; u_2^*)=(u_1 \; u_2) \;\; \bigg(
\matrix 2a  & b \cr b & 2c \cr \endmatrix \bigg)^{-1}  \;\;.
\Eqno{\uToDualu}
$$
The discriminant group is the abelian group generated by $u_1^*, u_2^*$
modulo the lattice $\Gamma\subset\Gamma^*$. The integral bilinear form on
$\Gamma$
extends to a rational bilinear form $(\,,\,)$ on $\Gamma^*$.
Since $\Gamma$ is even, we have a
natural quadratic form 
$q_\Gamma: \Gamma^*/\Gamma \rightarrow
\bold Q/2\bold Z$, called the {\it discriminant form} of $\Gamma$,
$$
q_\Gamma(v \text{ mod } \Gamma)
:= (v,v) \text{ mod } 2\bold Z \;\;.
$$
If $v=m_1 u_1^* + m_2 u_2^*$,
then $(v,v)$ can evaluated using the linear relation (\uToDualu).
Associated to the quadratic form $q_\Gamma$, we have a 
$\bold Q/\bold Z$-valued bilinear form
$(w,v)_{q}:=
\frac{1}{2}(q_\Gamma(w+v) - q_\Gamma(w)-q_\Gamma(v))$.
We denote the pair $(\Gamma^*/\Gamma, q_\Gamma)$ by $A_\Gamma$, and call it
the {\it discriminant} of $\Gamma$.

Consider the discriminants $A_\Gamma, A_{\Gamma'}$ of two
lattices $\Gamma$, $\Gamma'$. The group isomorphisms which preserve the
discriminant form will be called {\it isometries} of the discriminants.
Clearly an isomorphism of lattices induces an isometry of their discriminants.
(But the converse is not true.)
We denote the set of isometries by
$$
\text{Isom}(A_\Gamma,A_{\Gamma'}):=
\{ \varphi: \Gamma^*/\Gamma \;\simrightarrow\; {\Gamma'}^*/\Gamma' \; | \;
q_{\Gamma'}(\varphi(v)) \equiv q_{\Gamma}(v) \text{ mod } 2\bold Z \;\}
$$
Clearly, the orthogonal group $O(\Gamma)$ acts naturally on the
set $\text{Isom} (A_\Gamma,A_{\Gamma'})$ from the left, and
$O(\Gamma')$ from the right.

\noindent
{\bf Remark.} The set $\text{Isom}(A_\Gamma,A_{\Gamma'})$
is nonempty if and only if the two discriminants are isomorphic.
This is the case exactly when the discriminant groups are isomorphic,
$\Gamma^*/\Gamma \cong {\Gamma'}^*/\Gamma' \cong \bold Z_{d_1} \oplus
\bold Z_{d_2}$ $(d_1|d_2)$, and the isomorphism preserves the discriminant
forms. To spell this out explicitly, let us fix the isomorphisms;
$\Gamma^*/\Gamma \cong  \bold Z_{d_1}
\oplus \bold Z_{d_2}$
(respectively ${\Gamma'}^*/\Gamma' \cong  \bold Z_{d_1} \oplus
\bold Z_{d_2}$) by choosing a basis $w_1,w_2$ (respectively $w_1',w_2'$).
With respect to these bases, we represent an
isomorphism $\varphi:\Gamma^*/\Gamma \,\simrightarrow\, {\Gamma'}^*/\Gamma'$
by a matrix: $(\varphi(w_1)\,\varphi(w_2)) =(w_1'\, w_2')\left(
\smallmatrix \alpha & \beta \cr \gamma & \delta \cr \endsmallmatrix
\right)$ where $\alpha, \beta$ (respectively $\gamma, \delta$) are integers
considered mod $d_1$ ($d_2$). Let
$(\;,\;)_q$, $(\;,\;)_{q'}$ denote
the bilinear forms of $A_\Gamma,A_{\Gamma'}$.
That $\varphi$ is an isometry means that
$(\varphi(w_i),\varphi(w_j))_{q'}=(w_i,w_j)_{q}$, i.e.
$$
\bigg(^{\hskip-0.3cm t} \hskip0.2cm
\matrix \alpha  & \beta  \cr
        \gamma  & \delta \cr
\endmatrix \bigg)
\left( \matrix
(w'_1,w'_1)_{q'}  & (w'_1,w'_2)_{q'} \cr
(w'_2,w'_1)_{q'}  & (w'_2,w'_2)_{q'} \cr \endmatrix \right)
\bigg(
\matrix \alpha  & \beta  \cr
        \gamma  & \delta \cr
\endmatrix \bigg)
=
\left( \matrix
(w_1,w_1)_q  & (w_1,w_2)_q \cr
(w_2,w_1)_q  & (w_2,w_2)_q \cr \endmatrix \right) \;.
\Eqno{\discIso}
$$
\rqed

Discriminants of even, positive definite lattices of rank two and the
isometries between them are central objects in our description of
rational conformal field theory on $T^2$.

\vskip0.7cm
\noindent
{\bf (3-3) $c=2$ RCFT.}
If $\Gamma_l,\Gamma_r$ are even, positive definite lattices of
determinant $-D$ and $\varphi:A_{\Gamma_l}\rightarrow A_{\Gamma_r}$ is
an isometry, then we call $(\Gamma_l,\Gamma_r,\varphi)$ {\it a triple}
of determinant $-D$. We say that the triple is primitive
if $\Gamma_l,\Gamma_r$ are primitive lattices.
We define an equivalence relation on triples as follows:
$$
\align
&(\Gamma_l,\Gamma_r,\varphi)\sim(\Gamma_l',\Gamma_r',\varphi')\cr
&\Leftrightarrow\exists\text{ isomorphisms } g:\Gamma_l\rightarrow\Gamma_l',
\;\; h:\Gamma_r\rightarrow\Gamma_r' \text{ such that }
 \varphi'=\bar h\cdot\varphi\cdot\bar g^{-1}.
\endalign
$$
Here $\bar g:A_{\Gamma_l}\rightarrow A_{\Gamma_l'}$ is the isometry
induced by $g$ and similarly for $\bar h$. We call an equivalence class
of triples $[(\Gamma_l,\Gamma_r,\varphi)]$ an {\it RCFT data}.
For our description of the RCFT data, we define the following sets
of equivalence classes:
$$
\align
&RCFT_D:=\{\text{ triples of determinant } -D\}/\sim \;\;,\cr
&RCFT^p_D:=\{\text{ primitive triples of determinant } -D\}/\sim\;\;, \cr
\endalign
$$
and also
$$
\align
RCFT_D(\GGamma_l,\GGamma_r):=
\{ & \text{ triples } (\Gamma_l',\Gamma_r',\varphi')
     \text{ of determinant } -D \cr
   &\text{ with } [\Gamma_l']= \GGamma_l,\;[\Gamma_r']= \GGamma_r\}/\sim
\;\;,\cr
RCFT^p_D(\GGamma_l,\GGamma_r):=
\{ & \text{ primitive triples } (\Gamma_l',\Gamma_r',\varphi')
     \text{ of determinant } -D\cr
   &\text{ with } [\Gamma_l']=\GGamma_l,\;[\Gamma_r']=\GGamma_r\}/\sim.
\endalign
$$
Note that we have the obvious disjoint union:
$$
RCFT_D = \bigsqcup RCFT_D(\GGamma_l,\GGamma_r)\;,
$$
where the $\GGamma_l,\GGamma_r$ range over all classes in
$\widetilde{CL}(D)\equiv \Cal L(D)$.
Likewise, we have a similar decomposition for $RCFT_D^p$
with $\widetilde{Cl(D)}\equiv \Cal L^p(D)$.
Note that $RCFT_D(\GGamma_l,\GGamma_r)$ is nonempty if and only if
$\Gamma_l,\Gamma_r$ are in the same genus [Ni].

\Proclaim{Lemma}{\RCFTtoCoset}
Fix lattices $\Gamma_l, \Gamma_r$
in respective classes $\GGamma_l, \GGamma_r \in \Cal L(D)$. Then
the following map is well-defined and bijective:
$$
\matrix
B_{\Gamma_l,\Gamma_r}:
& RCFT_D(\GGamma_l,\GGamma_r) &  \rightarrow &
O(\Gamma_l)\setminus Isom(A_{\Gamma_l},A_{\Gamma_r})/O(\Gamma_r)  \cr
&[(\Gamma_l,\Gamma_r,\varphi)]& \mapsto & \;\;  [\varphi]\hfill  \cr
\endmatrix
\Eqno{\RtoCo}
$$
\endproclaim

\demo{Proof} By definition, an arbitrary class in
$RCFT_D(\GGamma_l,\GGamma_r)$ may be represented by a triple
$(\Gamma_l,\Gamma_r,\varphi)$ with some $\varphi \in
Isom(A_{\Gamma_l},A_{\Gamma_r})$.
The triples having this shape
$(\Gamma_l,\Gamma_r,*)$ are preserved exactly by
the group $O(\Gamma_l) \times O(\Gamma_r)$
acting on triples by
$$
(g \Gamma_l,h \Gamma_r, \bar h \varphi \bar g^{-1})
=
(\Gamma_l,\Gamma_r, \bar h  \varphi \bar g^{-1}) \; .
$$
This shows that the map is well-defined.
Now the bijectivity of the map is also clear.
\rqed
\enddemo

This map will be used in the next section.

\Proclaim{Definition}{\RCFTdef} {\bf ($c=2$ RCFT)}
Given a triple $(\Gamma_l,\Gamma_r,\varphi)$ of determinant $-D$, we
define the partition function
$$
Z^{\Gamma_l,\Gamma_r,\varphi}(q,\bar q)=
\frac{1}{|\eta(q)|^4} \sum_{a \in \Gamma_l^*/\Gamma_l}
\theta_a^{\Gamma_l}(q)
\overline{{\theta}_{\varphi(a)}^{\Gamma_r}(q)} \;,
$$
and its parity invariant form
$$
\tilde Z^{\Gamma_l,\Gamma_r,\varphi}(q,\bar q)=
\frac{1}{|\eta(q)|^4} \sum_{a \in \Gamma_l^*/\Gamma_l}
\frac{1}{2}\left(
\theta_a^{\Gamma_l}(q)
\overline{{\theta}_{\varphi(a)}^{\Gamma_r}(q)}+
\overline{\theta_a^{\Gamma_l}(q)}
{\theta}_{\varphi(a)}^{\Gamma_r}(q)
\right) ,
$$
where $\theta_a^{\Gamma}(q)
:=\sum_{v \in \Gamma} q^{\frac{1}{2}(v+a)^2}$, for $a\in \Gamma^*/\Gamma$,
is a theta function of the lattice $\Gamma$.
\endproclaim

\noindent
{\bf Remark.}
1) It is clear that equivalent triples have the same partition function, and
thus the partition function is defined for a class
$[(\Gamma_l,\Gamma_r,\varphi)] \in RCFT_D$.

\noindent
2) The involution
$(\Gamma_l,\Gamma_r,\varphi)\mapsto(\Gamma_r,\Gamma_l,\varphi^{-1})$
on triples is obviously compatible with the equivalence relation on triples.
In particular, the involution acts on the sets $RCFT_D, RCFT_D^p$. The
parity invariant form $\tilde Z^{\Gamma_l,\Gamma_r,\varphi}(q,\bar q)$
is defined so that it is invariant under this involution.

\noindent
3) Note that $Z^{\Gamma_l,\Gamma_r,\varphi}(q,\bar q)$ is nothing but
a finite linear combination of products $ch_{L,a}(q)\overline{ch_{R,b}(q)}$,
where
$$
ch_{L,a}(q)=\frac{\theta^{\Gamma_l}_a(q)}{\eta(q)^2}\;\;,\;\;
ch_{R,b}(q)=\frac{\theta^{\Gamma_r}_b(q)}{\eta(q)^2}\;\;,\;\;
$$
with $a \in \Gamma^*_l/\Gamma_l, b\in \Gamma_r^*/\Gamma_r$, are
characters of  representations of  certain chiral algebras
$\Cal A_L$ and $\Cal A_R$. They can also be viewed as characters of
certain unitary
reducible representations of the $c=2$ Virasoro algebra (see e.g. [KP]).
\rqed

\Proclaim{Proposition}{\modularinv}
The partition function for a triple is modular invariant.
\endproclaim

\demo{Proof} The theta function of an even lattice $\Gamma$
$(s:=\text{rk }\Gamma)$ has the transformation property, $
\theta^\Gamma_a(q)\vert_{t\rightarrow t+1}=
e^{\pi\sqrt{-1}(a,a)} \theta^\Gamma_a(q)$. Also, by the
Poisson resummation formula, we have (see e.g. [KP, Proposition 3.4]);
$$
\theta^\Gamma_a(q)\big|_{t\rightarrow -1/t} =
\frac{1}{|\Gamma^*/\Gamma|^{\frac{1}{2}}} (-\sqrt{-1}t)^{\frac{s}{2}}
\sum_{b \in \Gamma^*/\Gamma} e^{-2\pi \sqrt{-1} (a,b)}
\theta^{\Gamma}_b(q) ,
$$
where $(a,b) \in \bold Q/\bold Z$ is the bilinear form on the
discriminant group $\Gamma^*/\Gamma$ (see section (3-2)).
Using these relations
and also $\frac{1}{|\Gamma^*/\Gamma|} \sum_{a \in \Gamma^*/\Gamma}
e^{2\pi \sqrt{-1}(a,b-b')} = \delta_{b b'}$, it is straightforward
to verify that $Z^{\Gamma_l,\Gamma_r,\varphi}(q,\bar q)$
is invariant under $t \rightarrow t+1, t\rightarrow -1/t$. In the
calculation, we use the fact that $\varphi$ is an isometry and thus
we have $(\varphi(a),b)=(a,\varphi^{-1}(b))$ for $a \in \Gamma_l^*/\Gamma_l,
b\in \Gamma_r^*/\Gamma_r$. \rqed
\enddemo

We now state our problem precisely as follows:

\vskip0.5cm

\noindent
{\bf Classification Problem.}
{\it
(1) Formulate a correspondence between the triples
$(\Gamma_l,\Gamma_r,\varphi)$ and the Narain lattices
$\Gamma(\tau,\rho)$.
(2) Describe the values $\tau,\rho \in \bold H_+$ for the
equivalence classes of Narain lattices $\Gamma(\tau,\rho)$ which
correspond to the classes of triples in $RCFT_D$, and count them. }

\vskip0.5cm

Prop. 1.6.1 of Nikulin [Ni] gives a correspondence between
primitive embeddings in a fixed unimodular lattice
and isometries of discriminant groups by means of over-lattices.
An equivalent approach using {\it gluing theory} can be found in
[CS, Chapter 4] and references therein.
This correspondence will be needed to do (1) (see below) where we seek
a correspondence between abstract
triples and the parameters $\tau,\rho$.
The importance of discriminant groups in
the study of toroidal RCFTs was first observed in [Mo].
In the next two sections we will carry out (2).
The precise counting of both Narain lattices and triples will
involve taking into account a certain natural involution,
$(\tau,\rho) \mapsto (\tau,-\bar\rho)$, which represents the
worldsheet parity involution.
We will find that the Gauss product on the class
group $Cl(D)$ and its extension to $CL(D)$ play a central
role.

\vskip0.7cm
\noindent{\bf (3-4) Over-lattices.}
Our aim here is to define a map from $RCFT_D$ to
the Narain moduli space $\Cal M_{2,2}$.
The main idea is to look at {\it over-lattices}
of $\Gamma_l\oplus \Gamma_r(-1)$.

Given a triple $(\Gamma_l,\Gamma_r,\varphi)$, put
$$
\Gamma^\varphi:=\{\; (x,y)\in\Gamma_l^*\oplus\Gamma_r^*(-1)
\,|\,\varphi(x~mod~\Gamma_l)=y~mod~\Gamma_r(-1) \;\}.
\Eqno{\DefGamma}
$$
It is easy to check that
this is a sublattice of $\Gamma_l^*\oplus\Gamma_r^*(-1)$
containing $\Gamma_l\oplus\Gamma_r(-1)$, such that
$$
\Gamma^\varphi\big/(\Gamma_l\oplus \Gamma_r(-1)) = \{ \;
a \oplus \varphi(a) \,|\, a \in \Gamma_l^*/\Gamma_l \;\} \;\;,
\Eqno{\OverL}
$$
or equivalently,
$$
\Gamma^\varphi=\bigcup_{a \in \Gamma_l^*/\Gamma_l} \left(
a \oplus \varphi(a) + \Gamma_l \oplus \Gamma_r(-1) \right) \;\;.
\Eqno{\OverLexp}
$$

Fix a pair of isometric embeddings
$\iota_1:\Gamma_l\rightarrow\bold R^{2,0}$,
$\iota_2:\Gamma_r(-1)\rightarrow\bold R^{0,2}$.
Extend $\iota_1,\iota_2$ over $\bold R$,
and denote their extensions by $\iota_1,\iota_2$ also.

\Proclaim{Proposition}{\OLproperty}
Let $(\Gamma_l,\Gamma_r,\varphi)$ be a triple, and $\iota_1,\iota_2$ be
isometric embeddings as above. Then
\item{1)} $\Gamma^\varphi$ is an even, unimodular, integral 
lattice of signature $(2,2)$. Hence the image
$(\iota_1\oplus\iota_2)\Gamma^\varphi\subset\bold R^{2,2}$ is a Narain lattice.
\item{2)} The equivalence class $[(\iota_1\oplus\iota_2)\Gamma^\varphi]$ of
Narain lattices
is independent of the choices of $\iota_1,\iota_2$.
Moreover $[(\iota_1\oplus\iota_2)\Gamma^\varphi]$
depends only on the equivalence class of
the triple $(\Gamma_l,\Gamma_r,\varphi)$.
\endproclaim

\demo{Proof}
1) This follows from Proposition 1.6.1 in [Ni].

\noindent
2)
Suppose we have an equivalence $(\Gamma_l,\Gamma_r,\varphi)
\sim (\Gamma_l',\Gamma_r', \varphi')$ given by isomorphisms
$g:\Gamma_l \rightarrow \Gamma_l',\; h:\Gamma_r \rightarrow \Gamma_r'$.
As above, we fix choices of isometric embeddings
$\iota_1,\iota_2$ and $\iota_1',\iota_2'$ for the two triples.
Extend $g,h$ over $\bold R$,
and denote their extensions by $g,h$ also.
By (\DefGamma), we have
$(g\oplus h)\Gamma^\varphi=\Gamma^{\varphi'}$.
Clearly, the isometry
$f:=(\iota_1'\oplus \iota_2')\circ(g\oplus h)\circ(\iota_1\oplus\iota_2)^{-1}
=(\iota_1' \cdot g \cdot \iota_1^{-1})\oplus
(\iota_2' \cdot h \cdot \iota_2^{-1}):\bold R^{2,2}\rightarrow\bold R^{2,2}$
preserves the decomposition $\bold R^{2,0}\oplus\bold R^{0,2}$, hence
is an element of $O(d;\bold R) \times O(d;\bold R)$.
But $f\circ(\iota_1\oplus\iota_2)\Gamma^\varphi=
(\iota_1'\oplus\iota_2')\Gamma^{\varphi'}$.
It follows that $[(\iota_1\oplus\iota_2)\Gamma^{\varphi}]=
[(\iota_1'\oplus\iota_2')\Gamma^{\varphi'}]$. This proves 2).
\rqed
\enddemo

\vskip0.5cm

Now by Propositions {\OLproperty} and {\ModuliNarain}, we have
a well-defined map:
$$
F: \; \bigcup_D RCFT_D \longrightarrow \Cal M_{2,2}, \;\;
[(\Gamma_l,\Gamma_r,\varphi)]
\longmapsto [(\iota_1\oplus\iota_2)\Gamma^{\varphi}]\;\;.
\Eqno{\CFTtoNarain}
$$

\noindent
{\bf Remark.} 1) In the construction of $\Gamma^\varphi$
in the proof, if we replace $\Gamma_l,\Gamma_r$
by their images under the chosen embeddings $\iota_1,\iota_2$,
then the resulting $\Gamma^\varphi$ actually sits inside $\bold R^{2,2}$,
rather than being just an abstract lattice.

\noindent
2) Using the relation (\OverLexp), it is a simple
exercise to show that the partition function of
the Narain lattice $\Gamma=(\iota_1\oplus\iota_2)\Gamma^\varphi$
(Definition {\CFTonTd})) is given by
$$
Z^\Gamma(q,\bar q)
= Z^{\Gamma_l,\Gamma_r,\varphi}(q,\bar q).
\Eqno{\NarainToRational}
$$

\vskip0.7cm
\noindent
{\bf (3-5) RCFTs in $\Cal M_{2,2}$.}
Consider the Narain lattice $\Gamma(\tau,\rho)$ and
its equivalence class  $[\Gamma(\tau,\rho)]\in\Cal M_{2,2}$.
For generic $\tau,\rho \in \bold H_+$, the class $[\Gamma(\tau,\rho)]$
does not correspond to an RCFT, i.e. is not in the image
of $RCFT_D$ under the map (\CFTtoNarain) for any $D$.

\Proclaim{Definition}{\RatNarain}
We call a Narain lattice $\Gamma(\tau,\rho)$ {\it rational} if
its class $[\Gamma(\tau,\rho)]$ is in the image of the ``over-lattice
map'' $F$ given by (\CFTtoNarain).
\endproclaim

To study the rationality condition for $\Gamma(\tau,\rho)$,
we consider
$$
\Pi_l:=\Gamma(\tau,\rho) \cap \bold R^{2,0} \;\;,\;\;
\Pi_r:=\Gamma(\tau,\rho) \cap \bold R^{0,2} \;\;.
\Eqno{\PiLR}
$$
We call them {\it the momentum lattices of (left and right chiral
algebras corresponding to) $\Gamma(\tau,\rho)$}.
Note that:
1) The lattices $\Pi_l,\Pi_r$ are sublattices
of $\Gamma(\tau,\rho)$, which are zero for
generic $\tau,\rho$. 2) The natural embedding
$\Pi_l \hookrightarrow \Gamma(\tau,\rho)$ is a {\it primitive
embedding}, i.e. the natural map $\Gamma(\tau,\rho)^*=
\Gamma(\tau,\rho) \rightarrow \Pi_l^*$ is surjective, where
$x \in  \Gamma(\tau,\rho)^*$ is mapped to $\langle x,*\rangle:
\Pi_l \rightarrow \bold Z$ under the natural map. Likewise for $\Pi_r$.
The following properties (Propositions 3.8 and 3.10) are well-known
in the literatures [DHMV][Mo][Wa].

\Proclaim{Proposition}{\ratI} The Narain lattice $\Gamma(\tau,\rho)$ is
rational if and only if
$$
\text{rk}\, \Pi_l = \text{rk}\, \Pi_r =2 \;.
\Eqno{\rkcond}
$$
\endproclaim
\demo{Proof} $\Rightarrow)$
Suppose $\Gamma(\tau,\rho)$ is rational,
i.e. $[\Gamma(\tau,\rho)]=[\Gamma^\varphi]$ for some triple
$(\Gamma_l,\Gamma_r,\varphi)$
with $\Gamma_l\subset\bold R^{2,0}$,
and $\Gamma_r(-1)\subset\bold R^{0,2}$.
This means that $\Gamma^\varphi=f\Gamma(\tau,\rho)$
for some $f\in O(d;\bold R)\times O(d;\bold R)$.
So we have
$\Pi_l:=\Gamma(\tau,\rho)\cap\bold R^{2,0}=
f\Gamma(\tau,\rho)\cap\bold R^{2,0}=\Gamma^\varphi\cap\bold R^{2,0}=\Gamma_l$,
which has rank 2. Likewise for $\Pi_r$.

$\Leftarrow)$
When the both sublattices $\Pi_l$ and $\Pi_r$ have rank 2, they are
orthogonal complements of each other in $\Gamma(\tau,\rho)$
(since $\Pi_l, \Pi_r$ are primitive in $\Gamma(\tau,\rho)$).
Thus, for $x \in \Gamma(\tau,\rho)$, we may consider the
orthogonal decomposition
$x=x_l+x_r$ $(x_l \in \Pi_l \otimes \bold Q, x_r \in \Pi_r\otimes \bold Q)$
over $\bold Q$.  The natural (surjective) map
$\Gamma(\tau,\rho) \rightarrow \Pi_{l}^*$ may be given by
$x \mapsto \langle x,* \rangle=\langle x_l,* \rangle$, and has the kernel
exactly equal to $\Pi_r$.
Similarly we
have the natural surjective map $\Gamma(\tau,\rho) \rightarrow \Pi_{r}^*$
with the kernel $\Pi_l$. From this, as in [Ni], we obtain
$$
\Pi_l^*/\Pi_l
  \; \simmapleftarrow{\iota_L} \;
\Gamma(\tau,\rho)/(\Pi_l\oplus \Pi_r)
  \; \simmaprightarrow{\iota_R} \;
\Pi_r^*/\Pi_r \;\;.
\Eqno{\isoDis}
$$
We have now $\iota_R\circ \iota_L^{-1}: \Pi_l^*/\Pi_l \; \simrightarrow \;
\Pi_r^*/\Pi_r$, which implies that $\Pi_l$ and
$\Pi_r$ have the same determinant $|\Pi_l^*/\Pi_l|=|\Pi_r^*/\Pi_r|=:-D'$.
Also, over the discriminant groups $\Pi_l^*/\Pi_l$ and $\Pi_r^*/\Pi_r$,
we have natural quadratic forms $q_l$ and $q_r$ given  by
$$
q_l(x_l \text{ mod } \Pi_l )= \langle x_l, x_l \rangle \text{ mod } 2 \bold Z
\;\;,\;\;
q_r(x_r \text{ mod } \Pi_r )= \langle x_r, x_r \rangle \text{ mod } 2 \bold Z ,
$$
taking values in $\bold Q/2\bold Z$.
Since $\Gamma(\tau,\rho)$ is even, integral, we have
$$
q_l(x_l)+q_r(x_r) \equiv \langle x_l, x_l \rangle + \langle x_r, x_r \rangle
=  \langle x_l+x_r, x_l+x_r \rangle
= \langle x, x \rangle  \equiv 0 \text{ mod } 2 \bold Z.
$$
Therefore the isomorphisms in (\isoDis)
give rise to an isometry of the quadratic forms:
$\iota_R\circ
\iota_L^{-1}: (\Pi_l^*/\Pi_l,q_l) \rightarrow (\Pi_r^*/\Pi_r, -q_r)$.
By construction, the over-lattice determined by the triple
$(\Pi_l,\Pi_r(-1), \iota_R\circ \iota_L^{-1})$ is
$\Gamma(\tau,\rho)$ (see Proposition 1.6.1 in [Ni]).
Therefore $[\Gamma(\tau,\rho)]$ is in the image of the map $F$, i.e.
$\Gamma(\tau,\rho)$
is rational.
\rightline{\rqed}
\enddemo

For a rational Narain lattice $\Gamma(\tau,\rho)$, the correspondence
$$
[\Gamma(\tau,\rho)] \mapsto [(\Pi_l,\Pi_r(-1), \iota_R\circ \iota_L^{-1})]
\Eqno{\inverseFD}
$$
defines the inverse of the map 
$$
F: \bigcup_D RCFT_{D}
\rightarrow \Cal N:=\{[\Gamma(\tau,\rho)]\,|\,
\Gamma(\tau,\rho): \text{rational} \,\} \;.
$$

\Proclaim{Proposition}{\RCFTtoNbyF}
The ``over-lattice map'' given by (\CFTtoNarain) defines a
bijection
$$
F: \bigcup_D RCFT_D \; \rightarrow\;  \Cal N  \;\;.
\Eqno{\mapF}
$$
\endproclaim

For later use, we present equivalent
characterizations given in the literatures [DHMV][Mo][Wa]:

\Proclaim{Proposition}{\QD} Consider (\PiLR).
The following statements are equivalent;
\item{1)} $\text{rk}\, \Pi_l  =2 \;$ or $\;\text{rk}\, \Pi_r  =2$,
\item{2)} $\text{rk}\, \Pi_l = \text{rk}\, \Pi_r =2 \;$,
\item{3)} $\tau,\rho \in \bold Q(\sqrt{D})$ for some integer $D<0$.
\endproclaim
\demo{Proof}
$1) \Rightarrow 2)$:  By symmetry, we consider only the case
$\text{rk}\, \Pi_l  =2$. By definition of $\Pi_l$ and $\Pi_r$,
they are sublattices of the Narain lattice $\Gamma(\tau,\rho)$. Furthermore
we have $ \Pi_r \subset \Pi_l^\perp$, where $\Pi_l^\perp$ is the
orthogonal complement of $\Pi_l$ in $\Gamma(\tau,\rho)$.
Let $x \in \Pi_l^\perp$. Then we have the unique decomposition
$x=x_l+x_r$ over $\bold R$ with some $x_l \in \bold R^{2,0}$ and
$x_r \in \bold R^{0,2}$.
Since $\Pi_l$ has rank 2, we see that $x_l \in \Pi_l \otimes \bold R =
\bold R^{2,0}$.
It follows that $\langle x_l,x\rangle=\langle x_l,x_l\rangle=0$,
so that $x_l=0$.
Therefore $x=x_r\in\Pi_r$. This shows that $\Pi_r \supset \Pi_l^\perp$,
hence $\Pi_r=\Pi_l^\perp$ has rank 2.

\noindent
$2) \Rightarrow 1)$ is obvious.

\noindent
$1) \Leftrightarrow 3)$:
Fix a basis of $\Gamma(\tau,\rho)$ as in (\NarainEs).
Then an arbitrary element of $\Gamma(\tau,\rho)$ may be represented by
$$
p(m,n) =
m_1 \bold e_1(\tau,\rho) +
m_2 \bold e_2(\tau,\rho) +
n_1 \bold e_3(\tau,\rho) +
n_1 \bold e_4(\tau,\rho) \;\;.
\Eqno{\Pmn}
$$
By (\fixRdd) we have a unique decomposition
$p(m,n)=p_l(m,n)+p_r(m,n)$ over $\bold R$ where
$p_l(m,n)=\,^t(*,*,0,0)$ and $p_r(m,n)=\,^t(0,0,*,*)$.
We have
$$
p(m,n) \in \Pi_l=
\Gamma(\tau,\rho)\cap \bold R^{2,0}  \; \Leftrightarrow \; p_r(m,n)=0.
$$
This is equivalent to
$$
\cases
\;\;\,\tau_2 m_1  \hskip0.9cm -\rho_2 n_1 -(\rho_1\tau_2+\tau_1\rho_2)n_2&=0
       \cr
-\tau_1 m_1 + m_2 + \rho_1 n_1 +(\rho_1\tau_1-\tau_2\rho_2)n_2&=0 \;\;.
\endcases
\Eqno{\LeftProjEq}
$$
The integral solutions to these linear equations determine the
lattice $\Pi_l$. The lattice of integral solutions $(\cong \Pi_r)$
is of maximal rank, i.e. $2$, if and only if they can be solved over
$\bold Q$, or in other words, the normal vectors of the hyperplanes
(\LeftProjEq) are in $\bold Q^4$. Therefore we have
$$
\text{rk }\Pi_l=2
\; \Leftrightarrow \;
\tau_1,\rho_1,\rho_1\tau_1-\tau_2\rho_2, \frac{\rho_2}{\tau_2} \in \bold Q
\; \Leftrightarrow \;
\tau_1, \tau_2^2, \rho_1,\rho_2^2, \frac{\rho_2}{\tau_2} \in \bold Q .
\Eqno{\Qcond}
$$
The last condition says that $\tau=\tau_1+\sqrt{-1}
\tau_2$ satisfy a quadratic equation $a \tau^2 + b \tau + c =0$ for some
integers $a,b,c \; (b^2-4ac<0)$ and that the same holds for $\rho=\rho_1+
\sqrt{-1}\rho_2$. By the condition $\frac{\rho_2}{\tau_2}
\in \bold Q$, we see that $\tau, \rho$ must be in the same
quadratic imaginary fields $\bold Q (\sqrt{D})$. Conversely, it is
straightforward to see that for $\tau, \rho \in \bold Q (\sqrt{D})$
the last condition of (\Qcond) holds. Thus we obtain
the equivalence of 1) and 3).
\rqed
\enddemo

\noindent
{\bf Remark.} Condition 3) above was found
by Moore [Mo] to characterize rationality of CFTs on $T^2$.
It is also pointed out by him that the condition that $\tau \in
\bold Q(\sqrt{D})$ is equivalent to that the elliptic curve $E_\tau$
is of CM type, namely $E_\tau$ has non-trivial endmorphisms (see e.g.
[Ha,IV.4]).  A generalization to higher dimensional tori $T^{2d} (d\geq 1)$
has been done in [Wa, Theorem 4.5.5], where it was shown that rationality
implies that $T^{2d}$ is isogeneus to a product
$E_{\tau_1}\times E_{\tau_2} \times \cdots \times E_{\tau_d}$
of elliptic curves of CM types. (More precisely, it is shown that
rationality is equivalent to that $T^{2d}$ is isogeneous to a
product as above with each $\tau_i,\rho_i$ in $\bold Q(\sqrt{D_i})$
for some $D_i<0$ ($i=1,\cdots,d)$.)  As shown in [Mo][Wa],
these points corresponding to RCFTs are dense in the Narain moduli space. It
is argued in [Mo] that a similar density property holds for string
compactifications on K3 surfaces, where RCFTs correspond to
$\sigma$-models on singular
K3 surfaces, i.e. K3 surfaces with $\rho(X)=20$ (see e.g. [SI]).

In case of Calabi-Yau compactifications in dimension three, the connection
to CFTs is less clear except at some special points such as the Gepner
points, where one has a precise dictionary comparing CFTs and the geometry
of Calabi-Yau threefolds (see e.g. [Gr]).
Recently, Gukov-Vafa [GV] have proposed a criterion for
rationality of sigma models
on Calabi-Yau threefolds $X$.
They conjecture that RCFT occurs if and only if both $X$ and
its mirror manifold $X^\vee$ are of CM type. Roughly, a Calabi-Yau
threefold is called of CM type if its (Weil and Griffiths) intermediate
Jacobians are of CM type. See [Bo, Theorem 2.3], for more precise definitions
and several other equivalent definitions of CM type Calabi-Yau threefolds.

\vskip1cm
\head
{\bf \S 4. Classification of $c=2$ RCFT --- primitive case --- }
\endhead

\global\secno=4
\global\propno=4
\global\eqnum=4

In this section, we classify the $c=2$ RCFT data
$[(\Gamma_l,\Gamma_r,\varphi)]$
of primitive triples.
More precisely we describe the image of the set of $RCFT_D^p$
under the ``over-lattice map'' $F$ in (\mapF),
and determine its cardinality (up to an involution).

\vskip0.7cm
\noindent
{\bf (4-1) Narain lattices parameterized by $Cl(D)$.}
Given a positive
definite quadratic form $Q(a,b,c): f(x,y)=a x^2 + b xy + c y^2$, we can write
$$
f(x,y)=a |x+ \tau_{Q(a,b,c)} y |^2 \;\;, \;\;
\tau_{Q(a,b,c)}:=\frac{b+\sqrt{b^2-4ac}}{2a},
$$
where $D=b^2-4ac <0$. So we have a map
$$
Q(a,b,c) \mapsto \tau_{Q(a,b,c)} \in \bold H_+:=\{ x+\sqrt{-1} y \,|\, y>0 \}.
\Eqno{\tauQ}
$$
It is easy to verify that the $SL_2\bold Z$ action on quadratic forms
is compatible with the $PSL_2\bold Z$ actions on $\bold H_+$ under this map.
This shows that the $PSL_2\bold Z$ orbit of
$\tau_{Q(a,b,c)} \in \bold H_+$ depends only on the class
$C=[Q(a,b,c)]\in CL(D)$.
We denote the orbit by $\tau_C$.
We also write $\rho_{Q(a,b,c)}:=\tau_{Q(a,b,c)}, \;\rho_{C}:=\tau_{C}$, and put
$$
[\Gamma(\tau_{C},\rho_{C'})]:=[\Gamma(\tau_Q,\rho_{Q'})]
$$
for any $Q\in C$, $Q'\in C'$. This makes sense
since the equivalence class $[\Gamma(\tau,\rho)]$ is
invariant under $PSL_2\bold Z \times PSL_2 \bold Z$ acting on $\tau,\rho$,
by Proposition {\DualityInvariant} 1). Note that if
$Q(a_1,b_1,c_1),Q(a_2,b_2,c_2)$ belong to the same class $C\in CL(D)$,
then $Q(a_1,-b_1,c_1)$, $Q(a_2,-b_2,c_2)$ also belong to the same class.
Since $-\bar\rho_{Q(a,b,c)}=\rho_{Q(a,-b,c)}$, it follows that
the involution $(\tau,\rho)\mapsto (\tau,-\bar\rho)$
on $\bold H_+\times \bold H_+$, induces an involution on
the classes $[\Gamma(\tau_C,\rho_{C'})]$. If $C,C'$ are primitive,
then $[\Gamma(\tau_C,\rho_{C'})]\mapsto[\Gamma(\tau_C,\rho_{{C'}^{-1}})]$
under this involution,
because $[Q(a,-b,c)]=[Q(a,b,c)]^{-1}$ in the group $Cl(D)$ (see
Proposition A.7 in Appendix A).

\noindent
\Proclaim{Definition}{\NDprimeDef} For negative integers $D < 0$,
we put
$$
 \Cal N_D^p
:=\{ \; [\Gamma(\tau_{\Cal C},\rho_{\Cal C'})] \;\;|\;\;
    \Cal C, \Cal C' \in Cl(D) \; \} \;\;,\;\;
 \widetilde{\Cal N}_D^p
:= \Cal N_D^p /\bold Z_2\;,
\Eqno{\NDprime}
$$
where $\bold Z_2$ denotes the involution on $\widetilde{\Cal N}_D^p$
induced by $(\tau,\rho)\mapsto (\tau,-\bar\rho)$
on $\bold H_+\times \bold H_+$.
\endproclaim

\Proclaim{Proposition}{\cardN} We have a canonical bijection
$\widetilde{\Cal N}_D^p\cong Sym^2\widetilde{Cl}(D)$. In particular, we have
$$
|\widetilde{\Cal N}_D^p | =\frac{1}{2}\widetilde h(D)(\widetilde h(D)+1) .
\Eqno{\numberND}
$$
\endproclaim

\demo{Proof}
Obviously, we have a surjective map
$$
Cl(D)\times Cl(D)\rightarrow\widetilde{\Cal N}_D^p,\hskip.2in
(C,C')\mapsto[\Gamma(\tau_C,\rho_{C'})] \, mod \,\bold Z_2.
$$
Since $\bold Z_2:[\Gamma(\tau_C,\rho_{C'})]\mapsto
[\Gamma(\tau_C,\rho_{{C'}^{-1}})]$,
the pairs $(C,C')$ and $(C,{C'}^{-1})$ have the same image.
Since $[\Gamma(\tau,\rho)]$ is invariant under
duality transformations, we have
$
[\Gamma(\tau,\rho)]=[\Gamma(\rho,\tau)]=[\Gamma(-\bar\rho,-\bar\tau)],
$
by Proposition {\TdualG}. It follows that the pairs
$(C,C'),(C',C),({C'}^{-1},C^{-1})$ also have the same image
under the map. Note that $\widetilde{Cl}(D)=Cl(D)/\approx$,
where $\approx$ is the relation $C\approx C^{-1}$.
This shows that our map
descends to $Sym^2\widetilde{Cl}(D)\rightarrow\widetilde{\Cal N}_D^p$.
It remains to show that this is injective.

Let $(C_1,C_1'),(C_2,C_2')\in Cl(D)\times Cl(D)$ be two pairs having
the same image.
Then $[\Gamma(\tau_{Q_1},\rho_{Q_1'})]=[\Gamma(\tau_{Q_2},\rho_{Q_2'})]$ or
$[\Gamma(\tau_{Q_2},-\bar\rho_{Q_2'})]$, where $Q_i\in C_i$, $Q_i'\in C_i'$.
By Proposition {\DualityInvariant} 1), it follows that
$(\tau_{Q_1},\rho_{Q_1'})$ transforms,
by a duality transformation on $\bold H_+\times\bold H_+$, to
either $(\tau_{Q_2},\rho_{Q_2'})$ or $(\tau_{Q_2},-\bar\rho_{Q_2'})$.
By Proposition {\TdualG}, $(\tau_{Q_1},\rho_{Q_1'})$ transforms, by some
$g\in P(SL_2\bold Z\times SL_2\bold Z) \rtimes (\bold Z_2 \times \bold Z_2)$,
to one of the following:
$$
(\tau_{Q_2},\rho_{Q_2'}),~~ (\rho_{Q_2'},\tau_{Q_2}),~~
(-\bar\rho_{Q_2'},-\bar\tau_{Q_2}),~~ (\tau_{Q_2},-\bar\rho_{Q_2'}),~~
(-\bar\rho_{Q_2'},\tau_{Q_2}),~~
(\rho_{Q_2'},-\bar\tau_{Q_2}).
$$
It follows that $(C_1,C_1')$ is equal to one of the following:
$$
(C_2,C_2'),~~(C_2',C_2),~~({C_2'}^{-1},C_2^{-1}),~~
(C_2,{C_2'}^{-1}),~~({C_2'}^{-1},C_2),~~(C_2',C_2^{-1}).
$$
Therefore $(C_1,C_1')$ and $(C_2,C_2')$ represent
the same element in $Sym^2\widetilde{Cl}(D)$. This shows that
the map $Sym^2\widetilde{Cl}(D)\rightarrow\widetilde{\Cal N}_D^p$ is injective.
\rqed
\enddemo

\vskip0.7cm
\noindent
{\bf (4-2) Key lemma.}
If $\Gamma(\tau,\rho)$, $\Gamma(\tau',\rho')$ are equivalent Narain lattices,
then their left momentum lattices (\PiLR) are isomorphic; likewise for
their right momentum lattices. By Proposition (\DualityInvariant) 1),
this happens if $(\tau,\rho)$ transforms to $(\tau',\rho')$
by a duality transformation on $\bold H_+\times\bold H_+$.
In particular, the isomorphism class of the momentum lattices
of $\Gamma(\tau_Q,\rho_{Q'})$ depends only on
the equivalence classes $[Q]=\Cal C,[Q']=\Cal C'\in Cl(D)$. We denote the
isomorphism classes of the left and right momentum lattices by
$$
\Gamma(\tau_{\Cal C}, \rho_{\Cal C'}) \cap \bold R^{2,0} \;\;, \;\;
\Gamma(\tau_{\Cal C}, \rho_{\Cal C'}) \cap \bold R^{0,2} \;\; \;\;.
\Eqno{\projCC}
$$

\Proclaim{Lemma}{\KeyLemma} {\bf (Key lemma)}
For $\Cal C_1, \Cal C_2 \in Cl(D)$,
we have
$$
\Gamma(\tau_{\Cal C_1},\rho_{\Cal C_2})\cap \bold R^{2,0}
= q(\Cal C_1 * \Cal C_2^{-1}) \;\;,\;\;
\Gamma(\tau_{\Cal C_1},\rho_{\Cal C_2})\cap \bold R^{0,2}
= q(\Cal C_1 * \Cal C_2)(-1) \;\;,\;\;
\Eqno{\KeyProj}
$$
where $*$ is the Gauss product on $Cl(D)$ and $q$ is the natural map
$Cl(D) \rightarrow \widetilde{Cl}(D)$.
\endproclaim
\demo{Proof}
To evaluate the isomorphism classes of the left and right
momentum lattices (\projCC), we choose
forms $Q(a,b,c)\in \Cal C_1$, $Q(a',b',c') \in \Cal C_2$ which
are {\it concordant}. In fact, by Lemma A.3 we can arrange that
$$
(i) \; aa'\not=0,\;(a, a')=1  \qquad
(ii) \;  b=b'  \qquad
(iii) \; \frac{b^2-D}{4 a a'} \in \bold Z \;,
\Eqno{\concord}
$$
where $D=b^2-4ac={b'}^2-4a'c'$.
By definition, the Gauss product $\Cal C_1 * \Cal C_2$ is
$$
[Q(a,b,c)]*[Q(a',b',c')]:=[Q(aa',b,\frac{b^2-D}{4 a a'})] \;\;.
$$
(See Appendix A for details.)
We will compute a $\bold Z$-basis for the right momentum lattice $\Pi_r$
of $\Gamma(\tau,\rho)$, $\tau:=\tau_{Q(a,b,c)},\rho:=\rho_{Q(a',b',c')}$,
and compare
the resulting quadratic form of $\Pi_r$ with $Q(aa',b,\frac{b^2-D}{4 a a'})$.
Likewise for $\Pi_l$.

\noindent
$\bullet$ $\bold Z$-basis for $\Pi_r$.
A vector $p\in\Gamma(\tau,\rho)$ lies in $\Pi_r$
iff $p_l=0$, where $p=p_l+p_r$ is its decomposition under (\fixRdd).
But it is difficult to find a 
$\bold Z$-basis by solving $p_l=0$ directly.
A better way is to first find a $\bold Q$-basis, as follows.
Note that $(\Pi_r)^\perp=\Pi_l$ in $\Gamma(\tau,\rho)$.
Since $\Pi_l$ is a rank two sublattice in a rank four lattice,
it must be defined by
two independent linear integral equations whose coefficients
necessarily form a $\bold Q$-basis of $\Pi_r$. Thus to
find a $\bold Q$-basis of $\Pi_r$, we can write down
defining equations for $\Pi_l$ and read off the coefficients.
As in the proof of Proposition {\ratI},
we denote an arbitrary vector in $\Gamma(\tau,\rho)$
by $p(m,n)=m_1 \bold e_1 + m_2 \bold e_2 + n_1 \bold e_3 + n_2 \bold e_4$.
Then
$$
\aligned
p(m,n)\in\Pi_l
& \Leftrightarrow p_r(m,n)=0  \cr
& \Leftrightarrow
\cases
\quad\;  2a'm_1  \hskip1.5cm - 2 a \;\; n_1 \hskip0.9cm -2 b n_2 &=0 \\
-2a'bm_1+4 a a' m_2 + 2a b n_1+(b^2+D)n_2 &=0  \;\;. \\
\endcases
\endaligned
\Eqno{\przero}
$$
The last two equations read $\langle w_1,p(m,n)\rangle=
\langle w_2,p(m,n)\rangle=0$,
where
$$
w_1 =-a \,\bold e_1-b \,\bold e_2+a' \,\bold e_3  \;\;,\;\;
w_2 =ab \,\bold e_1+(b^2-2ac)\,\bold e_2-a'b\,\bold e_3+2a a' \,\bold e_4.
$$
So $w_1,w_2$ form a $\bold Q$-basis of $\Pi_r\otimes \bold Q$.
Consider the new $\bold Q$-basis
$$
w_1'=w_1= -a \,\bold e_1-b \,\bold e_2+a' \,\bold e_3 \;\;,\;\;
w_2'=\frac{1}{2aa'}(w_2 + b w_1)
= -\frac{c}{a'} \,\bold e_2 + \,\bold e_4  \;\;.
$$
Note that $w_1'$ is an integral primitive vector because
$(a,b)=(a,a')=1$ and so is $w_2'$ by 
$\frac{c}{a'}=\frac{b^2-D}{4aa'} \in \bold Z$.
By looking at the coefficient of $\bold e_4$, we may claim that
they form a $\bold Z$-basis of $\Pi_r$.

Computing the quadratic form, we obtain
$$
\left(
\matrix
(w_1',w_1') & (w_1',w_2') \cr
(w_2',w_1') & (w_2',w_2') \cr
\endmatrix
\right)
= \left(
\matrix
-2aa'  &  -b \cr
-b     & -\frac{2c}{a'} \cr
\endmatrix
\right) \;\;,
$$
which coincides with $-Q(aa',b,\frac{c}{a'})\in\Cal C_1*\Cal C_2$.
It follows that the isomorphism class of the lattice $\Pi_r(-1)$ coincides
with the improper equivalence class of $\Cal C_1*\Cal C_2$, i.e.
$[\Pi_r(-1)]=q(\Cal C_1*\Cal C_2)$.

\noindent
$\bullet$ $\bold Z$-basis for $\Pi_l$. As before, first we find
a $\bold Q$-basis of $\Pi_l\otimes \bold Q$:
$$
u_1=2a \,\bold e_1+ 2 a' \,\bold e_3 \;\;,\;\;
u_2=2ab \,\bold e_1 +(b^2-D) \,\bold e_2 -2 a' b \,\bold e_3
+ 4 a a' \,\bold e_4 \;.
$$
Put
$$
u_1'=\frac{1}{2}u_1=a \,\bold e_1+  a' \,\bold e_3 \;\;,\;\;
u_2'=\frac{1}{4a}(u_2+b u_1)=b \,\bold e_1+c \,\bold e_2 + a' \,\bold e_4 \;.
$$
Since $(a,a')=1$, we have integers $k,l$ satisfying
$ka+la'=1$. Now do a further change of basis to
$$
u_1''=a \,\bold e_1+  a' \,\bold e_3 \;\;,\;\;
u_2''=\frac{1}{a'}(u_2'-kb u_1')=
l b \,\bold e_1 +\frac{c}{a'} \,\bold e_2 - k b \,\bold e_3 + \,\bold e_4 \;.
$$
As before, the coefficients of
$u_1'', u_2''$ allow us to conclude that
they form an integral basis for $\Pi_l$. Computing its quadratic form, we
obtain
$$
\left(
\matrix
(u_1'',u_1'') & (u_1'',u_2'') \cr
(u_2'',u_1'') & (u_2'',u_2'') \cr
\endmatrix
\right)
= \left(
\matrix
2aa'  &  -b+2l ba' \cr
-b+2l ba'     & -2klb^2+\frac{2c}{a'} \cr
\endmatrix
\right) \;\;,
\Eqno{\leftGproduct}
$$
which coincides with $Q(aa',-b+2 l ba',-klb^2+\frac{c}{a'})$.

Now note that
$$
\align
&Q(a,b,c) \simunder{ SL_2\bold Z } Q(a,b-2(kb)a, c_k) = Q(a,-b+2lba',c_k) \cr
&Q(a',b',c')=Q(a',b,c') \simunder{GL_2\bold Z} Q(a',-b,c')
\simunder{SL_2\bold Z}
Q(a',-b+2(lb)a', c_l') \;\;,\cr
\endalign
$$
where $c_k:=c-(kb)b+(kb)^2a$ and $c_l':=c'+(lb)(-b)+(lb)^2a'$.
It is easy to verify that $ Q(a,-b+2lba',*)$ and $Q(a',-b+2(lb)a',*)$
are concordant
forms whose Gauss product coincides with
(\leftGproduct).
Notice that we have used the improper equivalence
once in the equivalence calculation above. So we conclude that
$[\Pi_l]=q(\Cal C_1 * \Cal C_2^{-1})$.
\rqed
\enddemo

\vskip0.2cm

Motivated by the key lemma, we introduce

\Proclaim{Definition}{\NGGdef}
For classes $\GGamma_l,\GGamma_r \in \Cal L^p(D) \equiv
\widetilde{Cl}(D)$, we set
$$
\G_D^p(\GGamma_l,\GGamma_r):=
\{\; (\Cal C_1,\Cal C_2) \in \text{Sym}^2 Cl(D) \;|\;
q(\Cal C_1*\Cal C_2^{-1})=\GGamma_l, \;
q(\Cal C_1*\Cal C_2)=\GGamma_r \;  \}
$$
$$
\G_D^p:=\bigsqcup_{\GGamma_l,\GGamma_r \in \Cal L^p(D)}
\G_D^p(\GGamma_l,\GGamma_r)
\;\;(=\text{Sym}^2 {Cl}(D) ) \;\;.\;\;
\Eqno{\NGG}
$$
Let $\widetilde \G_D^p(\GGamma_l,\GGamma_r)$ and $\widetilde \G_D^p$,
respectively, be the images
under the natural map $\text{Sym}^2 Cl(D) \rightarrow \text{Sym}^2
\widetilde{Cl}(D)$
induced by $q:Cl(D) \rightarrow \widetilde{Cl}(D)$.
\endproclaim

\vskip0.2cm

Note that $\G_D^p(\GGamma_l,\GGamma_r)$ is well-defined since
$q(\Cal C_1*\Cal C_2)=q(\Cal C_2*\Cal C_1)$ and also
$q(\Cal C_1*\Cal C_2^{-1})=q(\Cal C_2*\Cal C_1^{-1})$.
Note also that
$\widetilde \G^p_D(\GGamma_l,\GGamma_r)=
 \widetilde \G^p_D(\GGamma_r,\GGamma_l)$, and that
$\tilde \G_D^p(\GGamma_l,\GGamma_r)=
 \tilde \G_D^p(\GGamma_l',\GGamma_r')$ holds if and only if
$\{ \GGamma_l,\GGamma_r \}=\{ \GGamma_l',\GGamma_r' \}$.
{}From this, we have:
$$
\widetilde \G_D^p =
\bigsqcup_{(\GGamma_l,\GGamma_r) \in \text{Sym}^2 \widetilde{Cl}(D)}
\tilde \G^p_D(\GGamma_r,\GGamma_l) \;\;(=
\text{Sym}^2 \widetilde{Cl}(D)) \;\;.\;\;
\Eqno{\cardNGG}
$$

\vskip0.7cm
\noindent
{\bf (4-3) Classification.}
Now we
complete our classification.

\Proclaim{Proposition}{\PropFf}
The inverse of the ``over-lattice map''
$F^{-1}: \Cal N \rightarrow \cup_D RCFT_D$ (see (\mapF))
restricts to an injective map $f: \Cal N_D^p \rightarrow RCFT_D^p$, i.e.
$$
\matrix F^{-1}: & \Cal N & \rightarrow & \cup_D RCFT_D \cr
                &  \cup  &             & \cup  \cr
        f:      & \Cal N_D^p & \rightarrow & RCFT_D^p \cr
\endmatrix
$$
\endproclaim

\demo{Proof} By (\inverseFD), we have $F^{-1}:
[\Gamma(\tau,\rho)]\mapsto[(\Pi_l,\Pi_r(-1),\iota_R\circ\iota_L^{-1})]$.
Let $[\Gamma(\tau,\rho)] \in \Cal N_D^p$, i.e.
$\tau=\tau_{Q_1}, \rho=\rho_{Q_2}$ for some forms $Q_i$ of discriminant $D$.
Then $[\Pi_l]=q([Q_1]*[Q_2]^{-1}), [\Pi_r]=q([Q_1]*[Q_2])$,
by Lemma {\KeyLemma}.
It follows immediately that $\Pi_l,\Pi_r$ are both primitive
lattices of determinant $-D$.
This shows that $F^{-1}(\Cal N_D^p)\subset RCFT_D^p$.
\rqed
\enddemo

Consider the quotient
$$
 \widetilde{RCFT_D^p} := RCFT_D^p/(\pi_2:
[(\Gamma_l,\Gamma_r,\varphi)] \mapsto
[(\Gamma_r,\Gamma_l,\varphi^{-1})] ) ,
$$
and the decomposition
$$
\widetilde{RCFT_D^p}=
\bigsqcup_{(\GGamma_l,\GGamma_r) \in \text{Sym}^2 \widetilde{Cl}(D)}
\widetilde{RCFT^p_D}(\GGamma_l,\GGamma_r) \;\;,
\Eqno{\RCFTdecomp}
$$
with
$$
\widetilde{RCFT^p_D}(\GGamma_l,\GGamma_r):=
\{ \;[(\Gamma_l,\Gamma_r,\varphi)]\,mod\,(\pi_2) \; | \;
[\Gamma_l]=\GGamma_l,\; [\Gamma_r]=\GGamma_r \;\} .
$$
where $[(\Gamma_l,\Gamma_r,\varphi)]\,mod\,(\pi_2)$ represents the
$\bold Z_2$-orbit;
$[(\Gamma_l,\Gamma_r,\varphi)] \sim [(\Gamma_r,\Gamma_l,\varphi^{-1})]$.

\Proclaim{Lemma}{\ZtwoInvs} The involutions
$\pi_1: (\tau,\rho)\mapsto (\tau,-\bar\rho)$ and
$\pi_2: [(\Gamma_l,\Gamma_r,\varphi)]
\mapsto [(\Gamma_r,\Gamma_l,\varphi^{-1})]$
are compatible with the map $f: \Cal N_D^p \rightarrow RCFT_D^p$,
i.e. we have the diagram
$$
\matrix
[\Gamma(\tau,\rho)] &\mapright{f}
&[(\Pi_l,\Pi_r(-1),\iota_R\circ \iota_L^{-1})]&\\
 \mapdown{\pi_1} &       &   \mapdown{\pi_2}  &\\
[\Gamma(\tau,-\bar\rho)] &\mapright{f} &
                     [(\Pi_r(-1),\Pi_l,\iota_L\circ \iota_R^{-1})] &\;\;.\\
\endmatrix
\Eqno{\ZtwoComp}
$$
\endproclaim

\demo{Proof}
Consider the explicit basis $\bold e_1(\tau,\rho),\cdots,\bold e_4(\tau,\rho)$
of $\Gamma(\tau,\rho)$ in (\NarainEs).
The involution $\pi_2: (\tau,\rho)\mapsto (\tau,-\bar\rho)$ exchanges
the upper two components and the lower
two components of each $\bold e_j(\tau,\rho)$
(up to signs). Let $\pi_0:\bold R^{2,2}\rightarrow\bold R^{2,2}$
be the linear map defined by $\,^t(1,0,0,0)\leftrightarrow\,^t(0,0,1,0)$,
$\,^t(0,1,0,0)\leftrightarrow\,^t(0,0,0,1)$. Then we have
$\pi_0\Gamma(\tau,\rho)=\Gamma(\tau,-\bar\rho)$.
Clearly $\pi_0$ is an involutive anti-isometry of $\bold R^{2,2}$, i.e.
$\langle \pi_0(x),\pi_0(x)\rangle=-\langle x,x\rangle$,
which exchanges the two subspaces $\bold R^{2,0}$ and $\bold R^{0,2}$.
Let $\Pi_l,\Pi_r$ be the left and the right momentum lattices,
and $\iota_L,\iota_R$ the isomorphisms (\isoDis) for $\Gamma(\tau,\rho)$.
Then we have
$$
\pi_0\Pi_l=\Gamma(\tau,-\bar\rho)\cap\bold R^{0,2}=:\Pi_r',\hskip.2in
\pi_0\Pi_r=\Gamma(\tau,-\bar\rho)\cap\bold R^{2,0}=:\Pi_l',
$$
where $\Pi_l',\Pi_r'$ are the left and the right
momentum lattices of $\Gamma(\tau,-\bar\rho)$. This shows that
$\pi_0:\Pi_l\rightarrow\Pi_r'(-1)$, $\pi_0:\Pi_r(-1)\rightarrow\Pi_l'$,
define lattice isomorphisms.
Let $\iota_L',\iota_R'$
be the isomorphisms (\isoDis) for $\Gamma(\tau,-\bar\rho)$.
Then it is easy to check that
$\iota_R'\circ{\iota_L'}^{-1}=\bar\pi_0\circ\iota_L\circ\iota_R^{-1}\circ
\bar\pi_0$.
It follows that the triples
$(\Pi_l',\Pi_r'(-1),\iota_R'\circ{\iota_L'}^{-1})$,
$(\Pi_r(-1),\Pi_l,\iota_L\circ\iota_R^{-1})$ are equivalent.
This shows that $f:[\Gamma(\tau,-\bar\rho)]\mapsto
[(\Pi_r(-1),\Pi_l,\iota_L\circ\iota_R^{-1})]$.
\rqed
\enddemo

Now we can state our main theorems:

\Proclaim{Theorem}{\firstThm} Consider the map
$g: \G_D^p \rightarrow \Cal N^p_D$ defined by
$(\Cal C_1,\Cal C_2) \mapsto [\Gamma(\tau_{\Cal C_1},\rho_{\Cal C_2})]$, and
the injective map $f:\Cal N^p_D \rightarrow RCFT_D^p$ in
Proposition {\PropFf}. Then there exist corresponding induced maps
$\tilde g: \widetilde \G_D^p  \rightarrow
\tilde \Cal N^p_D$ and
$\tilde f: \tilde \Cal N^p_D \rightarrow \widetilde{RCFT_D^p}$ such that
the following diagram commutes:
$$
\CD
\G_D^p  @>g>> \Cal N^p_D @>f>> RCFT^p_D \\
@VV{q}V  @VV{\pi_1}V   @VV{\pi_2}V  \\
 \widetilde \G_D^p @>{\tilde g}>>
\widetilde{\Cal N}^p_D @>{\tilde f}>> \widetilde{RCFT_D^p} \\
\endCD
$$
Moreover $\tilde g$ is bijective and $\tilde f$ is injective.
\endproclaim

\demo{Proof}
In Proposition {\cardN}, we saw that
the map
$$
Cl(D)\times Cl(D)\rightarrow\Cal N_D^p,\hskip.2in (C_1,C_2)
\mapsto[\Gamma(\tau_{C_1},\rho_{C_2})],
$$
induces the bijection
$\widetilde\G_D^p=\text{Sym}^2\widetilde{Cl}(D)\rightarrow
\tilde\Cal N_D^p$ (as well as the surjection $g: \Cal G_D^p=\text{Sym}^2
Cl(D) \rightarrow \Cal N_D^p$).
This is the bijection $\tilde g$ we seek.
The commutative diagram (\ZtwoComp) implies that $f$ induces
a map $\tilde f$, as required. That $\tilde f$ is injective
follows immediately from the fact that $f$ is injective.
\rqed
\enddemo

\Proclaim{Theorem}{\secondThm}
\item{1)} The map $f: \Cal N_D^p \rightarrow RCFT_D^p$ is bijective.
Hence $\tilde f$ is also bijective.
\item{2)} The composition $\tilde f\circ \tilde g$ is a bijection with
$$
\tilde f\circ \tilde g\left(
\widetilde \G_D^p(\GGamma_l,\GGamma_r)\right)= 
\widetilde{RCFT_D^p}(\GGamma_l,\GGamma_r) \;\;.
\Eqno{\EqthmII}
$$
\endproclaim

\demo{Proof} 1) By the preceding theorem,
both $f$ and $\tilde f$ are injective.
Surjectivity of $f$ will be proved in section (5-2).
Thus $\tilde f$ is also surjective by (\ZtwoComp).

\noindent
2) That $\tilde f\circ\tilde g$ is bijective follows from 1)
and the preceding theorem.
Let $(C_1,C_2)\in\G^p_D(\GGamma_l,\GGamma_r)$, i.e.
$q(C_1*C_2)=\GGamma_l$ and $q(C_1*C_2^{-1})=\GGamma_r$.
By the key lemma (Lemma {\KeyLemma}), we have
$f\circ g((C_1,C_2))=f([\Gamma(\tau_{C_1},\rho_{C_2})])
\in RCFT_D^p(\GGamma_l,\GGamma_r)$. This shows that
$$
\tilde f\circ\tilde g\left(\widetilde\G^p_D(\GGamma_l,\GGamma_r)\right)
\subset\widetilde{RCFT_D^p}(\GGamma_l,\GGamma_r).
$$
The reverse inclusion follows from the fact that
$\tilde f\circ\tilde g$ is a bijection.
\rqed
\enddemo

\Proclaim{Proposition}{\PropAfterThm} For primitive classes $\GGamma_l,
\GGamma_r \in \tilde Cl(D)$, the bijection
$B_{\Gamma_l,\Gamma_r}: RCFT_D(\GGamma_l,\GGamma_r) \rightarrow
O(\Gamma_l)\setminus \text{Isom}(A_{\Gamma_l},A_{\Gamma_r}) /
O(\Gamma_r)$ in (\RtoCo) induces a bijection
$$
\widetilde{RCFT^p_D}(\GGamma_l,\GGamma_r) \; \leftrightarrow \;
O(\Gamma_l)\setminus
\text{Isom}(A_{\Gamma_l},A_{\Gamma_r}) / O(\Gamma_r) \;.
$$
\endproclaim

\demo{Proof}
We only need to show that the surjective map
$RCFT^p_D(\GGamma_l,\GGamma_r)\rightarrow
\widetilde{RCFT^p_D}(\GGamma_l,\GGamma_r)$,
$[(\Gamma_l,\Gamma_r,\varphi)]\mapsto[(\Gamma_l,\Gamma_r,\varphi)]mod(\pi_2)$
is also injective. This is a restriction of the map
$\pi_2:RCFT^p_D\rightarrow \widetilde{RCFT^p_D}$. The preimage of a given
$[(\Gamma_l,\Gamma_r,\varphi)]mod(\pi_2)$ consists of the two RCFT data
$[(\Gamma_l,\Gamma_r,\varphi)], [(\Gamma_r,\Gamma_l,\varphi^{-1})]$.
If $\GGamma_l \not=\GGamma_r$, then only the first RCFT
data lie in $RCFT^p_D(\GGamma_l,\GGamma_r)$.
If $\GGamma_l=\GGamma_r$, then the two RCFT data are of the shape
$[(\Gamma,\Gamma,\psi)]$, $[(\Gamma,\Gamma,\psi^{-1})]$ for
some $\psi\in O(A_\Gamma)$.
In Appendix B, we show that $\psi=\psi^{-1}$.
Thus, in either case, the preimage of
$[(\Gamma_l,\Gamma_r,\varphi)]mod(\pi_2)$ in $RCFT^p_D(\GGamma_l,\GGamma_r)$
contains just one RCFT data.
\rqed
\enddemo

As a corollary to Theorem {\secondThm}
and Proposition {\PropAfterThm}, we have the following equality.

\Proclaim{Corollary}{\CorToThm}
For $\GGamma_l,\GGamma_r \in \widetilde{Cl}(D)$, and for arbitrary
choices of lattices $\Gamma_l \in \GGamma_l$ and  $\Gamma_r \in \GGamma_r$,
we have
$$
|\widetilde \G_D^p(\GGamma_l,\GGamma_r)|=
|O(\Gamma_l)\setminus
\text{Isom}(A_{\Gamma_l},A_{\Gamma_r})/O(\Gamma_r)| \;\;.
\Eqno{\GaussCoset}
$$
\endproclaim

\noindent
{\bf Remark.}
1) It is interesting to note that this equality connects
the double coset space, which is group theoretical and arithmetical
in nature, to the Gauss product, which is algebraic.
\par\noindent
2) The average one formula in Theorem {\averageOne} follows
immediately from the
above equality and (\cardNGG).
\par\noindent
3) As pointed out earlier, if $\Gamma_l$ is not isogeneous
to $\Gamma_r$, then the set $\text{Isom}(A_{\Gamma_l},A_{\Gamma_r})$ is empty
(see e.g. [Ni, Corollary 1.9.4]), which implies
$|\widetilde \G_D^p(\GGamma_l,\GGamma_r)|=0$.
This can also be shown directly from the properties of Gauss product
(see [Ca]).

\vskip1cm
\head
{\bf \S 5. Classification of $c=2$ RCFT --- non-primitive case --- }
\endhead

\global\secno=5
\global\propno=1
\global\eqnum=1

We now consider the general case of
even, positive definite, not necessarily primitive
lattices $\GGamma_l, \GGamma_r \in
\Cal L(D)$ of fixed determinant $-D$. As one might expect, most of the
arguments in the preceding section apply to this case,
once the Gauss product is extended to $CL(D)$. This
extension is done in Appendix A.

\vskip0.7cm
\noindent
{\bf (5-1) Refinement of key lemma.} Let us look more closely at
the proof of the key lemma (Lemma \KeyLemma).
The proof relies only on the existence of concordant forms satisfying
properties (\concord).  Using those properties,
we found a $\bold Z$-basis for the left and right momentum
lattices $\Pi_l=\Gamma(\tau_{Q_1},\rho_{Q_2})\cap \bold R^{2,0}$,
$\Pi_r=\Gamma(\tau_{Q_1},\rho_{Q_2})\cap \bold R^{0,2}$
for $Q_1 \in \Cal C_1, Q_2 \in \Cal C_2$, and
identified their isomorphism classes $[\Pi_l]$ and $[\Pi_r]$
with the Gauss products
$q(\Cal C_1 * \Cal C_2^{-1})$ and $q(\Cal C_1 * \Cal C_2)$, respectively.
Thus in the case of non-primitive forms,
if (\concord) holds for $C_1,C_2\in CL(D)$ and
the composition $C_1*C_2$ is defined, then we expect that
the same lemma holds for $C_1,C_2$, as we now show.

We recall some elementary notions introduced
in Appendix A. We define
the gcd $\lambda:=gcd(a,b,c)$ of
a class $[Q(a,b,c)] \in CL(D)$, and the scalar multiple of a class
in an obvious manner.
Clearly a class is primitive iff its gcd is 1. In particular,
if $\lambda$ is the gcd of $C\in CL(D)$, then ${1\over\lambda}C$ is primitive.
Two classes are {\it coprime} if their gcd's are coprime.
It is shown in Appendix A that there is
a composition law $C_1*C_2$ for coprime classes
$C_1, C_2 \in CL(D)$, that generalizes the Gauss product for primitive classes.

\Proclaim{Lemma}{\refKeyLemma} {\bf (Refinement of the key lemma) }
For coprime quadratic forms $C_1, C_2 \in
CL(D)$, the following relations hold;
$$
\Gamma(\tau_{C_1},\rho_{C_2}) \cap \bold R^{2,0} = q(C_1*\sigma C_2)
\;\;,\;\;
\Gamma(\tau_{C_1},\rho_{C_2}) \cap \bold R^{0,2}= q(C_1*C_2)(-1) \;\;,
\Eqno{\refKeyProj}
$$
where $*$ is the composition law for coprime classes in $CL(D)$, and $q$ is the
natural map $q: CL(D) \rightarrow \widetilde{CL}(D) \equiv \Cal L(D)$.
\endproclaim

\demo{Proof} As before, we can choose quadratic forms
$Q(\alpha,\beta,\gamma) \in C_1$, $Q(\alpha',\beta',\gamma')$ $\in C_2$
to evaluate the left hand sides of the equations (\refKeyProj).
By Lemma A.3, we can arrange that
$$
(i) \;\;\alpha \alpha'\not=0,\;\; (\alpha,\alpha')=1 \qquad
(ii) \;\;  \beta=\beta'  \qquad
(iii) \;\; \frac{\beta^2-D}{4 \alpha \alpha'} \in \bold Z \;.
\Eqno{\concordRef}
$$
Now we see that the proof of Lemma {\KeyLemma} is valid verbatim,
by respectively replacing $a,b,c; a',b',c'$ there by
$\alpha,\beta,\gamma; \alpha',\beta',\gamma'$ here,
and $C^{-1}$ there by $\sigma C$ here (cf. Proposition A.7, 4)).
\rqed
\enddemo

For the composition law on $CL(D)$, the following property
will be useful (see Appendix A, Remark after Definition A.6 for a proof):

\Proclaim{Lemma}{\CCprimitive} For coprime classes $C_1, C_2 \in CL(D)$,
if $C_1*C_2 \in Cl(D)$ then $C_1,C_2\in Cl(D)$.
\endproclaim

\vskip0.7cm
\noindent
{\bf (5-2) Proof of surjectivity in Theorem {\secondThm}.}
We now proceed to proving the surjectivity part of Theorem {\secondThm}.

\Proclaim{Lemma}{\refineLemma}
Let $(\Gamma_l,\Gamma_r,\varphi)$ be a (not necessarily primitive) triple
of determinant $-D$ with
$\Gamma_l\subset\bold R^{2,0}$ and $\Gamma_r(-1)\subset\bold R^{0,2}$,
and consider the corresponding over-lattice $\Gamma^\varphi$.
There exist coprime quadratic forms $Q_1,Q_2$ of discriminant $D$
such that $\Gamma^\varphi$ is equivalent to $\Gamma(\tau_{Q_1},\rho_{Q_2})$.
\endproclaim

\demo{Proof}
By construction, as in section (3-4), $\Gamma^\varphi\subset\bold R^{2,2}$
is a Narain lattice. By Proposition {\DualityInvariant} 2),
$\Gamma^\varphi=h\Gamma(\tau,\rho)$ for some $\tau,\rho\in\bold H_+$
and $h\in O(2,\bold R)\times O(2,\bold R)$. So we have
$$
\Gamma_l=\Gamma^\varphi\cap \bold R^{2,0} = h\Pi_l \;\;,\;\;
\Gamma_r(-1)=\Gamma^\varphi\cap \bold R^{0,2} =h\Pi_r  \;\;,\;\;
\Eqno{\surjProofI}
$$
where $\Pi_l,\Pi_r$ are the momentum lattices of the left and right
chiral algebras of $\Gamma(\tau,\rho)$.
This shows that $\text{rk }\Pi_l=\text{rk }\Pi_r=2$.
By Proposition \QD, we have $\tau,\rho \in \bold Q(\sqrt{D'})$
for some $D'<0$.
So explicitly, $\tau$ and $\rho$ may be written
$$
\tau=\frac{b+c \sqrt{D'}}{2a} = \frac{bc'+\sqrt{(cc')^2 D'}}{2a c'} \;\;,\;\;
\rho=\frac{b'+c' \sqrt{D'}}{2a'} = \frac{b'c+\sqrt{(cc')^2 D'}}{2a' c} \;\;,
$$
where $(a,b,c)=(a',b',c')=1$ and $a,a',c,c'>0$.
Now we express these
values in terms of the following quadratic forms
of discriminant $(K cc')^2D'$, i.e. we have
$\tau=\tau_{Q_1}, \rho=\rho_{Q_2}$ with
$$
\align
&Q_1= Q(K ac', K bc', K \frac{(bc')^2-(cc')^2 D'}{4ac'}) \;\;,\;\; \\
&Q_2= Q(K a'c, K b'c, K \frac{(b'c)^2-(cc')^2 D'}{4a'c}) \;\;,\;\;
\endalign
$$
where a positive integer $K$ is chosen so that both
$Q_1$ and $Q_2$ are integral.
Clearly, we can choose such an integer $K$ so that $Q_1$ and $Q_2$
are coprime as well. Thus the composition $[Q_1]*[Q_2]$ makes sense.
Now by Lemma {\refKeyLemma}, we find
that $\text{det}\,\Pi_l =\text{det}\,\Pi_r=(K cc')^2D'$.
It follows that
$D=(K cc')^2D'$, since we have $\text{det}\,\Pi_l =\text{det}\,\Gamma_l=D$ by
(\surjProofI). Since
$\Gamma^\varphi=h\Gamma(\tau,\rho)=h\Gamma(\tau_{Q_1},\rho_{Q_2})$,
this completes the proof.
\rqed
\enddemo

\vskip0.2cm
\demo
{\it Proof of the surjectivity in Theorem {\secondThm}}
We want to show that the map $f: \Cal N_D^p \rightarrow RCFT_D^p$,
$[\Gamma(\tau,\rho)]\mapsto[(\Pi_l,\Pi_r(-1),\iota_R\circ\iota_L^{-1})]$
is surjective.
Recall that every element in $RCFT_D^p$ has the shape
$[(\Gamma_l,\Gamma_r,\varphi)]$, where $\Gamma_l\subset\bold R^{2,0}$
and $\Gamma_r(-1)\subset\bold R^{0,2}$ are of determinant $-D$ and also 
have primitive bilinear forms. 
By the preceding lemma, we have
$[\Gamma^\varphi]=[\Gamma(\tau,\rho)]$, where $\tau=\tau_{Q_1}$,
$\rho=\rho_{Q_2}$,
for some coprime forms $Q_1,Q_2$ of discriminant $D$.
In particular, we have
$$
\Pi_l=\Gamma(\tau,\rho)\cap\bold R^{2,0}=h\Gamma^\varphi\cap\bold
R^{2,0}=h\Gamma_l
$$
for some $h\in O(2;\bold R)\times O(2;\bold R)$. This shows that
$\Pi_l\cong\Gamma_l$;
likewise $\Pi_r\cong\Gamma_r(-1)$.
By Lemma {\refKeyLemma}, we have
$$
[\Gamma_l]=[\Pi_l]=q([Q_1]*\sigma[Q_2])\;\;,\;\;
[\Gamma_r(-1)]=[\Pi_r]=q([Q_1]*[Q_2])\;.
$$
Since the bilinear forms of $\Gamma_l,\Gamma_r$ are primitive, 
Lemma {\CCprimitive} implies $[Q_1],[Q_2]$ are also primitive, 
i.e. lie in $Cl(D)$.
This shows that $[\Gamma(\tau,\rho)]\in\Cal N_D^p$. Hence we have
$f:[\Gamma(\tau,\rho)]\mapsto[(\Pi_l,\Pi_r(-1),\iota_R\circ\iota_L^{-1})]$.
It remains to show that
$[(\Pi_l,\Pi_r(-1),\iota_R\circ\iota_L^{-1})]=[(\Gamma_l,\Gamma_r,\varphi)]$.
It is straightforward to check that the isomorphism
$h:\Gamma^\varphi\rightarrow\Gamma(\tau,\rho)$
induces an equivalence of triples
$(\Gamma_l,\Gamma_r,\varphi)\sim(\Pi_l,\Pi_r(-1),\iota_R\circ\iota_L^{-1})$.
\rqed
\enddemo

\vskip0.7cm
\noindent
{\bf (5-3) Classification.}  Using the refined key lemma
(Lemma \refKeyLemma), it is easy to generalize
Theorems {\firstThm}, {\secondThm}, and
Proposition {\PropAfterThm} to the non-primitive case, with only minor
modifications. The following definitions are, respectively, extensions
of Definitions {\NGGdef} and \NDprimeDef:

\Proclaim{Definition}{\NGGdefMulti} For $\GGamma_l,\GGamma_r \in \Cal L(D)
\equiv \widetilde{CL}(D)$, we set
$$
\G_D(\GGamma_l,\GGamma_r):=\{\; (C_1,C_2) \in \text{PSym}^2 CL(D) \;|\;
q(C_1*\sigma C_2)=\GGamma_l, \;
q(C_1*C_2)=\GGamma_r \;  \},
$$
$$
\G_D:=
\bigsqcup_{\GGamma_l,\GGamma_r \in CL(D)} \G_D(\GGamma_l,\GGamma_r)\;\;
(=\text{PSym}^2 CL(D) ) \;,
\Eqno{\NGG}
$$
where $\text{PSym}^2 CL(D):=\{ (C,C') \in \text{Sym}^2 CL(D) \,| \,
C,C': \text{coprime } \}$. Also we define
$\widetilde \G_D(\GGamma_l,\GGamma_r)$ and $\widetilde \G_D$
to be the image
under the natural map $\text{PSym}^2 CL(D) \rightarrow \text{PSym}^2
\widetilde{CL}(D)$ induced by $q:CL(D) \rightarrow \widetilde{CL}(D)$.
\endproclaim

As before, we have the following decomposition:
$$
\widetilde \G_D =
\bigsqcup_{(\GGamma_l,\GGamma_r) \in \text{PSym}^2 \widetilde{CL}(D)}
\tilde \G_D(\GGamma_r,\GGamma_l) \;\;(=
\text{PSym}^2 \widetilde{CL}(D)) \;\;.\;\;
\Eqno{\cardNGGExt}
$$

\Proclaim{Definition}{\NDdefMulti} We set:
$$
\Cal N_D:=\{\;[\Gamma(\tau_{C_1},\rho_{C_2})] \,|\, C_1, C_2 \in CL(D):
\text{ coprime} \;\} \;\;,
$$
and $\widetilde{\Cal N}_D$ be the image under $q: CL(D) \rightarrow
\widetilde{CL}(D)$.
\endproclaim

Let  $\widetilde{RCFT_D}$ be the quotient of $RCFT_D$ by the involution
$\pi_2$ (see Lemma \ZtwoInvs). Then we have the decomposition
$$
\widetilde{RCFT_D}=
\bigsqcup_{(\GGamma_l,\GGamma_r) \in \text{Sym}^2 \widetilde{CL}(D)}
\widetilde{RCFT_D}(\GGamma_l,\GGamma_r) \;\;,
\Eqno{\RCFTdecompExt}
$$
as in (\RCFTdecomp).

\Proclaim{Proposition}{\PropFfExt} The inverse of the ``over-lattice map''
$F^{-1}: \Cal N \rightarrow \cup_D RCFT_D$ (see (\mapF)) restricts to
an injective map $f: \Cal N_D \rightarrow RCFT_D$. Moreover $f$ is
compatible with the involutions $\pi_1: (\tau,\rho)
\mapsto (\tau,-\bar\rho)$ and $\pi_2:[(\Gamma_l,\Gamma_r,\varphi)]
\mapsto [(\Gamma_r,\Gamma_l,\varphi^{-1})]$ (see the diagram (\ZtwoComp)).
\endproclaim

\demo{Proof} If $F^{-1}: [\Gamma(\tau_{C_1},\rho_{C_2})] \in \Cal N_D
\mapsto [(\Pi_l,\Pi_r(-1), \varphi)]$, then
$[\Pi_l] =q(C_1*\sigma C_2), [\Pi_r]$ $= q(C_1*C_2)$ by
Lemma \refKeyLemma. From these relations, we conclude
$\text{det}\,\Pi_l = \text{det}\,\Pi_l=D$ and hence
$[(\Pi_l,\Pi_r(-1), \varphi)]\in RCFT_D$.
The compatibility with the involutions $\pi_1,\pi_2$
follows from the same argument
as in Lemma {\ZtwoInvs}.
\rqed
\enddemo

Now Theorem {\firstThm} generalizes to

\Proclaim{Theorem}{\firstThmMulti} Let $g$ be the map
$\G_D \rightarrow \Cal N_D$ defined by
$(C_1,C_2) \mapsto [\Gamma(\tau_{C_1},\rho_{C_2})]$, and
$f$ be the injective map $\Cal N_D \rightarrow RCFT_D$ given in
Proposition {\PropFfExt}. Then there exist corresponding induced maps
$\tilde g: \widetilde \G_D  \rightarrow \tilde \Cal N_D$ and
$\tilde f: \tilde \Cal N_D \rightarrow \widetilde{RCFT_D}$ such that
the following diagram commutes:
$$
\CD
\G_D @>g>> \Cal N_D @>f>> RCFT_D \\
@VV{q}V  @VV{\pi_1}V   @VV{\pi_2}V  \\
\widetilde \G_D @>{\tilde g}>>
\tilde{\Cal N}_D @>{\tilde f}>> \widetilde{RCFT_D} \\
\endCD
$$
Moreover $\tilde g$ is bijective and $\tilde f$ is injective.
\endproclaim

Since the proof of Theorem {\firstThm} carries over
with straightforward modifications, we do not
repeat it here. Theorem {\secondThm} now generalizes to

\Proclaim{Theorem}{\secondThmMulti}
\item{1)} The map $f: \Cal N_D \rightarrow RCFT_D$ is bijective. Hence
$\tilde f$ is also bijective.
\item{2)} The composition $\tilde f\circ \tilde g$ is a bijection with
$$
\tilde f\circ \tilde g\left(
\widetilde \G_D(\GGamma_l,\GGamma_r)\right)= 
\widetilde{RCFT_D}(\GGamma_l,\GGamma_r) \;\;.
\Eqno{\EqthmIIExt}
$$
\endproclaim

\demo{Proof} 1) By Proposition {\PropFfExt}, $f$ is injective.
The proof of surjectivity of $f$ is similar to that
of Theorem {\secondThm} in section (5-2).
Since $f$ is bijective, so is $\tilde f$ by Proposition {\PropFfExt}.

\noindent
2) Again, the proof of Theorem {\secondThm} 2) carries over
to the non-primitive case.
\rqed
\enddemo

Proposition {\PropAfterThm} and its corollary,
however, do not generalize immediately to non-primitive forms.
This is because the relation $\varphi^2=id$ for $\varphi \in O(A_\Gamma)$
no longer holds in general for a non-primitive lattice $\Gamma$.

\Proclaim{Proposition}{\PropAfterThmMulti}
\item{1)}
If $\GGamma_l\not=\GGamma_r$ $(\GGamma_l,\GGamma_r \in \widetilde{CL}(D))$,
the bijective map $B_{\Gamma_l,\Gamma_r}$ in (\RtoCo) defines
a natural one-to-one correspondence
$$
\widetilde{RCFT_D}(\GGamma_l,\GGamma_r) \; \leftrightarrow \;
O(\Gamma_l)\setminus
\text{Isom}(A_{\Gamma_l},A_{\Gamma_r}) / O(\Gamma_r) \;\;.\;
$$
\item{2)}
If $\GGamma_l=\GGamma_r=:\GGamma$, there is a
one-to-one correspondence
$$
\widetilde{RCFT_D}(\GGamma,\GGamma) \; \leftrightarrow \;
\left( O(\Gamma)\setminus
\text{Isom}(A_{\Gamma},A_{\Gamma}) / O(\Gamma) \right)/\sim \;,
$$
where $\sim$ represents the identification of
$[\varphi]$ with $[\varphi^{-1}]$.
The correspondence is given by mapping the class
$[(\Gamma,\Gamma,\varphi)] \sim [(\Gamma,\Gamma,\varphi^{-1})]$
to the class $[\varphi] \sim [\varphi^{-1}]$.
\endproclaim

The proof of Proposition {\PropAfterThm} now carries over
with some slight modifications in part 2), although we omit the details here.
As a corollary, we have

\Proclaim{Corollary}{\CorToThmMulti}
For $\GGamma_l,\GGamma_r \in \widetilde{CL}(D)$, and for any choice of
lattices $\Gamma_l \in \GGamma_l$ and $\Gamma_r \in \GGamma_r$,
the following equality holds:
$$
|\widetilde \G_D(\GGamma_l,\GGamma_r)|=
\cases
|O(\Gamma_l)\setminus \text{Isom}(A_{\Gamma_l},A_{\Gamma_r})/O(\Gamma_r)|
\;\;, &  \GGamma_l\not=\GGamma_r \cr
\frac{1}{2}
|O(\Gamma)\setminus \text{Isom}(A_{\Gamma},A_{\Gamma})/O(\Gamma)|
+ \frac{1}{2} n_\Gamma
& \GGamma_l=\GGamma_r=:\GGamma \cr
\endcases
\Eqno{\GaussCosetMulti}
$$
where $n_\Gamma:=\# \{ [\varphi] \in
O(\Gamma)\setminus Isom(A_\Gamma,A_\Gamma)/O(\Gamma)\;|\;
[\varphi^{-1}]=[\varphi] \,\}$.
\endproclaim

\vskip0.2cm

\noindent
{\bf Example.} Here for the reader's convenience, we present an example of
a non-primitive lattice $\Gamma$ which has an isometry $[\varphi]\not=
[\varphi^{-1}]$. The example has determinant $-D=236$ and bilinear
form
$$
\left( \matrix 12 & 2 \cr 2 & 20 \cr \endmatrix \right) \text{ with }
\Gamma=\bold Z v_1 \oplus \bold Z v_2.
$$
The discriminant group $\Gamma^*/\Gamma$ is isomorphic to $\bold Z_{d_1}
\oplus \bold Z_{d_2}=:\langle u_{d_1} \rangle \oplus \langle u_{d_2} \rangle$
with $d_1=2,\, d_2=118$. The generators may be chosen explicitly as $
u_{d_1}=\frac{1}{2}v_2,\; u_{d_2}=\frac{1}{118}(v_1-6v_2)$. The discriminant
form $q_\Gamma$ may be evaluated by using these expression for $u_{d_1},
u_{d_2}$. Now let us make the following orthogonal decomposition
of the discriminant form (cf. Appendix B):
$$
(\Gamma^*/\Gamma, q_\Gamma)=( \,
(\bold Z_2 \oplus \bold Z_2) \oplus \bold Z_{59}, q_1\oplus q_2\,) \;,
$$
where $q_1$ represents the discriminant form on the component
$\bold Z_2 \oplus \bold Z_2$.  If we write the generators for each
factor of $\bold Z_2 \oplus \bold Z_2 \oplus \bold Z_{59}$, respectively,
by $u, v, v^\perp$, they are given by
$u=\frac{1}{2}v_2,\,
v=\frac{1}{2} v_1,\,
v^\perp=\frac{1}{59}(v_1-6v_2)$.
With respect to these generators, the discriminant form $q_\Gamma$
may be represented by
$$
q_\Gamma=q_1\oplus q_2=\left(
\matrix 1 & \frac{1}{2} \cr
\frac{1}{2} & 1 \cr \endmatrix \right) \oplus (\frac{12}{59}) \;\;.
$$
It is rather straightforward to determine all isometries in $O(A_\Gamma)$.
The results are:
$$
\aligned
&
\left(
\matrix 1 & 0 \cr
        0 & 1 \cr \endmatrix \right) \oplus (\pm 1) \;,\;
\left(
\matrix 0 & 1 \cr
        1 & 0 \cr \endmatrix \right) \oplus (\pm 1) \;,\;
\left(
\matrix 1 & 1 \cr
        0 & 1 \cr \endmatrix \right) \oplus (\pm 1) \;,\;
\cr
&
\left(
\matrix 1 & 0 \cr
        1 & 1 \cr \endmatrix \right) \oplus (\pm 1) \;,\;
\left(
\matrix 0 & 1 \cr
        1 & 1 \cr \endmatrix \right) \oplus (\pm 1) \;,\;
\left(
\matrix 1 & 1 \cr
        1 & 0 \cr \endmatrix \right) \oplus (\pm 1) \;.\;
\endaligned
\Eqno{\exampleIso}
$$
As for the group $O(\Gamma)$, we see that it is trivial, i.e. $O(\Gamma)=
\{ \pm id \}$. Looking at the induced action on $\Gamma^*/\Gamma$, we see that
$O(\Gamma)$ acts on $O(A_\Gamma)$ by
$\{
\left(
\smallmatrix 1 & 0 \cr
             0 & 1 \cr \endsmallmatrix \right) \oplus (\pm 1)  \}$.
Hence each of the six pairs in (\exampleIso) represents the class
$O(\Gamma)\setminus O(A_\Gamma)/O(\Gamma)$. Now for the classes
$[\varphi_1]=
\{
\left(
\smallmatrix 0 & 1 \cr
             1 & 1 \cr \endsmallmatrix \right) \oplus (\pm 1)  \}$,
$[\varphi_2]=
\{
\left(
\smallmatrix 1 & 1 \cr
             1 & 0 \cr \endsmallmatrix \right) \oplus (\pm 1)  \}$,
we see $[\varphi_1^{-1}]=[\varphi_2],\, [\varphi_1]=[\varphi_2^{-1}]$.
This should be contrasted to the fact that we have $\varphi^2=id$ (and
hence $[\varphi^{-1}]=[\varphi]$) for
all $\varphi \in O(A_\Gamma)$ if $\Gamma$ is a primitive lattice
(see Appendix B).
\rqed

\vskip0.7cm
\noindent
{\bf (5-4) Diagonal RCFTs.} In their recent work [GV], Gukov and Vafa
obtained a characterization of diagonal rational conformal field theories.
We summarize
their result as follows:

\noindent
{\it (Diagonal RCFTs) Let $\Pi$ be
a lattice with quadratic form $\lambda \left(
\smallmatrix 2a & b \cr b & 2c \\ \endsmallmatrix \right)$ such that
$gcd(a,b,c)
=1$ and $D=\lambda^2(b^2-4ac)$. Then a necessary and sufficient
condition for a class of Narain lattices $[\Gamma(\tau,\rho)] \in \Cal N_D$ to
satisfy $f([\Gamma(\tau,\rho)])=[(\Pi,\Pi,id)]$ is that
$$
\tau=\frac{b+\sqrt{D/\lambda^2}}{2a} \;\;,\;\;
\rho=\lambda a \tau \;\;,
\Eqno{\GVdiagonal}
$$
where the equalities are understood to be
up to $PSL_2 \bold Z$ transformations.
}

This characterization can be seen as a special case
of our general classification scheme
as follows:

\Proclaim{Proposition}{\diagonalRCFT} {\bf (Diagonal triples)}
Under the bijection
$f: \Cal N_D \rightarrow {RCFT}_D$,
the `diagonal' triples $[(\Pi,\Pi,id)]$
correspond to Narain lattices of the shape $[\Gamma(\tau_{C}, \rho_{C_{e}})]$,
where $[\Pi]=q(C)$, and $C_{e}\in CL(D)$ is the unit class
(see Proposition A.7 in Appendix A). Moreover, we have
$\rho_{C_{e}}=\lambda a\tau_C$ if $C=[\lambda Q'(a,b,c)]$.
\endproclaim

\demo{Proof}
Put $Q:=\lambda Q'(a,b,c)$, and let $(\Pi_l,\Pi_r,\iota_R\circ\iota_L^{-1})$
be the triple
corresponding to the Narain lattice $\Gamma(\tau_Q,\rho_{Q_e})$, where
$Q_e\in C_e$ is the reduced form. Then
$f([\Gamma(\tau_{C}, \rho_{C_{e}})])=[(\Pi_l,\Pi_r,\iota_R\circ\iota_L^{-1})]$.
By Lemma {\refKeyLemma}, we have
$[\Pi_l]=[\Pi_r]=q(C*C_e)=q(C)$. In fact, by
computing $\Pi_l,\Pi_r$ explicitly, as in the proof of Lemma {\KeyLemma},
one finds that $\iota_R\circ\iota_L^{-1}$ is
induced by an isomorphism $\iota:\Pi_l\rightarrow\Pi_r$ (cf. [GV]).
It follows that
$$
f([\Gamma(\tau_{C}, \rho_{C_{e}})])=[(\Pi,\Pi,id)]
\Eqno{\tem}
$$
where $\Pi=\Pi_l$. Conversely, given any rank two lattice $\Pi$,
if $C\in CL(D)$ with $\Pi=q(C)$, then (\tem) holds.

Note that $\rho_{C_e}$ is the $PSL_2\bold Z$ orbit
of $\sqrt{D}/2$ if $D\equiv 0$ mod $4$, or
of $(1+\sqrt{D})/2$ if $D\equiv 1$ mod $4$.
Also we have $\tau_{\lambda Q(a,b,c)}=
((\lambda b)^2 +\sqrt{D})/(2\lambda a)$, and
$D\equiv (\lambda b)^2$ mod $4$. It is easy to see that
$\lambda a \, \tau_{\lambda Q(a,b,c)}=((\lambda b)^2 +\sqrt{D})/2$
is in the  $PSL_2\bold Z$ orbit of $\sqrt{D}/2$ if $D\equiv 0$ mod $4$, or
of $(1+\sqrt{D})/2$ if $D\equiv 1$ mod $4$. It follows that
$\rho_{C_e}=\lambda a \tau_{C}$.
\rqed
\enddemo

\vskip0.7cm
\noindent
{\bf (5-5) Summary and an example ($D=-144$).}

We summarize the main results of this paper a bit differently,
and illustrate them in an example as follows.

\Proclaim{Summary}{\summary} {}
Any two coprime classes of positive definite
 quadratic forms $A,B\in CL(D)$ yield an RCFT
(i.e. an element in $RCFT_D$)
with the momentum lattices of the left and right chiral algebras
$q(A*\sigma B)$, $q(A*B)\in\widetilde{CL}(D)$. Conversely,
all RCFTs arise this way. Moreover, the RCFT arising
from $A,B\in CL(D)$ in this way
corresponds to the class of Narain lattices $[\Gamma(\tau_A,\rho_B)]$.
Two pairs $(A',B')$ and $(A,B)$ yield the same RCFT iff
$[\Gamma(\tau_{A'},\rho_{B'})]=[\Gamma(\tau_A,\rho_B)]$.
\endproclaim

\vskip0.2cm

We will describe
the set $\widetilde{RCFT}_D$ for $D=-144$, in terms of the Gauss product
on quadratic forms. By Proposition {\reducedF}, we find
8 classes $C_i$ in $CL(D)$:
$$
\align
&C_1=[Q(1,0,36)]\;,\;
C_2=[Q(4,0,9)]\;,\;
C_3=[Q(5,4,8)]\;,\;
C_4=[Q(5,-4,8)]\;,\;  \cr
&
C_5=[Q(2,0,18)]\;,\;
C_6=[Q(3,0,12)]\;,\;
C_7=[Q(4,4,10)]\;,\;
C_8=[Q(6,0,6)]\;,\; \cr
\endalign
$$
where $C_4 = \sigma C_3$ and $C_5$ to $C_8$ are not primitive.
Their compositions are given in the following
table:

\vskip0.3cm
\centerline{
\vbox{\offinterlineskip
\def\vspa{\omit & \omit &height1pt& \omit && \omit  && \omit &\cr}
\halign{ \strut
       #&  $\;$  $#$  \hfil
&{\vrule# width1pt} &  $\;$ \hfil $#$ \hfil
&      #&  $\;$ \hfil $#$ \hfil
&      #&  $\;$ \hfil $#$ \hfil
&      #&  $\;$ \hfil $#$ \hfil
&      #&  $\;$ \hfil $#$ \hfil
&      #&  $\;$ \hfil $#$ \hfil
&      #&  $\;$ \hfil $#$ \hfil
&      #&  $\;$ \hfil $#$ \hfil
&      #
\cr
&   &&  C_1  && C_2 &&  C_3 && C_4 && C_5 && C_6 && C_7 && C_8 &\cr
\noalign{\hrule height1pt}
\vspa
&    C_1
  &&  C_1
  &&  C_2
  &&  C_3
  &&  C_4
  &&  C_5
  &&  C_6
  &&  C_7
  &&  C_8
  &\cr
\vspa
&  C_2
  &&  C_2
  &&  C_1
  &&  C_4
  &&  C_3
  &&  C_5
  &&  C_6
  &&  C_7
  &&  C_8
  &\cr
\vspa
&      C_3
  &&  C_3
  &&  C_4
  &&  C_2
  &&  C_1
  &&  C_7
  &&  C_6
  &&  C_5
  &&  C_8
  &\cr
\vspa
&   C_4
  &&  C_4
  &&  C_3
  &&  C_1
  &&  C_2
  &&  C_7
  &&  C_6
  &&  C_5
  &&  C_8
  &\cr
\vspa
&    C_5
  &&  C_5
  &&  C_5
  &&  C_7
  &&  C_7
  &&  -
  &&  C_8
  &&  -
  &&  -
  &\cr
\vspa
&    C_6
  &&  C_6
  &&  C_6
  &&  C_6
  &&  C_6
  &&  C_8
  &&  -
  &&  C_8
  &&  -
  &\cr
\vspa
&     C_7
  &&  C_7
  &&  C_7
  &&  C_5
  &&  C_5
  &&  -
  &&  C_8
  &&  -
  &&  -
  &\cr
\vspa
&      C_8
  &&  C_8
  &&  C_8
  &&  C_8
  &&  C_8
  &&  -
  &&  -
  &&  -
  &&  -
  &\cr
} }   }
\vskip0.2cm
{\leftskip1cm \rightskip1cm
\noindent
{\bf Table 3.}  Table of Gauss product on $CL(-144)$. The blanks ``$-$''
mean that the product is not defined.
 \par }

\vskip0.3cm

Let us denote by $\bar C_i=q(C_i)$ the $GL_2\bold Z$ equivalence classes
of $C_i$. Then the set $\widetilde{CL}(D)$ consists 7 classes, $\bar C_1,
\cdots, \bar C_3=\bar C_4, \cdots, \bar C_8$. From Table 3, it is easy to
determine the set $\G_D(\bar C_i, \bar C_j)$ defined in (\NGG). For example,
we have,
$$
\align
&
\G_D(\bar C_1,\bar C_2)=\{ (C_3,C_3) , (C_4,C_4) \} \;\;,\;\; \cr
&
\G_D(\bar C_3,\bar C_3)=\{ (C_1,C_3), (C_1,C_4), (C_2, C_3), (C_2,C_4) \}
\;\;,  \\
&
\G_D(\bar C_8,\bar C_8)=\{ (C_1,C_8), (C_2,C_8), (C_3,C_8), (C_4,C_8),
(C_5,C_6),(C_6,C_7) \} \;.
\endalign
$$
We also have
$$
\align
&
\widetilde \G_D(\bar C_1,\bar C_2)=\{ (\bar C_3,\bar C_3) \} \;\;, \cr
&
\widetilde \G_D(\bar C_3,\bar C_3)
=\{ (\bar C_1,\bar C_3), (\bar C_2,\bar C_3)  \}
 \\ \;\, \;\;
&
\widetilde \G_D(\bar C_8,\bar C_8)=\{ (\bar C_1,\bar C_8), (\bar C_2,\bar C_8),
(\bar C_3,\bar C_8), (\bar C_5,\bar C_6),(\bar C_6,\bar C_7) \} \;.
\endalign
$$
By Theorem {\secondThmMulti}, the set
$\widetilde{RCFT_D}(\GGamma_l,\GGamma_r)$ is in one to one correspondence
to the set $\tilde g (\widetilde \G_D(\GGamma_l,\GGamma_r))
\subset \tilde \Cal N_D$, i.e. Narain lattices up to the parity involution.
For the above examples, we have, respectively
$$
\align
&
\{ [\Gamma(\tau_{C_3},\rho_{C_3})]_{\bold Z_2} \}  \;\;,\;\;
\{ [\Gamma(\tau_{C_1},\rho_{C_3})]_{\bold Z_2},
[\Gamma(\tau_{C_2},\rho_{C_3})]_{\bold Z_2} \}  \;\;, \\
&
\{ [\Gamma(\tau_{C_1},\rho_{C_8})]_{\bold Z_2},
[\Gamma(\tau_{C_2},\rho_{C_8})]_{\bold Z_2},
[\Gamma(\tau_{C_3},\rho_{C_8})]_{\bold Z_2},
[\Gamma(\tau_{C_5},\rho_{C_6})]_{\bold Z_2},
[\Gamma(\tau_{C_6},\rho_{C_7})]_{\bold Z_2} \},
\endalign
$$
where $[\Gamma(\tau,\rho)]_{\bold Z_2}$
represents the $\bold Z_2$-orbit: $[\Gamma(\tau,\rho)] \sim
[\Gamma(\tau,-\bar\rho)]$. The diagonal triples correspond to the
Narain lattices $[\Gamma(\tau_{C_1},\rho_{C_k})]$ for $k=1,\cdots,8$.

A direct computation of the double coset spaces is not difficult, and it is
a good exercise to verify the equality obtained in Corollary {\CorToThmMulti}
in this case.

\vskip0.3cm

\vfill\eject

\head {\bf Appendix A. Gauss product on $CL(D)$ }
\endhead

\def\App{A}
\global\propno=1 \global\eqnum=1

\def\notdivide{\;\backslash \hskip-4pt | \hskip5pt}

In this appendix, we extend the Gauss product on $Cl(D)$ to a
composition law $C_1*C_2$ on $CL(D)$, which includes non-primitive
forms, but defined only when the classes $C_1, C_2 \in CL(D)$ are
coprime. The composition is commutative and
associative whenever defined. As seen in Lemmas \KeyLemma, and
\refKeyLemma, this extended composition law does arise naturally
in our description of RCFTs.

The observation here is that the classical composition law given
in terms of primitive concordant forms is valid verbatim for
non-primitive forms. We say that two forms $Q(a_1,b_1,c_1)$ and
$Q(a_2,b_2,c_2)$ are {\it concordant} if they satisfy the
following conditions;
$$
(1) \quad a_1 a_2 \not= 0 \;\;,\qquad (2) \quad b_1=b_2 (=:b)
\;\;,\qquad (3) \quad \frac{b^2-D}{4a_1a_2} \in \bold Z \;\;.
\EqnoApp{\defconcordant}
$$
Note that in case of $D<0$, the first condition (1) is void since $a_1
\not=0, a_2\not=0$ for $D=b_1^2-4a_1c_1=b_2^2-4a_2c_2<0$.
We say that the forms are coprime if
$gcd(a_1,b_1,c_1,a_2,b_2,c_2)=1$. It is clear that if
$Q(a_1,b_1,c_1)$ is coprime with one quadratic form in a class
$C\in CL(D)$, then it is coprime with all quadratic forms in $C$.
Thus it makes sense to speak of coprime classes. We write
$Q(a,b,*)$ to denote $Q(a,b,\frac{b^2-D}{4a})$. We call the number
$gcd(a,b,c)$ the gcd of the form $Q(a,b,c)$. Since equivalent
forms have the same gcd, we can speak of the gcd of a class.

The following construction parallels to that given in [Ca] in the
primitive case, and is valid for both $D<0$ and $D>0$. For the
modern definition using fractional ideals, see for example [Za].

\ProclaimApp{Lemma}{\lemmaAI} Consider a primitive class $\Cal C
\in Cl(D)$. For an arbitrary nonzero integer $M$, there is a
quadratic form $Q(a,b,c) \in \Cal C$ such that $(a,M)=1$.
\endproclaim

\demo{Proof} Take an quadratic form $Q(a',b',c'):
f(x,y)=a'x^2+b'xy+c'y^2 \in \Cal C$. Then $f$ represents an
integer which is coprime to $M$. To see this, let us first consider
the case where $M$ has only one prime factor, say $M=p^{e}$. In
this case the claimed integer $f(x,y)$ may be found by considering
the following four cases:

{\leftskip0.5cm
\item{(1)} When $p \notdivide a'$. Take $x,y$ so that
$p \notdivide x, p | y$, then we have $(f(x,y),p)=1$.
\item{(2)} When $p | a'$, $p \notdivide c'$. Take $x,y$ so that
$p | x, p \notdivide y$, then we have $(f(x,y),p)=1$.
\item{(3)} When $p| a', p | c'$. In this case $p \notdivide b'$ by
the condition $gcd(a',b',c')=1$ ($Q(a',b',c')$ is primitive). If we take
$p \notdivide x, p \notdivide y$, then we have $(f(x,y),p)=1$.
\par}

\noindent In the general case, let $M=p_0^{e_0}p_1^{e_1} \cdots
p_k^{e_k}$ be the prime factorization of $M$. Put
$$
 S_1=\{ p_i \,|\; p_i \notdivide a' \,\} \;,\;
 S_2=\{ p_i \,|\; p_i | a' , \; p_i \notdivide c' \,\} \;,\;
 S_3=\{ p_i \,|\; p_i | a' , \; p_i | c' \,\} \;,\;
$$
and define
$$
x=(\prod_{p \in S_2} p) \bar x \;\;,\;\;
y=(\prod_{q \in S_1} q) \bar y \;\;,\;\;
$$
with some integers $\bar x$ and $\bar y$ satisfying
$$
p \notdivide \bar x \;\; (p \in S_1 \cup S_3) \;\; \text{and} \;\;
p \notdivide \bar y \;\; (p \in S_2 \cup S_3).
$$
Then it is clear that for the all prime factors $p_0, \cdots, p_k$
of $M$, if $p_i \in S_r$ $(r=1,2,3)$,
then $x$ and $y$ have the properties ($r$),
and thus we have $(f(x,y), M)=1$.

In this way, we find $(x,y)=(n_1,n_2)$ such that $f(n_1,n_2)$ is
coprime to $M$. We may assume that $(x,y)=(n_1,n_2)$ is primitive in
$\bold Z^2$, since otherwise we may set $(x,y)=(\frac{n_1}{m},\frac{n_2}{m})
\in \bold Z^2$, with $m=gcd(n_1,n_2)$, preserving the property
$(f(\frac{n_1}{m},\frac{n_2}{m}),M)=1$. When $(n_1,n_2)$ is primitive
in $\bold Z^2$, there is an $SL_2\bold Z$ transformation
$g: (n_1,n_2) \mapsto(1,0)$.
Then the quadratic form $Q(a,b,c)=g\cdot Q(a',b',c')$ has
the desired property. \rqed
\enddemo

\ProclaimApp{Lemma}{\lemmaAII} Assume two primitive quadratic
forms are equivalent: $Q(a_1,b_1,c_1) \sim Q(a_2,b_2,c_2)$ and
$b_1=b_2=:b$. If there is an integer $l \in \bold Z$ such that
$l | c_1, l | c_2$ and $(a_1,a_2,l)=1$, then
$$
Q(l a_1,b, c_2/l) \; \sim \; Q(l a_2, b, c_2/l) \;\;.
$$
\endproclaim

\demo{Proof} Suppose $\left( \smallmatrix r  & s  \cr
        t  & u \cr
\endsmallmatrix \right)\in SL_2\bold Z$
transforms $Q(a_1,b_1,c_1)$ to $Q(a_2,b_2,c_2)$, i.e.
$$
\bigg(^{\hskip-0.3cm t} \hskip0.2cm \matrix r  & s  \cr
        t  & u \cr
\endmatrix \bigg)
\left(
\matrix 2a_1 & b    \\
         b   & 2c_1 \\
\endmatrix \right)
\left( \matrix r  & s  \cr
        t  & u \cr
\endmatrix \right)
= \left(
\matrix 2a_2 & b    \\
         b   & 2c_2 \\
\endmatrix \right)  \;\;,
\EqnoApp{\rstuMat}
$$
Eliminating $r,u$ from the resulting equations, we obtain $a_2 s +
c_1 t=a_1s+c_2t=0$. From these relations and our assumptions, we
see that $l|s$.  Then the matrix $\left( \smallmatrix r  & s/l
\cr
        l t  & u \cr
\endsmallmatrix \right)$ transforms
$Q(l a_1,b, c_2/l)$ to $Q(l a_2, b, c_2/l)$.  \rqed
\enddemo

\ProclaimApp{Lemma}{\lemmaAIII} Let $C_1$, $C_2$ be coprime
classes in $CL(D)$, and $\lambda_1,\lambda_2$ be their respective
gcd's.
For an arbitrary nonzero integer $M$, there exist quadratic forms
$Q_1 \in C_1, Q_2 \in C_2$ such that
$$
Q_1=Q(\lambda_1a_1,\beta_1,\gamma_1)\;\;,\;\;
Q_2=Q(\lambda_2a_2,\beta_2,\gamma_2)\;\;,\;\;
$$
with $(a_1,a_2)=(a_1,M)=(a_2,M)=1$ and also
$\beta_1=\beta_2=:\beta$. Furthermore $Q_1$ and $Q_2$ are
concordant.
\endproclaim

\demo{Proof} Applying Lemma {\lemmaAI} to the primitive class
$\frac{1}{\lambda_1} C_1$, we obtain a quadratic form $Q(a_1,b_1,c_1)
\in \frac{1}{\lambda_1} \; C_1$ with $(a_1,\lambda_2 M)=1$.
Likewise, we have $Q(a_2,b_2,c_2) \in \frac{1}{\lambda_2} \; C_2$
with $(a_2,\lambda_1 a_1M)=1$. Let
$$
Q_1:=Q(\lambda_1 a_1, \beta_1, \gamma_1) \;\;,\;\;
Q_2:=Q(\lambda_2 a_2, \beta_2, \gamma_2) \;\;,\;\;
$$
with $\beta_i:=\lambda_i b_i, \gamma_i:=\lambda_i c_i$. Since
$\lambda_1$ and $\lambda_2$ are coprime, we have $(\lambda_1
a_1,\lambda_2 a_2)=1$, hence there exist integers $A_1, A_2$
satisfying $\lambda_1a_1A_1 + \lambda_2a_2 A_2=1$. Then we have the
equivalence under $SL_2\bold Z$
$$
Q_1 \sim Q(\lambda_1a_1, \beta_1-2\lambda_1a_1
\frac{\beta_1-\beta_2}{2}A_1, *) \,,\;\; Q_2 \sim Q(\lambda_2a_2,
\beta_2-2\lambda_2a_2 \frac{\beta_2-\beta_1}{2}A_2, *) \;\;.
$$
Note that $\beta_1^2-\beta_2^2 \equiv 0$ mod $4$, so that
$(\beta_1-\beta_2)/2$ is an integer.
We see that the above
two quadratic forms are of the form;
$
Q_1 \sim Q(\lambda_1 a_1,B,C_1) \,,\,
Q_2 \sim Q(\lambda_2 a_2,B,C_2) \,,\,
$
with $B:=\lambda_2a_2 A_2 \beta_1+\lambda_1 a_1 A_1 \beta_2$, and
$C_1=(B^2-D)/(4\lambda_1 a_1)$ and $C_2=(B^2-D)/(4\lambda_2 a_2)$.
Since $(\lambda_1 a_1,\lambda_2 a_2)=1$, the quadratic forms
$Q(\lambda_1 a_1, B, C_1)$ and $Q(\lambda_2 a_2, B, C_2)$ are
concordant forms which have the asserted properties.
\rqed
\enddemo

\ProclaimApp{Definition}{\DefAI} For any concordant forms
$Q_1=Q(\alpha_1,\beta,\gamma_1), Q_2=Q(\alpha_2,\beta,\gamma_2)$
of a same discriminant $D$, we set the {\it composition} of
$Q_1, Q_2$ by
$$
Q_1 * Q_2 := Q(\alpha_1
\alpha_2,\beta,\frac{\beta^2-D}{4\alpha_1\alpha_2}) \;\;.
\EqnoApp{\composition}
$$
\endproclaim

\ProclaimApp{Proposition}{\propAI} Assume $C_1, C_2 \in CL(D)$ are
coprime. The class $C_3\in CL(D)$ of the composition $Q_1*Q_2$ is
independent of the choices of concordant forms $Q_1 \in C_1$, $Q_2
\in C_2$.
\endproclaim

\demo{Proof} Suppose that $ Q_1'=Q(\alpha_1',\beta',\gamma_1'),
\; Q_2'=Q(\alpha_2',\beta',\gamma_2')$ are concordant, that $
Q_1''=Q(\alpha_1'',\beta'',\gamma_1''), \;
Q_2''=Q(\alpha_2'',\beta'',\gamma_2'')$ are also concordant, and that
$Q_1'\sim Q_1'', Q_2'\sim Q_2''$. We will show that $Q_1'*Q_2'\sim
Q_1''*Q_2''$.

Let $\lambda_i$ be the gcd of $Q_i',Q_i''$. Put
$M=a_1'a_2'a_1''a_2'' \;\; (\alpha_i:=\lambda_i a_i',
\alpha_i'':=\lambda_i a_i'')$. Then by Lemma {\lemmaAIII}, there
exists another pair of concordant forms,
$$
Q_1=Q(\lambda_1a_1,\beta, \gamma_1) \in C_1 \;\;,\;\;
Q_2=Q(\lambda_2a_2,\beta, \gamma_2) \in C_2 \;\;,\;\;
$$
satisfying $(a_1,a_2)=(a_1,M)=(a_2,M)=1$. Since $(\lambda_1,
\lambda_2)=1$, $\lambda_1|\beta, \lambda_2|\beta$, we have
$\beta=\lambda_1\lambda_2 b$ for some integer $b$; likewise
$\beta'=\lambda_1\lambda_2 b'$, and $D=\lambda_1^2 \lambda_2^2
D_0$.
Therefore from (\composition), we have
$$
Q_1*Q_2=\lambda_1\lambda_2 Q(a_1a_2,b,\frac{b^2-D_0}{4a_1a_2}
)\;,\;\; Q_1'*Q_2'=\lambda_1\lambda_2
Q(a_1'a_2',b',\frac{{b'}^2-D_0}{4a_1'a_2'})\;.\;\;
$$
Since $(a_1a_2, a_1'a_2')=1$, there exist integers such that
$a_1'a_2'm_{12}' +a_1a_2 m_{12}=1$. From this, we get
$$
\aligned &Q_1*Q_2=\lambda_1\lambda_2 \, Q(a_1a_2,b,*) \sim
         \lambda_1\lambda_2 Q(a_1a_2,B,*)    \;\;, \\
&Q_1'*Q_2'=\lambda_1\lambda_2 \, Q(a_1'a_2',b',*) \sim
         \lambda_1\lambda_2 Q(a_1'a_2',B,*)  \;\;, \\
\endaligned
\EqnoApp{\QQprime}
$$
where $B:=b-2a_1a_2\frac{b-b'}{2}m_{12}=b'-2a_1'a_2'\frac{b'-b}{2}m_{12}'$.

Now from the shape of $B$, we find
$$
Q(a_1,\lambda_2 B,c_1) \sim Q(a_1,\lambda_2b,*) \sim
Q(a_1',\lambda_2b',*) \sim Q(a_1',\lambda_2 B,c_1') \;\;,\;\;
$$
where $c_1=\lambda_2^2 (B^2-D_0)/4a_1$,
$c_1'=\lambda_2^2(B^2-D_0)/4a_1'$ with $D=\lambda_1^2\lambda_2^2
D_0$. Then we see that $\lambda_2 a_2'|c_1$ and $\lambda_2
a_2'|c_1'$ since we have $(B^2-D_0)/4a_1 a_2,
 (B^2-D_0)/4a_1' a_2'  \in \bold Z $
from the equivalences in (\QQprime) and
$(a_1'a_2',a_1a_2)=1$. By Lemma {\lemmaAII}, we have
$$
\lambda_2 Q( a_1a_2', B, c_1/\lambda_2^2 a_2' ) \sim \lambda_2
Q(a_1'a_2', B, c_1'/\lambda_2^2 a_2' ) \;\;.\;\;
$$
In a similar way, starting from $ Q(a_2,\lambda_1B,c_2) \sim
Q(a_2,\lambda_1 b, * ) \sim  Q(a_2',\lambda_1 b',*) \sim
Q(a_2',\lambda_1 B, c_2')$ with $c_2=\lambda_1^2(B^2-D_0)/4a_2$
and $c_2'=\lambda_1^2(B^2-D_0)/2a_2'$, we obtain
$$
\lambda_1 Q(a_1a_2,B,c_2/\lambda_1^2a_1 ) \sim \lambda_1
Q(a_1a_2',B,c_2'/\lambda_1^2a_1 ) \;\;.
$$
Combining these two, and using
$c_1/\lambda_2^2a_2'=c_2'/\lambda_1^2a_1$, we obtain
$$
\lambda_1\lambda_2 Q(a_1a_2,B,c_2/\lambda_1^2a_1) \sim
\lambda_1\lambda_2 Q(a_1'a_2',B,c_1'/\lambda_2^2a_2' )  \;\;.
\EqnoApp{\ansI}
$$
Now from (\QQprime) and (\ansI), we see that $Q_1*Q_2 \sim
Q_1'*Q_2'$. Likewise, we get $Q_1*Q_2 \sim
Q_1''*Q_2''$. It follows that $Q_1'*Q_2'
\sim Q_1''*Q_2''$.  \rqed
\enddemo

By the proposition above, the
composition of coprime classes in $CL(D)$ now makes sense:

\ProclaimApp{Definition}{\DefAII} For coprime classes $C_1, C_2
\in CL(D)$, we define
$$
C_1*C_2=[Q_1*Q_2] \in CL(D)\;\;,
$$
for any choice of concordant forms $Q_1 \in C_1, Q_2 \in C_2$.
\endproclaim

\noindent {\bf Remark} 1) If both classes $C_1, C_2$ are
primitive, i.e. in $Cl(D)$, then the composition is nothing but
the Gauss product.

\noindent 2)  If $C_1,C_2 \in
CL(D)$ are coprime classes and $\lambda_1,\lambda_2$ their
respective gcd's, then
we have
$$
\frac{1}{\lambda_1 \lambda_2} \times  C_1* C_2  \in
Cl(D/(\lambda_1\lambda_2)^2) \;,
$$
i.e., it is a primitive class. This follows from
definition (\composition) and the fact that  $\lambda_1\lambda_2 |
\beta$ and $(\lambda_1\lambda_2)^2| (\beta^2-D)$ hold because
$(\lambda_1,\lambda_2)=1$.

\noindent 3) It follows from 2) that if either
$C_1$ or $C_2$ is not primitive, then the composition $C_1*C_2$
is not primitive. This is an important fact, used in our proof
of surjectivity in Theorem {\secondThm}.   \rqed

The following
properties of the composition law generalize the classical properties
(Theorem \gaussCl) of Gauss' group law on $(Cl(D),\,*\,)$.

\ProclaimApp{Proposition}{\propAII} For pairwise coprime classes
$C_1,C_2,C_3 \in CL(D)$, the following properties hold:
\item{1)} $C_1*C_2=C_2*C_1$,
\item{2)} $(C_1*C_2)*C_3=C_1*(C_2*C_3)$
\item{3)} Let $C_e$ be the class containing $Q(1,0,-\frac{D}{4})$ for
$D\equiv 0 \text{ mod } 4$ or $Q(1,1,\frac{1-D}{4})$ for $D\equiv
1 \text{ mod } 4$, then $C_e*C=C$ for any $C\in CL(D)$.
\item{4)} If $C_1 \in Cl(D)$, i.e. primitive class, then
$\sigma C_1*C_1=C_e$, where $\sigma$ is the involution
defined by $Q(a,b,c)\mapsto Q(a,-b,c)$ as in
(\invsigma).
\endproclaim

\demo{Proof} 1) This follows from Definition {\DefAII} based on
Proposition {\propAI} and (\composition).

\noindent 2) Let $\lambda_i$ be the gcd of $C_i$.
By Lemma {\lemmaAIII}, we
have concordant forms
$Q_1=Q(\lambda_1a_1,\beta_1,\gamma_1)$,
$Q_2=Q(\lambda_2a_2,\beta_2,\gamma_2)$ with $(a_1,a_2)=1$ and
$\beta_1=\beta_2=:\beta$. Then by Proposition {\propAI}, we have
$Q_1*Q_2=Q(\lambda_1\lambda_2a_1a_2,\beta,*) \in C_1*C_2$. Since
$C_1*C_2$ is a multiple by $\lambda_1\lambda_2$ of a primitive
class (see Remark 2)) and $(\lambda_1\lambda_2,\lambda_3)=1$ by
assumption, we may apply Lemma {\lemmaAIII} to coprime classes
$C_1*C_2$ and $C_3$ with $M=a_1a_2$. By this we see that there
exist a quadratic form $Q(\lambda_3a_3,\beta_3,*) \in C_3$ with
$(a_1a_2,a_3)=1$. Now we note the following equivalences;
$$
Q(\lambda_1\lambda_2a_1a_2,\beta, * ) \sim
Q(\lambda_1\lambda_2a_1a_2,b, * ) \;\; , \;\; Q(\lambda_3a_3,
\beta_3, * ) \sim Q(\lambda_3a_3, b, *) \;\;,
$$
where $b=\beta-2\lambda_1\lambda_2a_1a_2 \frac{\beta-\beta_3}{2}
m_{12} = \beta_3-2\lambda_3a_3\frac{\beta_3-\beta}{2}m_3$ with
integers $m_{12}, m_3$ satisfying $a_1a_2 m_{12}+a_3 m_3=1$.
{}From the shape of $b$, we find
$$
Q_1\sim Q_1'=Q(\lambda_1a_1,b, * ) \;,\; Q_2\sim
Q_2'=Q(\lambda_2a_2,b, * ) \;,\; Q_3\sim Q_3'=Q(\lambda_3a_3,b, *
) \;,\;
$$
with conditions $(a_1,a_2)=(a_1,a_3)=(a_2,a_3)=1$. It follows that
$$
(Q_1'*Q_2')*Q_3' =Q(\lambda_1\lambda_2\lambda_3 a_1a_2a_3,b, * )
=Q_1'*(Q_2'*Q_3') \;\;.
$$
Here we have used the fact that each
pair of quadratic forms being composed are concordant.
This yields the associativity on classes by Proposition {\propAI}.

\noindent 3) Any class $C \in CL(D)$ is coprime to the class $C_e$,
and so $C*C_e$ is defined. For any quadratic
form $Q(\lambda a,\beta,\gamma) \in C$, we can find a quadratic
form $Q(1,\beta,*) \in C_e$ (by using the equivalence
$Q(1,\beta',*)\sim Q(1,\beta'-2n,*)$). Then the composition
becomes $Q(\lambda a,\beta,\gamma)*Q(1,\beta,*)=Q(\lambda
a,\beta,\gamma)$, which implies $C*C_e=C$.

\noindent 4) Let $Q(a,b,c)$ with $ac\not=0$ be a quadratic form in
a primitive class $C$. (It is not hard to see that we can choose $Q(a,b,c)$
with $ac\not=0$. In fact, this is obvious if $D<0$.)
Then, since $\sigma\,Q(a,b,c)=Q(a,-b,c) \sim Q(c,b,a)$,
we have $[\sigma\,
Q(a,b,c)*Q(a,b,c)]=[Q(c,b,a)*Q(a,b,c)]=[Q(ac,b,1)]$. Depending on
the congruence of $D$ mod $4$, $Q(ac,b,1)\sim Q(1,-b,ac)$ is
equivalent either to $Q(1,0,-D/4)$ or to $Q(1,1,(1-D)/4)$. This
implies $\sigma\,C*C=C_e$. \rqed
\enddemo

\vfill\eject
\head
{\bf Appendix B. $O(A_\Gamma)$ for a primitive lattice $\Gamma$ }
\endhead

\def\App{B}
\global\propno=1
\global\eqnum=1

\vskip0.5cm

For a primitive lattice $\Gamma \in \Cal L^p(D)$, the group
$O(A_\Gamma)$, by definition,
consists of all isometries of the discriminant $A_\Gamma$. Here we prove
the following property:

\ProclaimApp{Proposition}{\propBI}
If $\Gamma \in \Cal L^p(D)$, then any element $\varphi \in O(A_\Gamma)$
satisfies $\varphi^2=id$.
\endproclaim

\demo{Proof} First note that, for a primitive lattice $\Gamma$,
the discriminant group $\Gamma^*/\Gamma$ is isomorphic to either
$\bold Z_{|D|}$ or
$\bold Z_2 \oplus \bold Z_{2d}$ with $2d=|D|/2$.
In the first case, the claim holds since we have the decomposition
$$
O(A_{\bold Z_{|D|}})=O(A_{\bold Z_{p_1^{e_1}}})\times \cdots \times
O(A_{\bold Z_{p_k^{e_k}}}) \;, \;\; (|D|=p_1^{e_1}\cdots p_k^{e_k}),
$$
and $O(A_{\bold Z_{p_i^{e_i}}})=\{ \pm 1 \}$ for each
prime factor $p_i$ of $D$.
In the second case, we have a group decomposition
$$
\Gamma^*/\Gamma \cong \bold Z_2 \oplus \bold Z_{2d} \cong
\bold Z_2 \oplus \bold Z_{2^l} \oplus \bold Z_{2s+1} \quad ,
$$
by writing $d=2^{l-1}(2s+1) \,(l\geq 1)$. More precisely, if we denote
$\bold Z_2=\langle u_2 \rangle$ and
$\bold Z_{2d}=\langle u_{2d} \rangle$, then $u:=u_2, \, v:=(2s+1)u_{2d}, \,
v':=2^l u_{2d}$ generate each group, $\bold Z_2, \,\bold Z_{2^l}, \,
\bold Z_{2s+1}$, respectively. Since $2^l(x,y)\equiv (2s+1)(x,y)\equiv 0$
$\text{mod }\bold Z$ holds for $x \in \bold Z_2 \oplus \bold Z_{2^l}\,, \,
y\in \bold Z_{2s+1}$ and $(2^l,(2s+1))=1$, we see $(x,y)\equiv 0 \,
\text{mod }\bold Z$ and have the following orthogonal decomposition:
$$
(\Gamma^*/\Gamma,q_\Gamma) \cong
( \bold Z_2\oplus \bold Z_{2^l} \oplus \bold Z_{2s+1},
q_{\bold Z_2\oplus \bold Z_{2^l}}\oplus q_{\bold Z_{2s+1}} ) \;\;,
\EqnoApp{\decompDis}
$$
hence $O(A_\Gamma)\cong O(A_{\bold Z_2\oplus \bold Z_{2^l}}) \times
O(A_{\bold Z_{2s+1}})$. Thus we only need to consider
$O(A_{\bold Z_2\oplus \bold Z_{2^l}})$.

Let $u$, $v$ be the respective generators of $\bold Z_2$ and $\bold Z_{2^l}$
as above. Then an element $\varphi \in
O(A_{\bold Z_2\oplus \bold Z_{2^l}})$ can be represented by
$$
\varphi: (u,v) \mapsto (u',v')=(u,v)
\left( \matrix \alpha & \beta \\
               \gamma &\delta \\ \endmatrix \right) \;\;.
$$
Since $2u'\equiv 0, 2^l v' \equiv 0$, it follows that
$\alpha=0,1$, $\beta=0,1$, $\gamma=0,
2^{l-1}$. Also $\delta$ must be zero or an odd integer.
First, note that if a matrix
$ \left( \smallmatrix 1 & 0 \\
                0 & \delta \\ \endsmallmatrix \right)$ represents an isometry
$\varphi \in O(A_{\bold Z_2\oplus \bold Z_{2^l}})$, then $\delta$
must be $\pm 1$. In the following, we condider two cases, $l \geq 2$
and $l=1$, seperately.

\noindent
$\bullet \;$ $l\geq 2$:
In this case, there are five possibilities for $\varphi$:
$$
\left( \matrix 1 & 0 \\
               0 & \pm 1 \\ \endmatrix \right) \;\;,\;\;
\left( \matrix 1 & 0 \\
               2^{l-1} & k \\ \endmatrix \right) \;\;,\;\;
\left( \matrix 1 & 1 \\
                0 & k \\ \endmatrix \right) \;\;,\;\;
\left( \matrix 1 & 1 \\
               2^{l-1} & k \\ \endmatrix \right)\;\;,
\left( \matrix 0 & 1 \\
               2^{l-1} & k \\ \endmatrix \right) \;\;,\;\;
\EqnoApp{\grTwo}
$$
where $k$ is odd.
First of all, we may exclude the last (the fifth) isometry from our
consideration. To see this, write the relations between the generators:
$u'=2^{l-1}v, \, v'=u+k v$. Then we have $2^{l-1}v'=k 2^{l-1} v
=2^{l-1}v$ and also $u'=k u'=k 2^{l-1}v=2^{l-1}v$. Hence we have
$u'=2^{l-1}v'$, which cannot be for the generators. Thus the fifth
isometry will never appear for $l\geq2$.

For the first four possible isometries, we
verify that each one is an involution. This is easy for the
first three cases, and proceeds as follows for the fourth case:
We have
$$
\left( \matrix 1 & 1 \\
               2^{l-1} & k \\ \endmatrix \right)^2
=
\left( \matrix 1 +2^{l-1} & k+1 \\
               2^{l-1}(k+1) & 2^{l-1}+k^2 \\ \endmatrix \right)
\equiv
\left( \matrix 1 & 0 \\
               0 & 2^{l-1}+k^2 \\ \endmatrix \right) \;\;.
$$
Since an isometry of the form
$\left( \smallmatrix 1 & 0 \\
               0 & m \\ \endsmallmatrix \right)$ exists only for $m=\pm1$,
we obtain $2^{l-1}+k^2=\pm1$. But $2^{l-1}+k^2\not\equiv-1~mod~2^l$
for $k$ is odd and $l\geq3$. This shows that $\varphi^2=id$ for $l\geq 3$.
When $l=2$, the isometry takes the following form:
$\varphi_{\pm}=\left(
\smallmatrix 1 & 1 \cr
             2 & \pm 1 \cr
\endsmallmatrix \right)$.
Now write the discriminant form
$q_2=\left( \smallmatrix
\frac{a}{2} & \frac{b}{2} \cr
\frac{b}{2} & \frac{c}{4} \cr
\endsmallmatrix \right)$
with respect to the generators $u, v$, and assume
that $q_2$ allows the isometry $\varphi_\pm$. Then we see that
the discriminat form $q_2$ is restricted to be
$$
q_2=\left( \matrix 1 & \frac{1}{2} \cr
 \frac{1}{2} & \frac{c}{4} \cr \endmatrix \right) \quad,\quad
(\frac{c}{4}=0,\frac{1}{2},1,\frac{3}{2})\;.
$$
This shows that we have for the generators;
$(u,u)\equiv 1 \text{ mod } 2\bold Z$,
$(u,v)\equiv \frac{1}{2} \text{ mod } \bold Z$,
$(v,v)\equiv \frac{c}{4} \text{ mod } 2 \bold Z$.
Note that $l_1:=2u, l_2:= 4v$ are elements in $\Gamma$, and in fact,
give a basis of $\Gamma$. (Precisely, $l_1, l_2$ are lattice vectors
chosen, respectively, from $2(u+\Gamma), 4(v+\Gamma)$.)
Then the following symmetric matrix represents the bilinear form
of the lattice $\Gamma$:
$$
\left( \matrix (l_1,l_1) & (l_1,l_2) \cr
(l_2,l_1) & (l_2, l_2) \cr \endmatrix \right) =
\left( \matrix 4 a_0 & 4b_0 \cr
4 b_0 & 8 c_0 \cr \endmatrix \right) \,
$$
where $a_0,b_0,c_0 \in \bold Z$. This shows that
$\Gamma$ is not primitive, since
the corresponding quadratic form becomes $Q(2a_0,4b_0,4c_0)$.
Therefore $q_2$ does not appear from a primitive lattice $\Gamma$,
and hence $\varphi_\pm$ may be excluded.

\noindent
$\bullet$ $l=1$:
In this case, there are six possibilities for $\varphi$:
$$
\left( \matrix 1 & 0 \cr 0 & 1 \cr \endmatrix \right) \;\;,\;\;
\left( \matrix 1 & 1 \cr 0 & 1 \cr \endmatrix \right) \;\;,\;\;
\left( \matrix 1 & 0 \cr 1 & 1 \cr \endmatrix \right) \;\;,\;\;
\left( \matrix 0 & 1 \cr 1 & 0 \cr \endmatrix \right) \;\;,\;\;
\left( \matrix 1 & 1 \cr 1 & 0 \cr \endmatrix \right) \;\;,\;\;
\left( \matrix 0 & 1 \cr 1 & 1 \cr \endmatrix \right) \;\;.
\EqnoApp{\isoTwoTwo}
$$
We verify $\varphi^2=id$ for the first four. For the rest,
$\varphi^2=id$ does not hold, but we may exclude these by a similar
argument above done for $l=2$: As above assume the following possible
forms for the discriminant form $q_{\bold Z_2 \times \bold Z_2}$:
$$
q=\left( \matrix \frac{a}{2} & 0 \cr 0 & \frac{b}{2} \cr \endmatrix \right)
\;(a,b=1,2,3) \;\;,\;\;
q'=\left( \matrix \frac{a}{2} & \frac{1}{2} \cr
                   \frac{1}{2} & \frac{b}{2} \cr \endmatrix \right)
\;(a,b=0,1,2,3) ,
$$
with respect to the generators $u$ and $v$.
Then it is straightforward to see that
a non-involutive isometry (i.e. the fifth or sixth of (\isoTwoTwo))
is  possible only for the discriminant form
$q_1:=\left( \matrix 1 & \frac{1}{2} \cr
\frac{1}{2} & 1 \cr \endmatrix \right)$.  We may claim that this discriminant
form $q_1$ never appears from a primitive lattice $\Gamma$ in the exactly
same way as above for $l=2$.

This completes the proof. \rqed
\enddemo

Note that for non-primitive lattices, the index $d_1$ in
$\Gamma^*/\Gamma \cong \bold Z_{d_1} \oplus \bold Z_{d_2} \; (d_1|d_2)$ can be
greater than two in general.
Even if we have $d_1=2$, the above proof shows that
$\varphi_\pm$ for $l=2$ case or the last two cases of (\isoTwoTwo) for $l=1$
can be possible for non-primitive lattices.  The latter case
has appeared in the example presented at the end of section (5-3).

\vfill\eject

\head
{\bf Appendix C. The coset space $O(d,\bold R)\times O(d,\bold R)\setminus
O(d,d;\bold R)$ and $O(d,d;\bold Z)$ }
\endhead

\def\App{C}
\global\propno=1
\global\eqnum=1

\vskip0.5cm

Here we will prove Proposition {\cosetpara}, and
claim 1) of Proposition {\TdualG}.

\vskip0.3cm

\noindent
{\it Proof of Proposition {\cosetpara}.}  Let $W'\in O'(d,d;\bold R)$,
and write the corresponding ``half conjugated matrix''
$\widetilde W$ as
$$
\widetilde W:=
\frac{1}{\sqrt{2}}
\left( \matrix \bold 1_d & \bold 1_d \cr
        \bold 1_d & - \bold 1_d \cr \endmatrix \right) W'
= \left( \matrix
X  &  X'  \cr
Y  &  Y'  \cr \endmatrix \right) .
$$
Then the condition that $W' \in O'(d,d;\bold R)$ becomes
$$
\;\; \,^tX X-\,^tY Y= \bold 0_d \;\;, \;\;
\,^tX' X-\,^tY' Y= \bold 1_d \;\;, \;\;
\,^tX' X'-\,^tY' Y'= \bold 0_d ,
\EqnoApp{\CeqI}
$$
where $\bold 0_d$ is the $d\times d$ zero matrix. Here we conclude that
$\text{det }X\not=0$. To see this, assume that the real symmetric
matrix $\,^tXX=\,^tYY$ has a (real) eigenvector $\bold v (\not=\bold 0)$
with zero eigenvalue. Then we have $\,^tXX\bold v= \,^tY Y \bold v
=\bold 0$, from which we have $||X\bold v||^2=||Y\bold v||^2=0$, i.e.
$X\bold v=Y\bold v=\bold 0$. However, by the second equation of (\CeqI),
we have $\bold v=(\,^t X'X-\,^t Y'Y)\bold v=\bold 0$, which
is a contradiction. Therefore we conclude that all eigenvalues are not
zero, which means $\text{det }X\not=0$, $\text{det }Y\not=0$.

Now consider the $O(d,\bold R)\times O(d,\bold R)$-orbit of the matrix $W'$:
$$
\frac{1}{2}
\left( \matrix \bold 1_d & \bold 1_d \cr
        \bold 1_d & - \bold 1_d \cr \endmatrix \right)
\left( \matrix g_L &  0 \cr
             0 & g_R \cr \endmatrix \right)
\left( \matrix \bold 1_d & \bold 1_d \cr
        \bold 1_d & - \bold 1_d \cr \endmatrix \right) W'
=
\frac{1}{\sqrt{2}}
\left( \matrix g_L X+g_R Y & g_L X' + g_R Y' \cr
              g_L X-g_R Y & g_L X' - g_R Y' \cr
  \endmatrix \right).
\EqnoApp{\gaugeI}
$$
Since $\text{det }X\not=0$, $g_R YX^{-1}$ makes sense. It's
easy to check that it lies in $O(d;\bold R)$. Thus by choosing
$g_L=g_R Y X^{-1}$,
we get $g_L X-g_R Y=\bold 0_d$, and
(\gaugeI) becomes
$$
\left( \matrix g & 0 \cr 0 & g \cr \endmatrix \right)
\frac{1}{\sqrt{2}}
\left( \matrix 2 Y  & YX^{-1}X'+Y' \cr
         0 &  YX^{-1}X'-Y' \cr
 \endmatrix \right) \;,
$$
where $g:=g_R$. Put $\Lambda:=\frac{1}{\sqrt{2}}(YX^{-1}X'-Y')$. By
(\CeqI),  we find that $\sqrt{2}Y = \,^t\Lambda^{-1}$ and that
$\,^t\Lambda \frac{1}{\sqrt{2}}( YX^{-1}X'+Y')$ $=-
\frac{1}{2}(\,^tY'YX^{-1}X'-\,^tX' \,^t X^{-1} \,^t YY')=:-B$
is antisymmetric. This shows that each orbit has the shape
$G_{diag}\cdot W'(\Lambda,B)$ as claimed.

Conversely, it is easy to check that each $G_{diag}\cdot W'(\Lambda,B)$
$(\Lambda \in GL(d,\bold R), B \in A(d,\bold R))$ is a
$O(d,\bold R)\times O(d,\bold R)$-orbit.
This proves the first assertion.

The last assertion of Proposition (\cosetpara) is straightforward.
\rqed

\vskip0.3cm

We now describe generators of the discrete
group $O(2,2;\bold Z)$, as in Proposition {\TdualG}.
In general, doing the same for
$O(d,d;\bold Z)$  $(d \geq3)$ is much harder.
However for $d=2$, we have the following nice description of $\bold R^{2,2}$.
Let $Mat_{2,2}(\bold R)$ be the space of $2\times 2$ real matrices,
equipped with
the quadratic form $||*||:=\text{det}(*)$. Its associated bilinear form
has the signature $(2,2)$, and therefore $Mat_{2,2}(\bold R)
\cong \bold R^{2,2}$ as quadratic spaces.
Consider an embedding of the hyperbolic lattice $U^{\oplus 2}$ defined by
$$
\matrix
\Phi_0: &\; U^{\oplus 2} & \hookrightarrow  & Mat_{2,2}(\bold R)
\cr
& e_1,e_2,f_1,f_2 & \mapsto & E_1,E_2,F_1,F_2 \;\;,\cr
\endmatrix
\EqnoApp{\embPhi}
$$
where
$$
E_1=\sqrt{2} \left(\matrix 1 & 0 \cr 0 & 0 \cr \endmatrix \right) \;,\;
E_2=\sqrt{2} \left(\matrix 0 & 1 \cr 0 & 0 \cr \endmatrix \right) \;,\;
F_1=\sqrt{2} \left(\matrix 0 & 0 \cr 0 & 1 \cr \endmatrix \right) \;,\;
F_2=\sqrt{2} \left(\matrix 0 & 0 \cr -1 & 0 \cr \endmatrix \right) \;.
$$
The embedding is isometric since
$\Phi_0(x_1e_1+x_2e_2+y_1f_1+y_2f_2)=\sqrt{2}\left( \smallmatrix
x_1 & x_2 \cr -y_2 & y_1 \cr \endsmallmatrix \right)$ has determinant
$2x_1y_1+2x_2y_2$.  We denote by
$\Phi_0^{\bold R}:U^{\oplus 2}\otimes \bold R=
\bold R^{2,2}\cong M_{2,2}(\bold R)$
the real scalar extension of $\Phi_0$.

\ProclaimApp{Proposition}{\appCI}
1) The linear action $X \mapsto g X h^{-1}$ of
$(g,h) \in SL_2\bold R \times SL_2 \bold R$ on $X \in Mat_{2,2}(\bold R)$
defines a group homomorphism
$$
\phi_{\bold R}: SL_2\bold R \times SL_2\bold R
\rightarrow O'(2,2;\bold R) \cong O(2,2;\bold R) \;\;.
$$
2) $\phi_{\bold R}$ maps surjectively onto $O'_0(2,2;\bold R)$, the
connected component of the identity, and has the kernel
$\{ (\bold 1_2,\bold 1_2),(-\bold 1_2,-\bold 1_2) \}$.
\endproclaim

\demo{Sketch of Proof} It is easy to verify claim 1). For 2),
we observe that the differential $d \phi_{\bold R}$ gives an
isomorphism of the Lie algebras, and therefore the image of $\phi_{\bold R}$
is given by the connected component of the identity. \rqed
\enddemo

\ProclaimApp{Proposition}{\appCII} The restriction of
$\phi_{\bold R}$ to $SL_2\bold Z \times SL_2 \bold Z$ defines a surjective
group homomorphism:
$$
\phi_{\bold Z}: SL_2 \bold Z \times SL_2\bold Z \;\; \rightarrow \;\;
O(2,2;\bold Z) \cap O'_0(2,2;\bold R) \;,
$$
with $\text{Ker}(\phi_{\bold Z})=
\{ (\bold 1_2,\bold 1_2),(-\bold 1_2,-\bold 1_2) \}$.
\endproclaim

\demo{Proof} It is clear that the image of $\phi_{\bold Z}$
lies in $O(2,2;\bold Z) \cap O'_0(2,2;\bold R)$ by Proposition {\appCI}.
To prove
surjectivity, let us solve for
$g=\left( \smallmatrix a & b \cr c & d \cr \endsmallmatrix \right),
h=\left( \smallmatrix a' & b' \cr c' & d' \cr
\endsmallmatrix \right)\in SL_2\bold Z$ in
$$
\phi_{\bold R}(g,h)
= \left( \matrix
ad' & -ac' & -bc' & -bd' \cr
-ab' & aa' & ba' & bb' \cr
-cb' & ca' & da' & db' \cr
-cd' & cc' & dc' & dd' \cr \endmatrix \right)=S,
\EqnoApp{\abcd}
$$
for any given $S \in O(2,2;\bold Z) \cap O'_0(2,2;\bold R)$.
By Proposition {\appCI}, there are solutions $g,h\in SL_2\bold R$.
To prove surjectivity, we will
show that they are integral. First note that
$
a^2=\det{S_{12;12}},\,
b^2=\det{S_{12;34}},\,
c^2=\det{S_{34;12}},\,
d^2=\det{S_{34;34}};\,
{a'}^2=\det{S_{23;23}},\,
{b'}^2=-\det{S_{23;14}},\,
{c'}^2=-\det{S_{14;23}},\,
{d'}^2=\det{S_{14;14}}$, where $S_{ij;kl}$ represents the minor made by
the $i,j$-th rows and the $k,l$-th columns. From this, we see that
the square of each variable $a,b,...$ is integer.
So we can write $a=x_a \sqrt{r_a}$, with some integer $x_a$
and a square free positive integer $r_a$.
Likewise, $b=x_b\sqrt{r_b}, c=x_c \sqrt{r_c}$
and so on. The 16 components of (\abcd) are required to be integers. For
example, we have $ad'=x_a x_{d'} \sqrt{r_a r_{d'}} \in \bold Z$. The
last expression becomes integer only if $r_a=r_{d'}$. Likewise
we have $r_a=r_b=\cdots=r_{d'}=:r$. Therefore we have,
$$
a,b,c,d,a',b',c',d' \in \sqrt{r} \bold Z \;,
$$
for some square free integer $r$. Since $ad-bc=a'd'-b'c'=1$, we see that
the only possibility is $r=1$, which means that $g,h\in SL_2\bold Z$. Thus
$\phi_{\bold Z}$ is surjective. It is easy to determine the kernel of
$\phi_{\bold Z}$. \rqed
\enddemo

Recall that $SL_2\bold Z$ is generated by
$
\bold S =\left( \smallmatrix 0 & -1 \cr 1 & 0 \cr \endsmallmatrix \right),
\bold T =\left( \smallmatrix 1 & -1 \cr 0 & 1 \cr \endsmallmatrix \right)
$. It is easy to see that $\phi_{\bold Z}$ maps
$
(\bold 1_2,\bold S),
(\bold 1_2,\bold T),
(\bold S,\bold 1_2),
(\bold T,\bold 1_2)$, respectively, to the matrices
$S_1,T_1,S_2,T_2$ given in 2) of Proposition {\TdualG}.

The two generators $R_1$ and $R_2$ are related to the other connected
components of $O'(2,2;\bold R)$. In fact, $O'(2,2;\bold R)$ consists
of the following four connected components:
$$
O'(2,2;\bold R)=
O_{+,+}'(2,2;\bold R) \sqcup
O_{+,-}'(2,2;\bold R) \sqcup
O_{-,+}'(2,2;\bold R) \sqcup
O_{-,-}'(2,2;\bold R) \; ,
$$
where, for example, $O_{-,+}'(2,2;\bold R)$ consists of those elements
which reverse the orientation of a positive definite two plane and preserve
that of a negative definite two plane. (Note that if $g$ reverses
the orientation of a single positive definite two plane, then it does
so for every
positive two plane. Similarly for negative definite two planes. Likewise
if $g$ preserves the orientation of a two plane.) Obviously,
$O_0'(2,2;\bold R)=O_{+,+}'(2,2;\bold R)$. Now giving the orientations
$(e_1+f_1)\wedge(e_2+f_2)$ and $(e_1-f_1)\wedge(e_2-f_2)$, respectively,
for positive and negative definite two planes, we see that $R_1$ belongs to
$O_{-,+}'(2,2;\bold R) \cap O(2,2;\bold Z)$ and $R_2$ belongs to
$O_{-,-}'(2,2;\bold R) \cap O(2,2;\bold Z)$. It is now clear that
we have the following decomposition of $O(2,2;\bold Z)$;
$$
O(2,2;\bold Z)=
O_{+,+}(2,2;\bold Z) \sqcup
O_{+,-}(2,2;\bold Z) \sqcup
O_{-,+}(2,2;\bold Z) \sqcup
O_{-,-}(2,2;\bold Z) \; ,
\EqnoApp{\Zdecomp}
$$
with $O_{+,+}(2,2;\bold Z)=O'_0(2,2;\bold R)\cap O(2,2;\bold Z)$ and
$O_{+,-}(2,2;\bold Z)=O_{+,+}(2,2;\bold Z)R_1R_2$,
$O_{-,+}(2,2;\bold Z)=O_{+,+}(2,2;\bold Z)R_1$,
$O_{+,-}(2,2;\bold Z)=O_{+,+}(2,2;\bold Z)R_2$. In particular, this yields
property 1) of Proposition {\TdualG}.

\vfill\eject


\vfill\eject

\def\fsize#1{{\eightpoint #1 }}
\def\efsize#1{{\eightpoint#1 }}
\def\spc{\hskip1cm}

\vskip1cm
\settabs 4 \columns
\+ \spc \fsize{Shinobu Hosono}         && \fsize{Bong H. Lian}  \cr
\+ \spc \fsize{Graduate School of }   &&\fsize{Department of mathematics} \cr
\+ \spc \fsize{Mathematical Sciences}       && \fsize{Brandeis University}  \cr
\+ \spc \fsize{University of Tokyo}      && \fsize{Waltham, MA 02154}  \cr
\+ \spc \fsize{Komaba 3-8-1, Meguroku}   && \fsize{U.S.A.}  \cr
\+ \spc \fsize{Tokyo 153-8914, Japan}    &&
                              \efsize{Email:}\efsize{lian\@brandeis.edu} \cr
\+ \spc \efsize{Email:}\efsize{hosono\@ms.u-tokyo.ac.jp}
                  &&  \cr

\vskip0.3cm
\settabs 4 \columns
\+ \spc \fsize{Keiji Oguiso}         & & \fsize{Shing-Tung Yau}  \cr
\+ \spc\fsize{Graduate School of }&&\fsize{Department of mathematics} \cr
\+ \spc \fsize{Mathematical Sciences}       && \fsize{Harvard University}  \cr
\+ \spc \fsize{University of Tokyo}      && \fsize{Cambridge, MA 02138}  \cr
\+ \spc \fsize{Komaba 3-8-1, Meguroku}   && \fsize{U.S.A.}   \cr
\+ \spc \fsize{Tokyo 153-8914, Japan}    &&
                  \efsize{Email:}\efsize{yau\@math.harvard.edu} \cr
\+ \spc \efsize{Email:}\efsize{oguiso\@ms.u-tokyo.ac.jp}
                  &&   \cr

\bye

\noindent
{\bf Example.} ($d=1$) We continue the previous example; $\varphi:
U \hookrightarrow \bold R^{1,1}$. For simplicity, let us define
$$
\bold e_1:=\varphi(e)={1\over\sqrt{2}}\;^t({1\over R},{1\over R})\;\;,\;\;
\bold e_2:=\varphi(f)={1\over\sqrt{2}}\;^t(R,-R) \;.
$$
Then the basis $\,^t(1,0), \,^t(0,1)$ of $\bold R^{1,0}$ and $\bold R^{0,1}$,
respectively, are related to $\bold e_1$ and $\bold e_2$ as follows:
$$
\sqrt{2}R\,^t(1,0)=R^2 \bold e_1 + \bold e_2 \;\;,\;\;
\sqrt{2}R\,^t(0,1)=R^2 \bold e_1 - \bold e_2 \;.
$$
In this form it is clear that $\bold R^{1,0}$ and $\bold R^{0,1}$ have
non-trivial intersections with $\Gamma(\varphi)=\bold Z \bold e_1 + \bold Z
\bold e_2$ iff $R^2=p/q$ for coprime positive integers. In that case
we have
$$
\Gamma_L=\bold Z (p \bold e_1 + q \bold e_2) \cong \langle 2pq \rangle
\;\;,\;\;
\Gamma_R=\bold Z (p \bold e_1 - q \bold e_2) \cong \langle -2pq \rangle .
$$

\noindent
{\bf Example.} ($d=1$) I continue the example of the Narain lattice
$\varphi_R: U \hookrightarrow \bold R^{1,1}$. The lattice $\Gamma(\varphi_R)$
defines a RCFT iff $R^2=p/q$. In this case we have
$$
\langle 2pq \rangle \oplus \langle -2pq \rangle \subset
\Gamma(\varphi_R) \subset
\langle 2pq \rangle^* \oplus \langle -2pq \rangle^* \;\;.
$$
If we write this as
$$
\langle 2n \rangle \oplus \langle -2n \rangle \subset
\Gamma(\varphi) \subset
\langle 2n \rangle^* \oplus \langle -2n\rangle^* \;\;,
$$
then the number of inequivalent even unimodular over-lattices is
$$
|O(A_{\bold Z/2n})|/2 = 2^{p(n)} \;\;.
$$
Physicists say that form the chiral algebras
$\Cal A_L (\langle 2n\rangle)$ and $\Cal A_R (\langle 2n \rangle)$
we obtain $2^{p(n)}$ non-isomorphic RCFT.

\bye